\documentclass[sigconf]{acmart}

\usepackage{subcaption}
\usepackage{booktabs}
\usepackage{geometry}
\usepackage{longtable}
\usepackage{caption}
\usepackage[utf8]{inputenc}
\usepackage{multicol}
\usepackage{makecell}    % For handling multi-line cells (optional)
\usepackage{amsmath}     % For mathematical formatting
\usepackage{float}
\usepackage{tabularx}
\usepackage{multirow} % For multi-column cells
\usepackage{wrapfig}
\usepackage{siunitx} 
\usepackage{makecell}  % only if you still need it
\usepackage{amsmath}
\usepackage{amssymb}% Optional: for aligning numbers in columns

\AtBeginDocument{%
  }

\begin{document}

\title{Exploring Language Patterns of Prompts in Text-to-Image Generation and Their Impact on Visual Diversity}


\renewcommand{\shortauthors}{tr. et al}
\author{Maria-Teresa De Rosa Palmini}
\affiliation{%
  \institution{University of Zurich}
  \city{Zurich}
  \country{Switzerland}}
\email{maria-teresa.derosa-palmini@uzh.ch}

\author{Eva Cetinic}
\affiliation{%
  \institution{University of Zurich}
  \city{Zurich}
  \country{Switzerland}}
\email{eva.cetinic@uzh.ch}

\settopmatter{printacmref=false} % Removes citation information below abstract
\renewcommand\footnotetextcopyrightpermission[1]{} % Removes footnote with conference info
\pagestyle{plain} % Removes running headers

\begin{abstract}

Following the initial excitement, Text-to-Image (TTI) models are now being examined more critically. While much of the discourse has focused on biases and stereotypes embedded in large-scale training datasets, the sociotechnical dynamics of user interactions with these models remain underexplored. This study examines the linguistic and semantic choices users make when crafting prompts and how these choices influence the diversity of generated outputs. Analyzing over six million prompts from the Civiverse dataset on the CivitAI platform across seven months, we categorize users into three groups based on their levels of linguistic experimentation: \emph{consistent repeaters}, \emph{occasional repeaters}, and \emph{non-repeaters}. Our findings reveal that as user participation grows over time, prompt language becomes increasingly homogenized through the adoption of popular community tags and descriptors, with repeated prompts comprising 40\%–50\% of submissions. At the same time, semantic similarity and topic preferences remain relatively stable, emphasizing common subjects and surface aesthetics. Using Vendi scores to quantify visual diversity, we demonstrate a clear correlation between lexical similarity in prompts and the visual similarity of generated images, showing that linguistic repetition reinforces less diverse representations. These findings highlight the significant role of user-driven factors in shaping AI-generated imagery, beyond inherent model biases, and underscore the need for tools and practices that encourage greater linguistic and thematic experimentation within TTI systems to foster more inclusive and diverse AI-generated content.

\end{abstract}




\maketitle

\section{Introduction}

\begin{figure*}[htbp]
    \centering
    \includegraphics[width=0.7\linewidth, trim=10 0 0 0, clip]{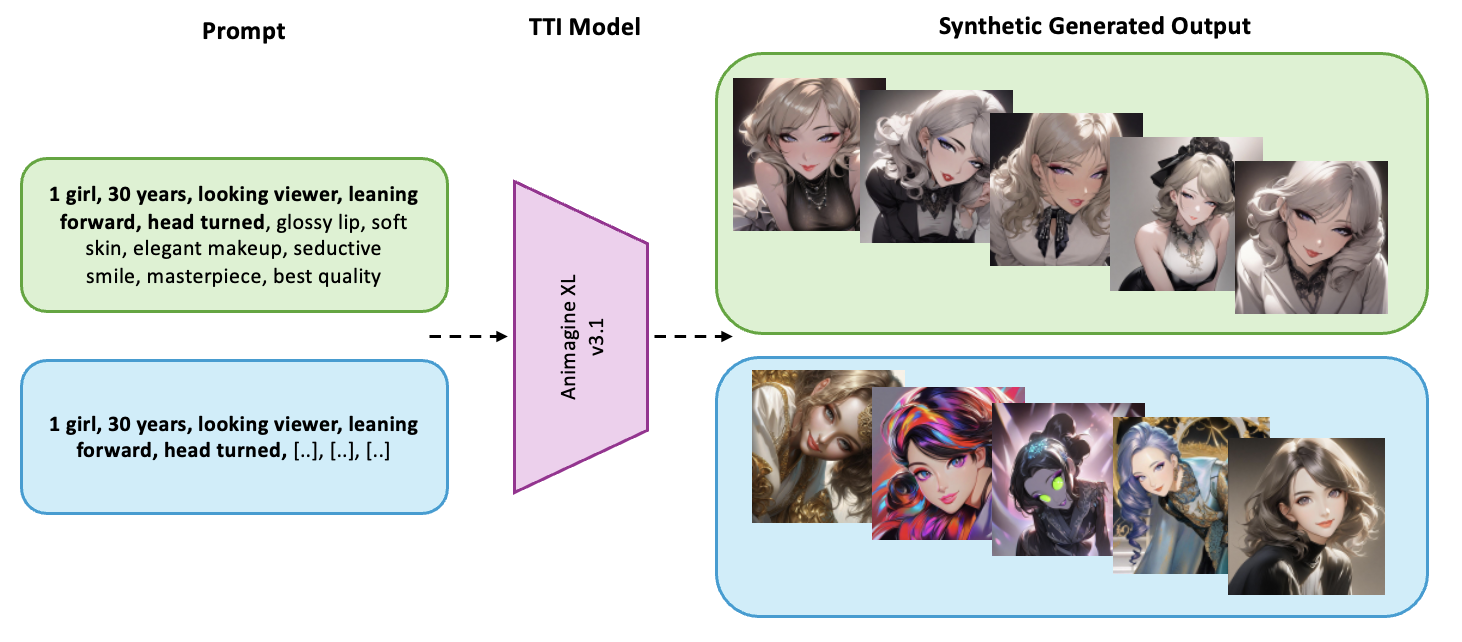} % Adjust width percentage
    \caption{Images from the Civiverse dataset, generated with the Animagine XL v3.1 model, using variations of a common prompt pattern: 1 girl, 30 years, looking at viewer, leaning forward, head turned, glossy lips, soft skin, elegant makeup, seductive smile, masterpiece, best quality. The first row shows images from prompts with 20 identical tokens, while the second row uses prompts with 9 identical tokens.}
    \label{fig:process} % Updated label for referencing
\end{figure*}

Recent advancements in Computer Vision and Natural Language Processing have significantly expanded the capabilities of generative models, particularly in text-guided image generation. Leading examples such as DALL-E 2 \cite{ramesh2022hierarchical} and Stable Diffusion \cite{rombach2022high} use sophisticated image-and-text embedding methods, most notably Contrastive Language and Image Pre-training (CLIP) \cite{radford2021learning}, to produce photorealistic and compelling images from textual prompts. The growing adoption of these Text-to-Image (TTI) models spans a variety of fields, from news media graphics \cite{liu2022opal} to product design \cite{ko2023large}, where they streamline and accelerate creative workflows.

Despite these technical successes, important questions have emerged regarding TTI's broader societal impact. For instance, large-scale training datasets often reflect Western-centric biases related to culture, race, and gender \cite{bianchi2023easily, cho2023dall}, sometimes yielding visually homogeneous outputs that lack diversity. Meanwhile, the practice of emulating recognized artistic styles or “style mimicry” has sparked debates about originality and ethics in digital art \cite{hoquet2023modern, shan2023glaze}. Additionally, some scholars argue that authorship is shifting, with prompt-makers' contributions often overlooked in favor of model developers \cite{wilde2023generative}.

A growing body of research also explores how \emph{users} interact with generative AI systems and how their choices can influence outputs. In TTI, for example, user prompt repositories like DiffusionDB \cite{wang2020minilm} and Civiverse \cite{palmini2024civiverse} reveal a lexicon of style tags, formulaic cues, and descriptors that users employ for targeted results. Overreliance on such “community-driven prompt formulas” can, however, restrict creative exploration and inadvertently marginalize alternative aesthetics or subjects \cite{mccormack2024no, sanchez2023examining}. A parallel trend is observed in Large Language Models (LLMs), where users often default to minimal prompts that yield repetitive or “safe” responses \cite{anderson2024homogenization}. This cyclical process, where effective prompts are shared, reused, and refined, can narrow aesthetic and lexical variety over time \cite{padmakumar2023does}.

Although evidence suggests that prompts tend to converge toward relatively restrictive patterns, it remains unclear whether this trend of linguistic uniformity might expand as TTI systems become adopted in broader and more diverse contexts. Currently, there is a lack of studies examining the large-scale evolution of prompts over an extended period, as well as those focusing on how similarities in prompt language translate into visual homogeneity and impact the creative potential of generated outputs. 

To address this gap, we analyze a seven-month record of user interactions from the Civiverse dataset \cite{palmini2024civiverse}, containing over six million prompts from the CivitAI platform. Our research uncovers how user prompting behavior evolves over time, revealing that increased participation leads to more homogenized prompt language. Additionally, we explore the implications of this trend on the visual diversity of generated imagery and reveal that increased lexical similarity in prompts correlates with reduced visual diversity in generated images. The paper is organized as follows: In Section ~\ref{sec:related_work}, we review the existing literature, focusing mainly on studies related to prompt analysis. Section ~\ref{sec:dataset_overview} describes the dataset used for our analysis. In Section ~\ref{sec:user_behavior} we provide an overview of our user engagement analysis and categorize users into three distinct groups based on their levels of linguistic experimentation. In Section \ref{ref:lexical_diversity_of_prompts}, we examine the evolution of lexical diversity in user prompts, detailing our methodology and findings on how language use among TTI users changes over time. Section~\ref{sec:semantic_diversity} shifts focus to semantic diversity, exploring how distinct concepts and themes in prompts evolve and vary across user categories. Building on these insights, Section~\ref{sec:visual_diversity} delves into visual diversity, analyzing the relationship between prompt language, both lexical and semantic, and the diversity of generated images, as well as the implications of dominant prompt templates on community norms and stylistic conventions.

Our results reveal that while the TTI user community is growing, the linguistic and thematic scope of prompts is contracting, leading to decreased visual diversity. Figure~\ref{fig:process} illustrates this finding, showing how images generated using variations of a lexically common prompt pattern demonstrate that even slight changes in linguistic structure affect visual consistency and aesthetics. This self-reinforcing cycle, driven by repeated tokens, stylistic phrases, and model refinements, often limits the emergence of novel styles and subjects. By examining the interplay between user inputs and model dynamics, we highlight the need to address both technical and social factors, such as diversifying training datasets and encouraging community-driven prompt practices, to foster generative AI systems that are fair, accountable, and conducive to creativity.

\section{Related Work} 
\label{sec:related_work}

\paragraph {Human Agency and Prompts in Generative AI} Human agency is widely regarded as central to creative processes, ensuring that innovation and aesthetic decisions remain in human hands rather than defaulting to automated or formulaic routines \cite{cetinic2022understanding, chiou2023designing, grba2024art, paananen2023using}. In the context of Generative AI and LLMs specifically, prompts play a pivotal role in preserving this agency, translating users' intent into textual instructions the model can interpret. Despite the technical sophistication of these systems, recent scholarship underscores that \emph{how} humans craft, iterate, and refine prompts profoundly influences the expressive qualities and novelty of generated outputs, whether text, images, or multimodal compositions \cite{grba2023renegade, audry2021art}. Consequently, prompts serve as a mechanism through which users can explore and shape the model’s latent space, exercising creative control over the output.

\paragraph{Prompt Diversity in LLMs} Much of the research on prompt diversity has centered on LLM-generated text. Several studies show that when users rely on minimalistic or template-based prompts, the resulting outputs may converge toward predictable patterns, limiting creative expression \cite{padmakumar2023does}. By contrast, even small changes to linguistic properties, such as word reordering or shifts in semantic focus, can substantially alter the generated text, highlighting the malleability of outputs under different prompt conditions \cite{moon2024homogenizing}. However, iterative prompt refinement can introduce new risks: frequent edits or additive instructions may lead to inconsistencies or unintended constraints on creativity \cite{tafreshipour2024prompting}. Prompt diversity research has also shed light on the cultural implications of homogenized writing. For example, users from non-Western backgrounds often adapt their style to align with Western-centric AI models, narrowing the scope of local cultural expression \cite{agarwal2024ai}. In response, several lines of inquiry have explored different strategies, ranging from paraphrasing to structural reordering to preserve output uniqueness, while tools such as compression ratios, self-repetition metrics, and self-BLEU/BERTScore facilitate a robust assessment of text diversity \cite{shaib2024standardizing}. These findings underscore the importance of prompt engineering to maintain both novelty and authenticity in LLM based applications.

\paragraph{Prompt Behavior in TTI Applications} Compared to text-focused LLM scenarios, fewer studies have explored how prompts influence TTI settings \cite{xie2023prompt, oppenlaender2023taxonomy}. Existing work typically examines the \emph{content} of user prompts at a single timepoint, revealing the prevalence of specific stylistic tokens, e.g., \emph{“8k,” “HDR,” “masterpiece”}, and domain‐specific descriptors such as “watercolor” or “macro lens” \cite{sanchez2023examining, wang2022diffusiondb}. Others have used topic modeling or cluster analysis (e.g., with CLIP or MiniLM embeddings) to map out major themes in prompts—portraits, fantasy worlds, cinematic scenes, and highlight the dominance of standard tokens designed to boost output resolution \cite{mccormack2024no, xie2023prompt}. Such convergence aids efficiency but raises concerns about the “erosion” of creative potential and the marginalization of underexplored topics \cite{sanchez2023examining}. Moreover, while guidelines on platforms like Stable Diffusion or DALL·E encourage detailed prompt construction, they can also codify formulaic practices and further homogenize image outputs \cite{xie2023prompt}.

\paragraph{Gaps in Understanding TTI Prompt Diversity}
While TTI researchers have made progress in mapping common stylistic tokens and thematically clustering prompts, most studies are limited to short-term data or single-timepoint snapshots. As a result, critical questions remain unanswered, such as \emph{how} TTI prompts evolve over time, what linguistic experimentation behaviors are prevalent among different user groups, and whether adherence to popular "best practices" ultimately reduces the visual diversity of outputs \cite{sanchez2023examining, xie2023prompt}. These challenges mirror concerns raised in LLM research, where certain user cohorts may unintentionally drive homogenization by over-relying on established prompt patterns \cite{anderson2024homogenization}. Furthermore, TTI studies have yet to adequately explore the cultural, linguistic, and exploratory dimensions of user-driven prompts, factors that significantly influence creativity, user agency, and the breadth of represented topics. Addressing these gaps is essential to better understand how user behaviors and community norms shape the potential and limitations of TTI systems.

\section{Dataset}
\label{sec:dataset_overview}

The Civiverse dataset \cite{palmini2024civiverse} consists of over 6.5 million images created by a database of 52,905 unique contributors from the CivitAI platform \cite{civitai}, where individuals collaboratively share and refine derivatives of Stable Diffusion models. Spanning October 2023 to April 2024, it collates metadata such as web URLs, image hashes, post IDs, timestamps, anonymized user identifiers, model information, and content ratings. By capturing both the frequency and evolution of user-submitted prompts, we enable the analysis of large-scale engagement and content-generation patterns over time.

All user prompts were standardized through a preprocessing pipeline. Emojis and special characters were removed, retaining only alphanumeric characters and underscores. The resulting text was then tokenized, with English stopwords removed using the NLTK library~\cite{bird2009natural}. Additionally, 3\% of the initial prompts were in languages other than English and were excluded prior to analysis. While this language limitation may affect the applicability of our findings to multilingual contexts, the small proportion of non-English prompts suggests a negligible impact on the overall analysis. A breakdown of the other languages present in the initial Civiverse dataset is provided in Appendix A. Finally, tokens were lemmatized with WordNetLemmatizer \cite{bird2009natural} to unify grammatical variants, yielding a uniform, content-rich dataset for further analysis.

To account for users iterating on the same prompt text multiple times, the dataset was divided into two subsets: one \textbf{with} duplicates (5,864,047 prompts) and one \textbf{without} duplicates (2,747,380 unique prompts). To ensure integrity, 8,000 identical images with matching timestamps were removed from both. We define duplicate prompts as those that match exactly, character-for-character, after cleaning and tokenization. The duplicate-inclusive subset captures how users refine or repeat prompts, while the unique-prompt subset provides a clearer view of linguistic and thematic diversity. All subsequent analyses in this paper rely on the unique-prompt subset, whereas the duplicate-inclusive subset is used exclusively to classify different user repetition behaviors.

\section{User Behavior Categorization}
\label{sec:user_behavior}

In this section, we examine the evolution of user engagement and repetition trends over time to assess whether TTI models encourage creative exploration or reinforce recurrent motifs.

In order to better understand the overall platform dynamics and engagement of users, we first analyse how the number of users and prompts changed over time. From October 2023 to April 2024, monthly prompting attempts increased from 356,368 to 1.7~million, while unique prompts rose from 213,779 to 764,838, in parallel with a growth in unique users from 8,320 to 21,129. Table~\ref{tab:monthly_growth} provides a detailed breakdown of these monthly figures, showing a steady increase in both user participation and the volume of submitted prompts. Notably, the proportion of duplicate prompts remained consistent at 40--50\% of all submissions across this period, indicating users' ongoing focus on refining or iterating upon specific ideas.

\begin{table}[htbp]
    \centering
    \small % Reduces font size
    \caption{Monthly Growth of Prompts and Users. The table summarizes the monthly count of unique and total prompts submitted, along with the number of unique users engaging with the Civiverse platform.}
    \label{tab:monthly_growth}
    \begin{tabular}{@{}lrrr@{}} % Removes extra spacing
        \toprule
        \textbf{Month} & \textbf{Unique Prompts} & \textbf{Total Prompts} & \textbf{Users} \\
        \midrule
        Oct 2023   & 213,779  & 356,368  & 8,320  \\
        Nov 2023   & 274,999  & 491,156  & 9,341  \\
        Dec 2023   & 309,649  & 585,146  & 10,552 \\
        Jan 2024   & 334,407  & 715,328  & 12,060 \\
        Feb 2024   & 362,012  & 806,377  & 12,771 \\
        Mar 2024   & 516,410  & 1,191,718 & 15,200 \\
        Apr 2024   & 764,838  & 1,716,634 & 21,129 \\
        \bottomrule
    \end{tabular}
\end{table}

\begin{figure*}[htbp]
    \centering
    \includegraphics[width=0.70\linewidth]{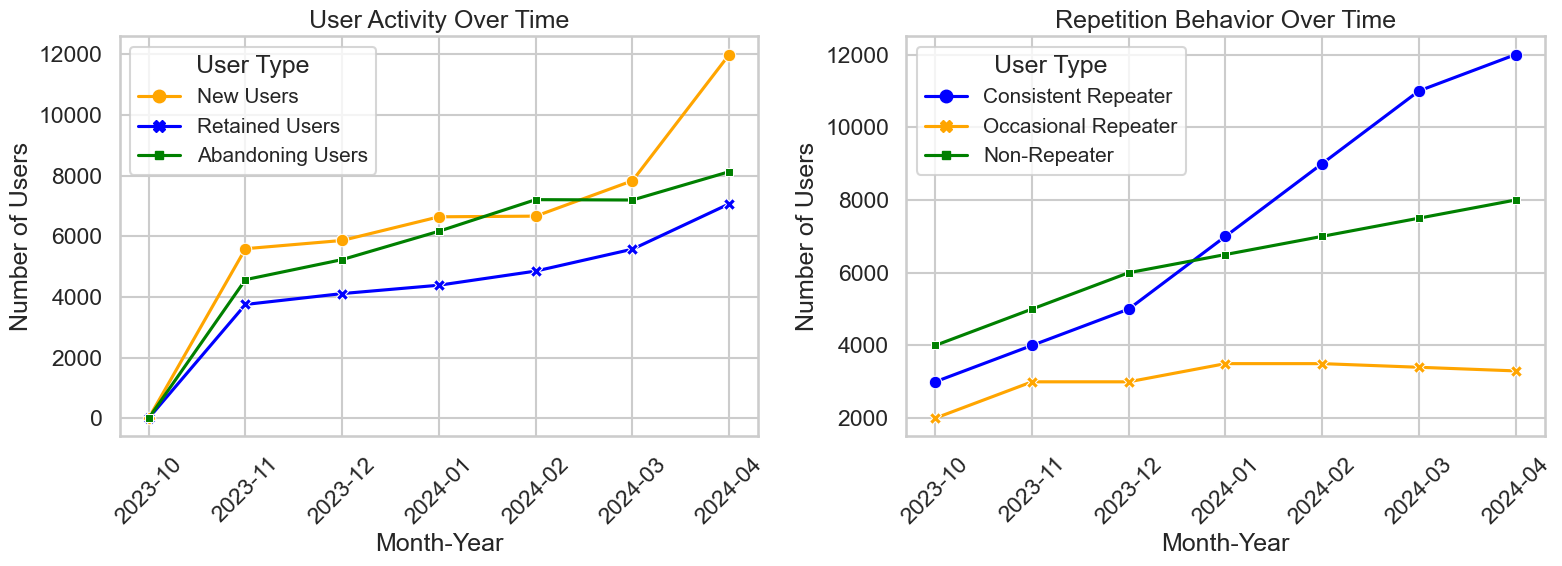}
    \caption{Monthly distribution of new, abandoning, and retained users (left) and monthly distribution of users based on their categorization into consistent, occasional or non-repeaters of prompts (right). }
    \label{fig:user_engagement}
\end{figure*}

To explore user engagement further, we classified users into three categories: \emph{new}, \emph{abandoning}, and \emph{retained}. User data was aggregated on a monthly basis, with sets of unique user IDs constructed based on the unique usernames in the Civiverse dataset. \textbf{New users} were defined as those whose unique usernames appeared for the first time in the current month when comparing sets of usernames sequentially on a month-to-month basis. Simultaneously, we identified \textbf{retained users} as those whose usernames appeared in both the current and preceding month, signifying ongoing engagement. In contrast, users who had been active in one month but failed to reappear in the subsequent month were categorized as \textbf{abandoning}. This sequential month-to-month analysis revealed that new users consistently represented the largest cohort, growing from 6,000 in November 2023 to 12,000 by April 2024, while abandoning users ranged between 4,000 and 7,000 monthly. Retained users, though constituting the smallest group, showed steady growth from 4,000 in November to 6,000 in April. Figure~\ref{fig:user_engagement} (left) visualizes the monthly distribution of new, abandoning, and retained users, highlighting how new users consistently dominate, while retained users steadily increase, demonstrating the platform’s growing engagement.

Additionally, categorized users into three distinct groups based on their level of linguistic experimentation and patterns in prompt submission throughout the overall dataset. Consistent repeaters are defined as users who consistently resubmitted identical prompts in every month they were active, exhibiting the lowest level of linguistic experimentation. This behavior resulted in nearly 2 million unique repeated prompts overall, averaging 76.6 repeats per user. In contrast, we define occasional repeaters as unique users who exhibited intermittent repetition behavior, reusing prompts in some months but not in others, with a reuse rate of 20\% or less and averaging 35.2 repeated prompts per user. Meanwhile, non-repeaters are defined as users who never resubmitted identical prompts, consistently favoring unique, one-off submissions or engaging in exploratory behavior with each prompt they created. As shown in Table \ref{tab:user_category_distribution}, consistent repeaters, while representing 47.36\% of the unique users, contributed a dominant 72.39\% of the total prompts, totaling 2,009,651 unique prompts. Occasional repeaters accounted for 13.40\% of the users and contributed 24.35\% of the prompts (676,046), whereas non-repeaters made up 39.23\% of the user base but only contributed 3.26\% of the prompts (90,521). Interestingly, despite consistent and non-repeaters together occupying approximately the same proportion of the user base, their contributions to the overall prompt activity are drastically disproportionate, highlighting how user behavior varies in terms of repetition versus exploration. Figure~\ref{fig:user_engagement} (right) illustrates the monthly distribution of each user category, showing how consistent repeaters steadily increased over time, while occasional repeaters and non-repeaters exhibited more stable or slower growth patterns (see Appendix B for exact numbers).

\begin{table}[htbp] % Standard table environment
    \centering
    \small % Reduce font size
    \caption{Distribution of user categories and their contributions to the unique user database and prompts.}
    \label{tab:user_category_distribution}
    \setlength{\tabcolsep}{3pt} % Reduce column padding
    \renewcommand{\arraystretch}{0.9} % Reduce row height
    \begin{tabular}{lrrrr}
        \toprule
        User Category & Unique Users & User \% & Total Prompts & Prompt \% \\
        \midrule
        Consistent    & 25,058  & 47.36 & 2,009,651 & 72.39 \\
        Non-Repeaters & 20,756  & 39.23 & 90,521    & 3.26  \\
        Occasional    & 7,091   & 13.40 & 676,046   & 24.35 \\
        \bottomrule
    \end{tabular}
\end{table}

\section{Lexical Diversity of Prompts}
\label{ref:lexical_diversity_of_prompts}

The vocabulary and phrasing in user-submitted prompts provide valuable insights into the evolution of language use among TTI users. This section outlines our approach to analyzing lexical diversity in the Civiverse dataset, focusing on whether prompts converge on a limited set of terms or expand to include a broader range of expressions. To capture various aspects of lexical variation and redundancy, we employed four key metrics: Type-Token Ratio (TTR), Self-Repetition Score (SRS), Compression Ratios (CR), and Effective Number of Words (ENW).

\subsection{Metrics for Lexical Diversity}
\label{sec:metrics_for_lexical_diversity}

A widely used measure for assessing lexical diversity is the \emph{TTR} \cite{richards1987type, hess1984type, cunningham2020measuring}, which calculates the ratio of unique words (types) to total words (tokens) in a dataset, where $\lvert V \rvert$ is the number of distinct word types and $N$ is the total number of tokens in dataset $D$ (see Equation~\ref{eq:TTR}). While TTR offers insight into the diversity of vocabulary, it is known to exhibit downward bias as corpus size increases, as repeated tokens accumulate more quickly than new word types \cite{mccarthy2010mtld,covington2010cutting}. To gain a better understanding, this study complements TTR with three additional metrics drawn from the literature. First, the \emph{SRS} captures the recurrence of longer $n$-grams (e.g., 4-grams) across the dataset \cite{salkar2022self}, thereby highlighting formulaic patterns that single-word metrics like TTR may overlook. Specifically, as defined in Equation~\ref{eq:SRS}, the SRS is calculated where $K$ represents the total number of distinct 4-grams, $N_i$ denotes the number of prompts containing the $i$th 4-gram, and $\lvert D \rvert$ signifies the total number of prompts in dataset $D$. Consequently, higher SRS values indicate an increased reliance on repeated multi-word phrases, suggesting a convergence toward more predictable and less varied language use. Second, CR quantifies redundancy by comparing the size of the uncompressed text to its compressed size using the \texttt{gzip} algorithm \cite{shaib2024standardizing}, as indicate in Equation ~\ref{eq:CR}, shown in Table \ref{tab:lexical_diversity_metrics}. This metric operates under the premise that lower CR values signify higher compressibility, which in turn suggests the presence of repetitive or predictable structures within the text. Therefore, a lower CR indicates that the dataset may contain less lexical diversity, as the compression algorithm can effectively reduce the dataset size by exploiting these redundancies. Finally, \emph{ENW} captures the distributional balance of the vocabulary by drawing from Shannon entropy \cite{nicolis1994toward, ebeling1991entropy}, as defined in Equations~\ref{eq:H} and~\ref{eq:ENW}. In this context, $p(w)$ represents the frequency of the word $w$ in the dataset. The ENW accounts for the balance of word usage by distinguishing datasets dominated by a few high-frequency terms from those with a more uniform word distribution. Consequently, a higher ENW reflects a more balanced and diverse vocabulary, whereas a lower ENW indicates that a limited set of words predominates, reducing overall lexical diversity.

To address the possibility that data volume alone might drive these observations, we evaluated both the \emph{full dataset} and a \emph{fixed-size sample} of 213,779 prompts per month, which corresponds to the total prompt count in October 2023, the earliest month in our dataset. This ensures that any trends observed are not merely artifacts of varying data volumes across months, but reflect underlying patterns in the dataset.

\begin{table}[!htbp]
    \centering
    \footnotesize % Further reduce font size
    \setlength{\tabcolsep}{2pt} % Reduce column padding
    \renewcommand{\arraystretch}{0.8} % Reduce row height
    \caption{Lexical Diversity Metrics and Their Equations}
    \label{tab:lexical_diversity_metrics}
    \begin{tabular}{@{}p{2cm} p{6cm}@{}}
        \toprule
        \textbf{Metric} & \textbf{Equation} \\
        \midrule
        \textbf{TTR} &  
        \makecell[l]{
            \refstepcounter{equation}\label{eq:TTR}%
            $\text{TTR}(D) = \dfrac{\lvert V \rvert}{N}$ \quad (\theequation)
        } \\
        \midrule
        \textbf{SRS} &  
        \makecell[l]{
            \refstepcounter{equation}\label{eq:SRS}%
            $\text{SRS}(D) = \dfrac{1}{\lvert D \rvert} \sum_{i=1}^{K} \log (1 + N_i)$ \quad (\theequation)
        } \\
        \midrule
        \textbf{CR} &  
        \makecell[l]{
            \refstepcounter{equation}\label{eq:CR}%
            $\text{CR}(D) = \dfrac{\text{size}_{\text{compressed}}(D)}{\text{size}_{\text{uncompressed}}(D)}$ \quad (\theequation)
        } \\
        \midrule
        \textbf{ENW} &  
        \makecell[l]{%
            \refstepcounter{equation}\label{eq:H}%
            $H(D) = - \sum_{w \in V} p(w) \ln p(w)$ \quad (\theequation) \\%
            \refstepcounter{equation}\label{eq:ENW}%
            $\text{ENW}(D) = \exp(H(D))$ \quad (\theequation)
        } \\
        \bottomrule
    \end{tabular}
\end{table}

\subsection{Lexical Diversity Over Time}
\label{sec:lexical_diversity_over_time}

\begin{figure*}[h]
    \centering
    % First row
    \begin{subfigure}[t]{0.45\textwidth}
        \centering
        \includegraphics[width=\linewidth]{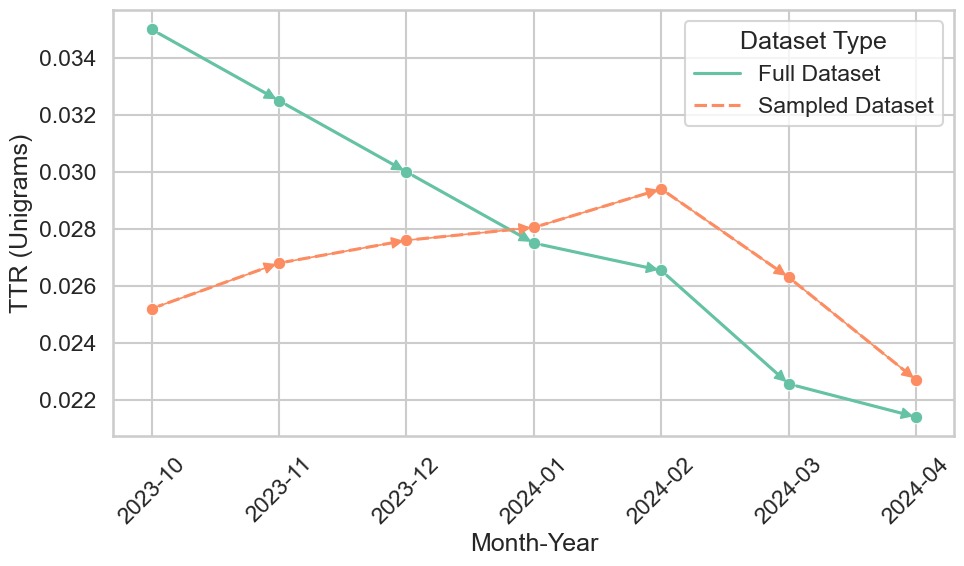}
        \caption{Type-Token Ratio (TTR).}
        \label{fig:ttr}
    \end{subfigure}
    \hfill
    \begin{subfigure}[t]{0.45\textwidth}
        \centering
        \includegraphics[width=\linewidth]{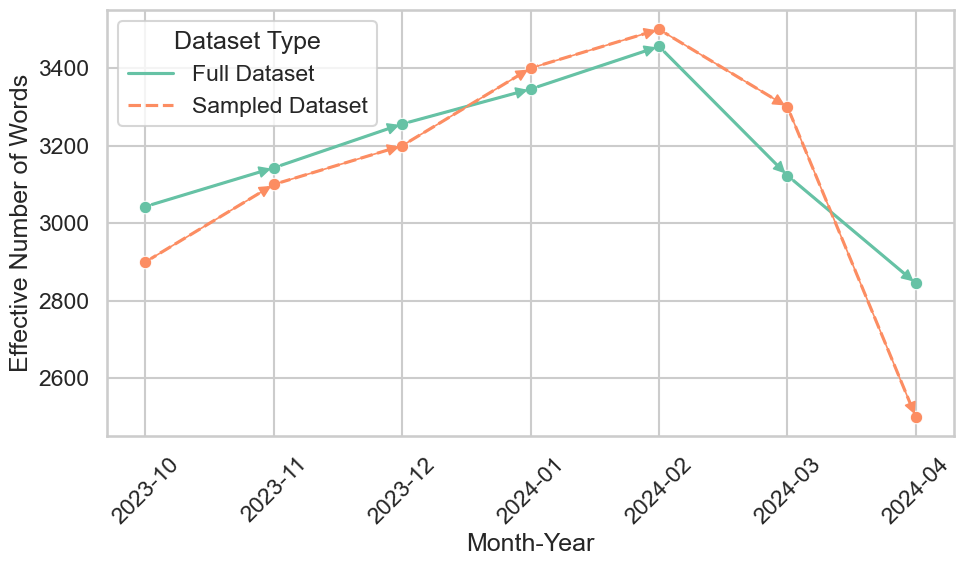}
        \caption{Effective Number of Words (ENW).}
        \label{fig:enw}
    \end{subfigure}
    % Second row
    \begin{subfigure}[t]{0.45\textwidth}
        \centering
        \includegraphics[width=\linewidth]{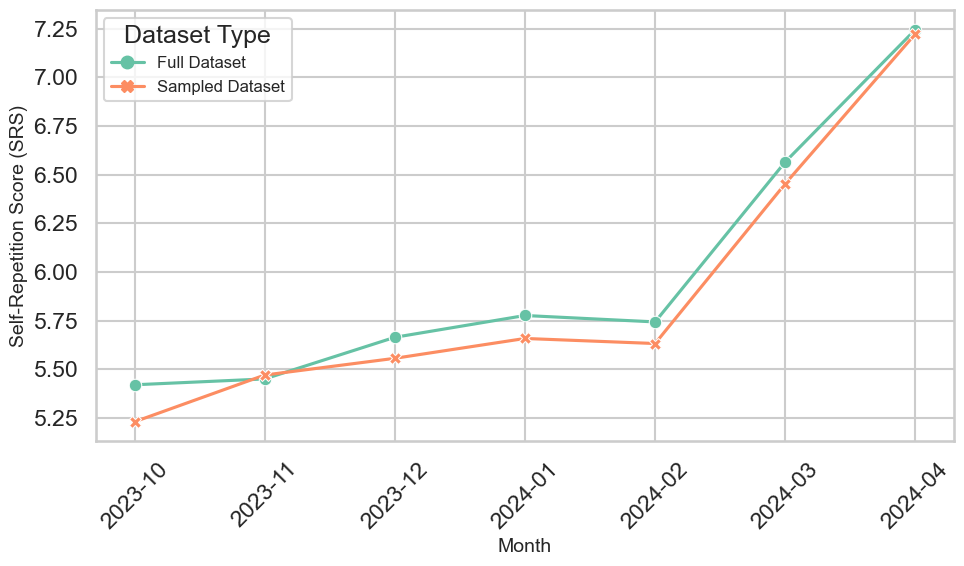}
        \caption{Self-Repetition Score (SRS).}
        \label{fig:srs}
    \end{subfigure}
    \hfill
    \begin{subfigure}[t]{0.45\textwidth}
        \centering
        \includegraphics[width=\linewidth]{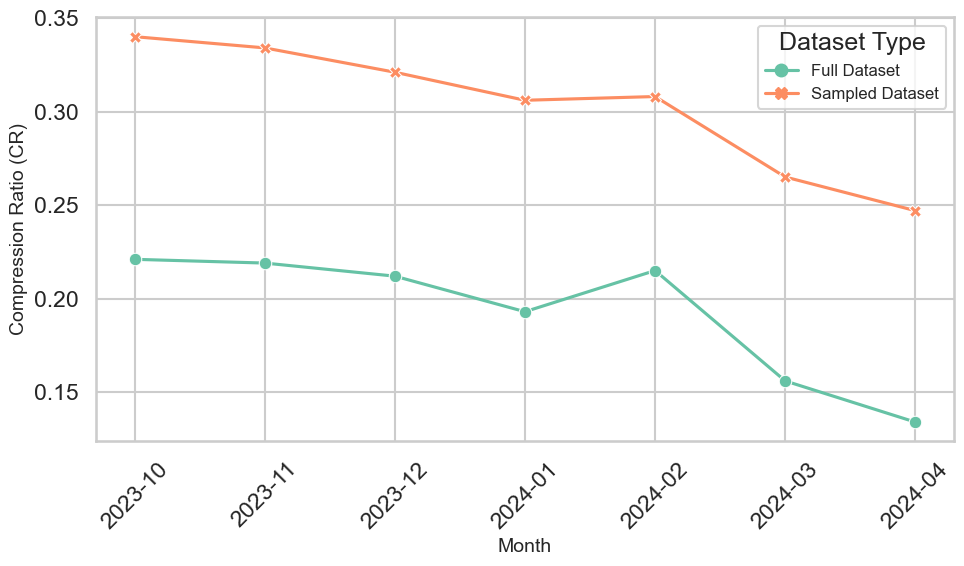}
        \caption{Compression Ratio (CR).}
        \label{fig:crs}
    \end{subfigure}
    \caption{Monthly changes of the values of TTR, ENW, SRS, and CR scores calculated from the prompts of Civiverse dataset.}
    \label{fig:comparison}
\end{figure*}

Across the full Civiverse dataset, a clear pattern emerges: language in prompts becomes increasingly formulaic and repetitive over time. This trend is reflected by the negative correlation between SRS and CR (e.g., $r = -0.55$), indicating that as texts become more repetitive (higher SRS), they also grow more compressible (lower CR). Notably, SRS exhibits near-perfect alignment between the full and sampled datasets (correlation coefficient of 0.9956), suggesting that these findings are not artifacts of corpus size. By contrast, TTR displays a steady decline. As an example, TTR for April 2024 is 0.017 in the full data but rises to 0.023 when analyzing the fixed-size monthly sample, highlighting TTR’s sensitivity to sheer volume. ENW, which accounts for the distributional balance of word usage, similarly decreases from around 3{,}500--3{,}600 in early 2024 to below 2{,}000 by April, indicating an increasing reliance on a smaller set of frequently used terms. Figure \ref{fig:comparison} shows in detail how the scores changed over time, with a clear decrease in lexical diversity after February 2024.

One possible explanation for this is the release of the Pony Diffusion XL\footnote{\url{https://huggingface.co/stablediffusionapi/pony-diffusion-v6-xl-2}} models in January 2024, which introduced a unique approach to TTI and demonstrated particular strengths in NSFW contexts compared to other models such as Stable Diffusion XL (SDXL) or Stable Diffusion 1.5 (SD1.5). These models rely on an extensive dataset of 2.5 million images evenly split among anime, furry, anthro, and cartoon styles, and employ a structured Danbooru-style tagging system\footnote{\url{https://github.com/KichangKim/DeepDanbooru}} for granular control over image attributes such as quality, style, and content ratings (e.g., \texttt{rating\_safe}, \texttt{rating\_questionable}, \texttt{rating\_explicit}).
For example, rating tag combinations like "score\_9, score\_8\_up, score\_7\_up, score\_6\_up," along with other rating tags (see Appendix C.2 for details), appeared in 39.58\% of prompts in February 2024, whereas they were absent in previous months. This formulaic tagging system simplifies prompt engineering and reduces vocabulary variability, contributing to the decline in lexical diversity as users increasingly rely on standardized tag combinations for optimal results.

A closer look at n-gram and 4-gram frequencies, which can be examined in more detail in Appendix C.2, reveals a progressive shift toward formulaic language and a narrowing range of lexical diversity in prompts. By April 2024, certain terms like "hair" appear in 93\% of prompts, up from 74\% six months earlier. Similarly, descriptors such as "detailed," "quality," and "masterpiece" become increasingly dominant, reflecting a shared focus on photorealism and adherence to widely recognized aesthetic principles. Many frequently used terms emphasize specific details, such as body parts (e.g., "face," "skin," "breast") or technical attributes like "8k" and "high quality." The evolution of 4-grams also highlights this trend. Early examples like "tom cruise tom cruise" or "high quality film grain" have given way to Pony Diffusion-related tags, such as "score9 score8up score7up score6up" and "style sdxllorapony diffusion v6".

\subsection{Lexical Diversity Across User Categories}

Beyond aggregate trends, user-level repetition patterns also appear to shape lexical diversity. We computed TTR, ENW, SRS, and CR for each category under two conditions: using all prompts and an equal sample of 90,521 prompts per group (matching the smallest contributor, non-repeaters). As shown in Table \ref{tab:equal_sample_minipage}, when controlling for prompt sample size, consistent repeaters exhibit the lowest TTR (0.065) and ENW (2,909) but the highest SRS (5.75), reflecting a narrower vocabulary and heavy reliance on repeated sequences. Occasional repeaters fall in the middle, with a TTR of 0.076, ENW of 3,483, and SRS of 5.83, showing moderate lexical diversity. Non-repeaters, with the highest TTR (0.10) and ENW (3,719) and lowest SRS (4.43), demonstrate broader lexical exploration. In the uncontrolled condition similar trends persist: consistent repeaters still show the lowest TTR (0.024) and ENW (2,946) but the highest SRS (5.96), while non-repeaters maintain the highest TTR (0.10) and ENW (3,719), alongside the lowest SRS (4.43). This suggests that even when consistent repeaters change their prompts, they do so in a very similar way, likely because they aim to achieve specific aesthetics in their generated outputs.

\begin{table}[h]
    \centering
    \caption{Lexical Diversity Metrics per User Category (Controlled Prompt Sample).}
    \label{tab:equal_sample_minipage}
    \begin{tabular}{@{}lcccc@{}}
        \toprule
        \textbf{User Category} & \textbf{TTR} & \textbf{ENW} & \textbf{CR} & \textbf{SRS} \\
        \midrule
        Consistent Repeaters   & 0.065 & 2909 & 0.26 & 5.75 \\
        Occasional Repeaters   & 0.076 & 3483 & 0.29 & 5.83 \\
        Non-Repeaters          & 0.10  & 3719 & 0.39 & 4.43 \\
        \bottomrule
    \end{tabular}
\end{table}

\subsection{Dominant Prompt Patterns}
\label{sec:dominant_prompt_patterns}

The lexical diversity metrics presented in ~\ref{sec:lexical_diversity_over_time} provide an overview of how TTI language evolves over time but do not reveal the $specific$ linguistic structures that dominate the dataset. To address this gap, we conducted a complementary analysis, MinHash-based prompt similarity detection, that identifies the most prevalent linguistic patterns in the dataset. These patterns often consist of recurring combinations of descriptive attributes and keywords (e.g., "masterpiece, ultradetailed, 8k HDR, best quality"), indicating a preference for detailed, attribute-heavy prompts over sentence-like constructions. Analyzing their frequency, user adoption, and monthly evolution provides deeper insights into community dynamics and linguistic convergence, especially after February~2024 when lexical diversity decreased.

MinHash-based Jaccard similarity \cite{kannan2016big} provides an efficient way to detect near-duplicate documents in large text corpora. This approach, originally popularized by Broder~\cite{broder1998min} and further refined in subsequent work~\cite{broder2000identifying, ji2013min, ertl2020probminhash}, is particularly valuable when dealing with sizeable collections of user-generated content, such as millions of TTI prompts. In these scenarios, traditional pairwise comparisons to compute exact similarities would require prohibitive $\mathcal{O}(n^2)$ operations, whereas MinHash allows approximate yet highly scalable estimation of similarity between documents. Following the standard framework, each prompt is transformed into a set of tokens and then reduced to a fixed-size MinHash \emph{signature}, which approximates its set membership in a compact vector. This technique leverages \emph{min-wise independent permutations}~\cite{broder1998min} (and extensions such as \emph{min-max hash}~\cite{ji2013min}), ensuring that token sets with high Jaccard similarity are likely to produce matching MinHash signatures. By inserting these signatures into a Locality-Sensitive Hashing (LSH) index, we can efficiently retrieve potentially near-duplicate prompts, those exceeding a specified Jaccard similarity threshold, without performing exhaustive pairwise comparisons. This approach operates in roughly linear or sub-quadratic time, making it well-suited for large-scale datasets. Similar methods have been applied in other domains, such as song lyrics, where MinHash-LSH effectively identifies repeated patterns and stylistic markers~\cite{moreno2022use}, enabling deeper analyses of recurrent formulas and stylistic trends across various contexts. To standardize comparisons, prompts are truncated to 20 tokens (the dataset's average length). Setting a Jaccard threshold of 0.8 (requiring at least 16 shared tokens) allows identification of near-duplicate clusters. This method not only highlights common patterns but also reveals lexical convergence trends, as an increasing proportion of near-duplicates over time indicates a shift toward standardized prompt language.

\begin{table}[h]
\centering
\small
\renewcommand{\arraystretch}{0.9} % Slightly reduce row height
\setlength{\tabcolsep}{5pt} % Adjust column spacing for readability
\caption{Distribution of Prompts Above and Below 0.8 Similarity Threshold.}
\label{tab:similarity_summary_prompts}
\begin{tabular}{lrr}
\toprule
\textbf{Metric} & \textbf{Above 0.8} & \textbf{Below 0.8} \\
\midrule
Total Prompts & 1,670,103 & 829,897 \\
\midrule
Unique Users & 34,034 & 43,587 \\
\midrule
Consistent Repeater Prompts & 1,289,758 (77.2\%) & 520,428 (62.7\%) \\
Consistent Repeater Users & 19,946 & 20,345 \\
\midrule
Occasional Repeater Prompts & 350,433 (21.0\%) & 258,076 (31.1\%) \\
Occasional Repeater Users & 6,412 & 6,876 \\
\midrule
Non-Repeater Prompts & 29,912 (1.8\%) & 51,393 (6.2\%) \\
Non-Repeater Users & 7,676 & 16,366 \\
\bottomrule
\end{tabular}
\end{table}

Table ~\ref{tab:similarity_summary_prompts} indicates that among the approximately 2.5 million sampled prompts (dataset without duplicate prompts), about 1.67 million (66.8\%) surpass the 0.8 similarity threshold, highlighting a prevalent trend of users recycling or adapting near-identical text. Notably, \emph{consistent repeaters} contribute over 77\% of these near-duplicate prompts, underscoring their significant role in propagating and refining “formulaic” language. In contrast, the remaining prompts below the 0.8 similarity threshold are more diverse, with a higher proportion of contributions from \emph{non-repeaters} (6.19\%) compared to the above-0.8 group (1.79\%), suggesting that non-repeaters are more inclined to produce varied and unique prompts.

Table ~\ref{tab:top3_prompt_clusters} lists the three largest “clusters” of near-duplicate prompts, collectively covering 20.53\% of the 1,670,103 prompts with a similarity score above 0.8. Each cluster is defined by a distinct combination of community tagging conventions, references to LoRA modules, and recurring thematic cues. The majority of these clusters emphasize human subjects, predominantly female, and frequently include body or feature focused tokens. Common attributes across these clusters include “quality” tags (e.g., \texttt{best quality12}), advanced descriptors (e.g., \texttt{ultra realistic32kraw}), and references to LoRA modules (e.g., \texttt{lora\_gothic\_outfit06}), with occasional explicit NSFW tokens. Photorealistic details are frequently prioritized through descriptors such as “hdr ultra detailed” or “masterpiece12.” Most submissions within these clusters come from consistent repeaters, who contribute over 70\% of the prompts. Occasional and non-repeat users are far less systematic in adopting these patterns; for instance, non-repeaters contribute no more than 3.03\% of prompts in any cluster. See Appendix C.5 for a detailed analysis of additional clusters.

\begin{table}[ht]
\centering
\footnotesize % Further reduce font size
\renewcommand{\arraystretch}{0.7} % Reduce row height
\setlength{\tabcolsep}{2pt} % Reduce column spacing
\caption{Top 3 Most Common Near-Duplicate Prompt Clusters. Each cluster represents a set of prompts where the listed tokens appear as frequently as indicated in the “\# Prompts” column, with minor variations that may include up to four additional unlisted tokens.}
\label{tab:top3_prompt_clusters}
\begin{tabular}{p{3.5cm}rccc}
\toprule
\textbf{Prompt Pattern} & \textbf{\# Prompts} & \textbf{Users} & \textbf{\% Cons.} & \textbf{Emergence} \\ 
\midrule
\textit{realistic, atmospheric scene, masterpiece, best quality, detailed face, ultradetailed, body cinematic light, 1girl, blonde hair, 8k} 
& 246,899 & 9,003 & 75.8\% & Oct 2023 \\ 
\midrule
\textit{ultra realistic 32k raw, photo, detailed skin, 8k uhd, dslr, high quality, film grain, 1girl blonde, huge breasts} 
& 61,143 & 4,301 & 99.3\% & Jan 2024 \\ 
\midrule
\textit{1girl, 30 years, looking viewer, leaning forward, head turned, glossy lip, soft skin, elegant makeup, seductive smile, masterpiece, best quality} 
& 34,786 & 928 & 99.97\% & Mar 2024 \\ 
\bottomrule
\end{tabular}
\end{table}

To analyze how near-duplicate prompts evolve over time, the data is divided into two groups: prompts above and below the 0.8 similarity threshold {(see Appendix C.6 for more details in numbers)}. From October 2023 to January 2024, high-similarity prompts account for approximately 68–73\% of submissions, allowing for some diversity in language use. However, starting in February 2024, the proportion of above-0.8 prompts increases markedly, rising from 72.27\% to 77.03\% and surpassing 83\% by April. This trend aligns with the decline in lexical diversity metrics discussed in Section~\ref{sec:lexical_diversity_over_time}, indicating that once certain “templates” or LoRA models gain popularity, they quickly dominate the dataset. In contrast, while the below-0.8 group grows in absolute numbers each month, its share shrinks proportionally, reflecting a reduced space for varied or experimental prompts. During this period, the proportion of consistent repeaters in this category rises significantly, from 54.93\% in October 2023 to 70.25\% by April 2024, suggesting that even within lower-similarity submissions, this user cohort increasingly dominates. Meanwhile, the contributions of occasional repeaters decline from 36.62\% to 23.39\%, and non-repeaters drop slightly from 8.44\% to 6.36\%; however, both groups still contribute more to this cohort compared to the above-0.8 group, where their combined share never exceeds 30\%. This dynamic highlights a self-reinforcing cycle where active users refine and propagate dominant patterns that are subsequently adopted by newcomers or less frequent participants, further narrowing the range of descriptive language and stylistic choices across the community.

\section{Semantic Diversity of Prompts}
\label{sec:semantic_diversity}

Apart from analyzing lexical diversity, we also focus on the exploration of semantic diversity. In this context, semantic diversity refers to the breadth and complexity of distinct concepts, themes, and ideas that are present in user-submitted prompts. Instead of focusing solely on individual words or phrases, our analysis examines how larger thematic structures evolve over time and vary across user categories, offering a complementary perspective to the lexical analyses discussed.

\subsection{Methodological Overview}
\label{sec:method_overview}

To investigate semantic diversity, we adopted a systematic approach grounded in topic modeling techniques, drawing inspiration from previous studies on TTI prompts \cite{sanchez2023examining, mccormack2024no} and related datasets, such as DiffusionDB \cite{wang2022diffusiondb}. Our methodology focuses on the extraction of individual \emph{prompt specifiers}, which are specific text fragments that describe or define desired characteristics of the generated image output. These specifiers, such as "shiny skin" or "high definition," play a critical role in shaping the semantic content of the prompt. To identify these specifiers, we divided the raw prompts into segments by splitting them at commas, a technique consistent with prior research aimed at isolating discrete descriptors.

Each retained specifier was transformed into a 384-dimensional embedding using the MiniLMv2 model \cite{wang2020minilm}, chosen for its efficiency and strong performance in resource-constrained scenarios \cite{reimers2019sentence}. For visualization and dimensionality reduction, embeddings were projected into two-dimensional space via Uniform Manifold Approximation and Projection (UMAP) \cite{mcinnes2018umap}, after which the HDBSCAN clustering algorithm \cite{campello2013density} was employed to identify coherent groups of semantically related specifiers. By integrating HDBSCAN with UMAP \cite{asyaky2021improving, suryadjaja2021improving}, we obtained clusters of concepts and subtopics. To interpret and assign labels to these clusters, we first extracted representative keywords using class-specific TF-IDF (c-TF-IDF), which highlights terms statistically significant within each topic, and then asked an LLM, specifically GPT-4o, to propose topic labels based on these keywords.

\subsection{Semantic Diversity Over Time}

\begin{figure}[h]
    \centering
    \includegraphics[width=0.6\linewidth]{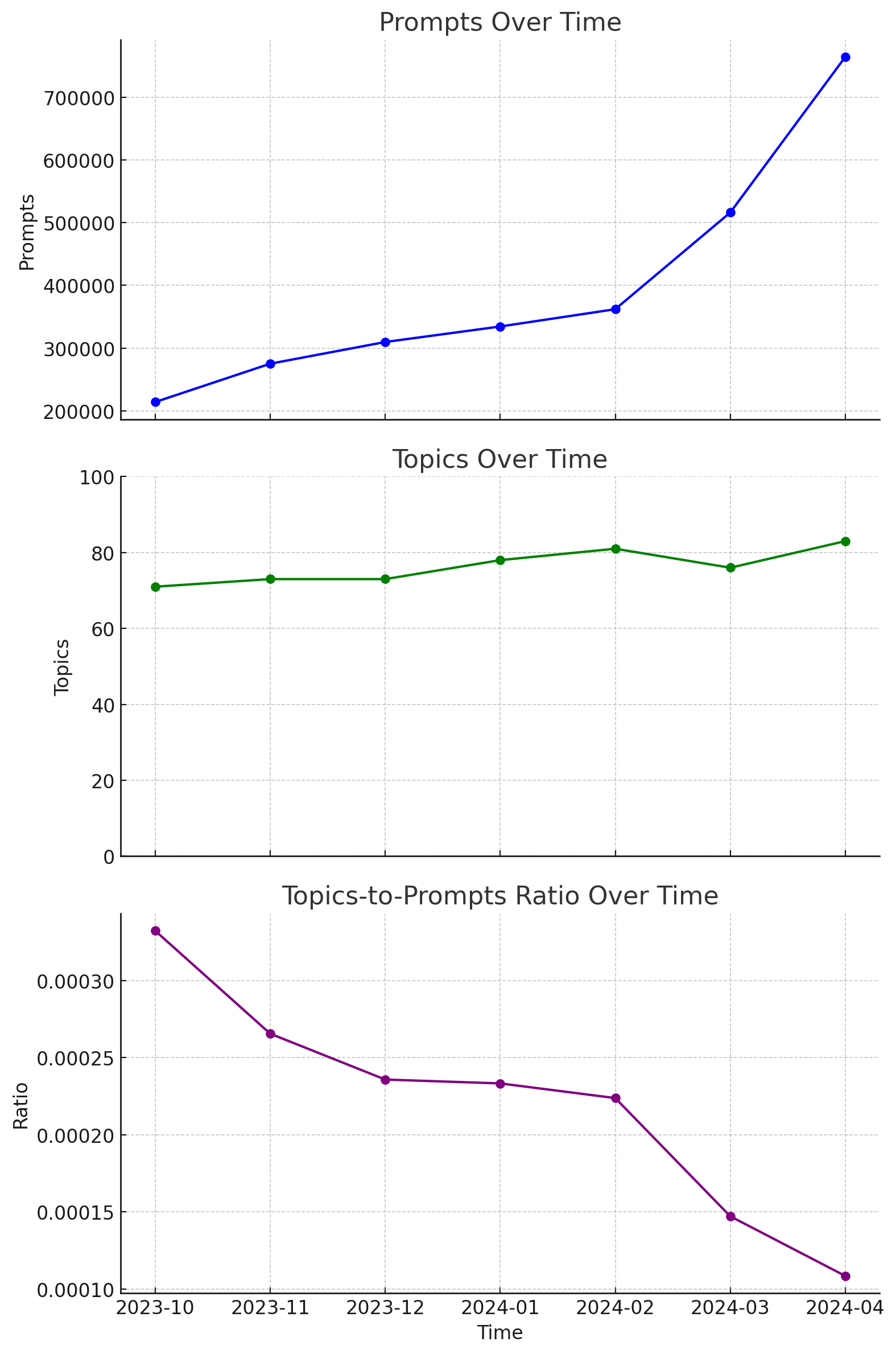} % Adjust width or use scale
    \caption{Semantic Trends Over Time: (a) Total Prompts, (b) Topics, and (c) Topics-to-Prompts Ratio.}
    \label{fig:semantic_diversity}
\end{figure}

Figure \ref{fig:semantic_diversity} illustrates the monthly trends from October 2023 to April 2024, focusing on prompt volume, the variety of semantic topics, and the topics-to-prompts ratio. Throughout this period, the number of prompts consistently rose, with a particularly sharp increase in February 2024 that signaled heightened user engagement. However, the number of distinct semantic topics grew only modestly—rising from 72 in October to 83 in December and 89 by February—and even declined in March 2024, indicating that while users were generating more prompts, their focus narrowed to a smaller range of themes. Consequently, the topics-to-prompts ratio steadily decreased over time, highlighting a reduction in semantic diversity as the rapid expansion of prompt volume outpaced the introduction of genuinely new topics.

This analysis focused on monthly trends, as an overall topic modeling of the Civierse dataset \cite{palmini2024civiverse} had already been conducted. The incremental increase in recognized topic clusters suggests that many of the so-called "new" topics introduced each month were not truly novel but rather small variations or subclusters of existing themes. This pattern was influenced by the rise of formulaic LoRA and style tokens (e.g., \emph{lora\_gothic\_outfit06}), modifiers (e.g., \emph{(full body shot:1.3)}), and identifiers (e.g., \emph{fashigirl\-v7}), which often focused on niche or hybrid subjects but rarely deviated from recurring motifs such as body and face details, clothing, or scenic elements. Although the overall number of topic clusters increased, these additions largely represented narrower permutations of established themes rather than a genuine broadening of thematic diversity, further underscoring the relative stagnation in semantic variety during this period.

Despite the growing diversity of prompts and the emergence of new clusters, certain themes and preferences dominated the dataset, reflecting established stylistic conventions and shared community interests. A central focus was the depiction of humans, particularly women, often emphasizing physical and facial features (e.g., \emph{“toned thighs,” “slim waist,” “big eyes,” “seductive smile”}). \emph{Anime and stylized characters} also remained prominent, with prompts referencing specific figures like \emph{“Mikasa Ackerman”} and \emph{“Hinata Hyuga”}. Many prompts highlighted body attributes or poses, often paired with perspective-driven phrases such as \emph{“looking directly at the viewer”} or \emph{“smiling seductively”}, fostering a sense of intimacy or engagement. In contrast, \emph{landscapes and cityscapes} offered a broader focus, showcasing vivid scenes like \emph{“misty mountain sunrise”}. \emph{Photographic and cinematic aesthetics} emerged as another dominant theme, with users specifying high-resolution details (e.g., \emph{“8k resolution,” “HDR lighting”}) and elements like \emph{“cinematic lighting”} and \emph{“lens flare”} to create striking visuals. Additionally, prompts frequently referenced \emph{artistic mediums and styles}, such as \emph{“watercolor”} or \emph{“digital illustration”}. \emph{Lighting and color} played a significant role in shaping ambiance, with phrases like \emph{“golden hour glow”} and \emph{“neon highlights”} appearing regularly. These themes and descriptors highlight a persistent preference for detailed, visually refined outputs, reflecting the community’s collective visual inclinations.

\subsection{Semantic Diversity Across User Categories}
\label{subsubsec:topic_modeling_categories}

A balanced subset of 90,521 unique prompts was analyzed to compare topic diversity across user categories. Non-repeaters engage with 89 topics (5,069 specifiers), slightly more than consistent repeaters, who focus on 71 topics (3,201 specifiers), and occasional repeaters, who cover 76 topics (4,120 specifiers). Figure~\ref{fig:consistent_repeaters} visualizes the MiniLM-L6-v2 embeddings of prompt specifiers for the \textbf{Consistent Repeaters} category using HDBSCAN clustering to reveal topic structure. The resulting scatterplot shows tightly grouped clusters centered around themes such as \textit{NSFW Tags}, \textit{Erotic Acts}, \textit{Body Parts}, and \textit{Textures}, highlighting this group’s focus on explicit and technically specific descriptors. Overlapping areas in themes like \textit{Lighting Effects} and \textit{Color Palettes} indicate an intentional blending of artistic and functional elements, reflecting a refined and specialized approach to prompt generation. For comparison, a broader and more diffuse topic landscape observed in the Non-Repeaters group is included in Appendix~\ref{sec:semantic_diversity_user_categories}.

\begin{figure}[h] % 'h' to place it here, you can also use 't' for top, 'b' for bottom
        \centering
        \includegraphics[width=1.1\linewidth]{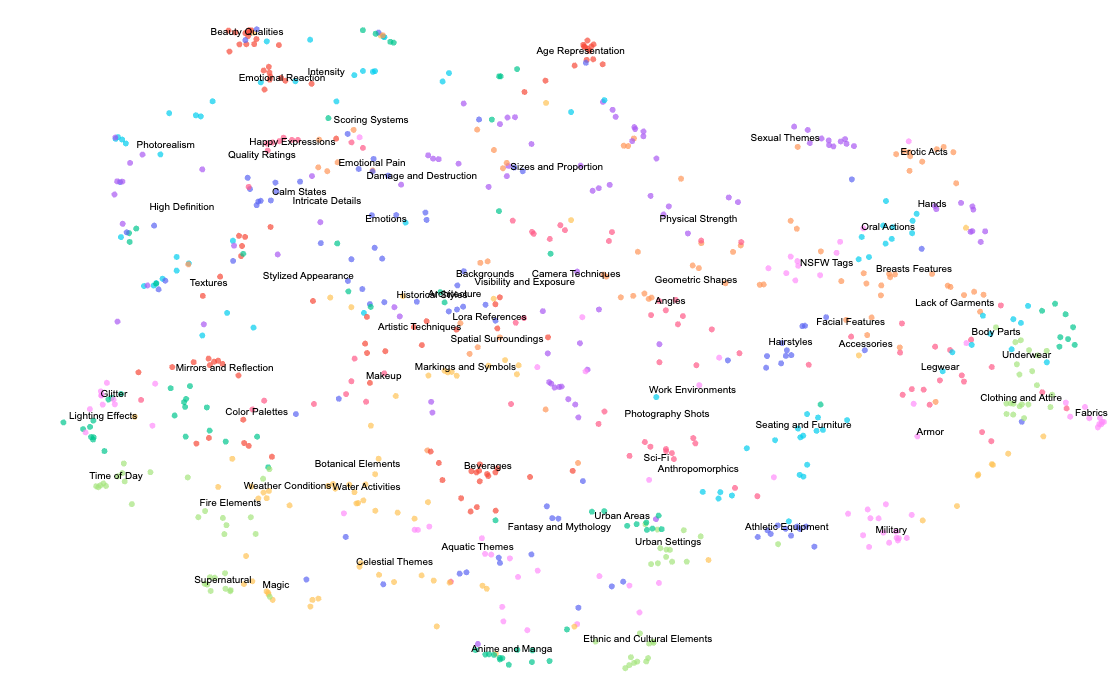} % Second image
        \caption{UMAP visualization of MiniLM-L6-v2 embeddings of prompt specifiers and HDBSCAN-identified topics for the \textbf{Consistent Repeaters}  user category}
        \label{fig:consistent_repeaters}
\end{figure}

\section{Visual Diversity of Prompts}
\label{sec:visual_diversity}

This section examines the temporal evolution of visual diversity, measured through Vendi scores, and investigates correlations between the lexical diversity metrics discussed in Section~\ref{sec:metrics_for_lexical_diversity} and the diversity of generated images. Additionally, it explores the relationship between token similarity and image similarity, as well as how the number of identical tokens influences the strength and significance of these trends.

\subsection{Incorporating Vendi Scores to Measure Visual Diversity}

To quantify the diversity of generated outputs, we used the Vendi Score, a metric designed to measure the "effective number" of distinct elements in a dataset by considering both \emph{richness} (the number of unique items) and \emph{evenness} (the balance of their distribution). This means it provides a more nuanced measure of diversity than a simple count of unique elements, as it reflects how evenly the elements are represented. The Vendi Score has been widely applied in TTI research to assess image diversity, with Zhang et al.~\cite{zhang2024partiality} using it to identify cultural biases in TTI models and Kannen et al.~\cite{kannen2024beyond} leveraging a quality-weighted variant to measure cultural diversity across different artifacts. 

\begin{figure}[h]
    \centering
    \includegraphics[width=0.95\linewidth]{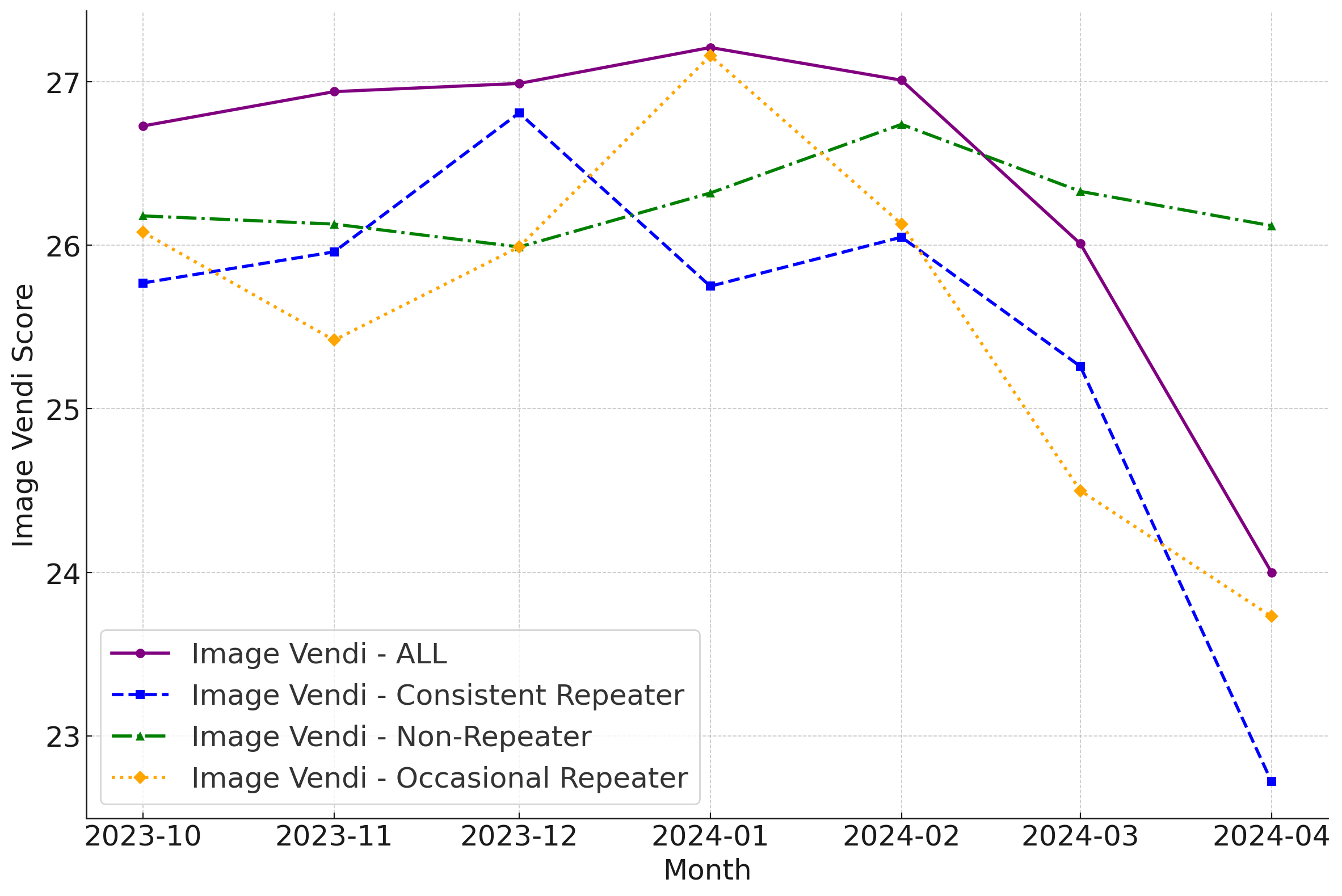}
    \caption{Values of image Vendi scores over time for each user category and combined.}
    \label{fig:vendi_scores}
\end{figure}

To investigate the evolution of visual diversity, we computed Vendi scores for images over time and across user categories. Due to computational constraints, we sampled 10,000 prompts and their corresponding images from each user category, consistent, occasional, and non-repeaters, per month, resulting in a total of 30,000 monthly samples. This approach enabled us to track temporal trends in diversity while ensuring the analysis remained computationally efficient. As shown in Figure~\ref{fig:vendi_scores}, visual diversity exhibited a general upward trend across most user categories from October 2023 to February 2024, while remaining relatively stable for the subset of the dataset. However, a marked decline in Vendi scores is observed after February 2024, particularly among the "Occasional Repeaters" and "Non-Repeaters" categories. This decline corresponds to a simultaneous drop in lexical diversity during the same period, suggesting a potential relationship between reduced textual variation in prompts and diminished visual diversity in generated outputs.

To examine the relationship between textual diversity, measured using metrics such as TTR, SRS, ENW, and CR, and visual diversity, assessed through Vendi scores, we computed Pearson correlation coefficients to explore the impact of lexical variation on visually diverse outputs, illustrated in Table \ref{tab:correlation_results}. TTR showed a weak and non-significant correlation with visual diversity ($\rho = 0.184$, $p = 0.349$), indicating its limited utility for predicting visual variation. In contrast, ENW exhibited a moderate positive correlation ($\rho = 0.521$, $p = 0.005$), suggesting that prompts with a more balanced word distribution tend to generate more visually diverse images. CR demonstrated the strongest positive correlation ($\rho = 0.620$, $p = 0.012$), highlighting that richer and less redundant prompts are associated with greater image diversity. On the other hand, SRS showed a significant negative correlation ($\rho = -0.536$, $p = 0.003$), indicating that formulaic or repetitive language patterns are linked to reduced visual diversity. These findings suggest that textual richness and variation align more closely with increased visual diversity, whereas repetitive phrasing limits it.

\begin{table}[h]
    \centering
    \caption{Pearson correlation coefficients (\( \rho \)) and p-values between lexical metrics with Vendi Scores.}
    \label{tab:correlation_results}
    \begin{tabular}{@{}lcc@{}}
        \toprule
        \textbf{Metric} & \textbf{Correlation (\( \rho \))} & \textbf{p-value} \\
        \midrule
        TTR & 0.184 & 0.349 \\
        ENW & 0.521 & 0.005 \\
        SRS & -0.536 & 0.003 \\
        CR  & 0.620 & 0.012 \\
        \bottomrule
    \end{tabular}
\end{table}

\subsection{Does Token Diversity Translate into Visual Diversity?}
\label{sec:lexical_visual_relationship}

To explore the relationship between prompt language and the similarity of generated images, we conducted a two-part experiment focusing on token and semantic similarities among prompts. First, we used the MinHash algorithm introduced in Section~\ref{sec:dominant_prompt_patterns} to cluster prompts based on token similarity, creating two types of clusters: those with high token similarity (\(\geq 0.8\)) and those with moderate token similarity (0.5--0.7). Prompts in high token similarity clusters contained at least 16 out of 20 identical tokens, while those in moderate token similarity clusters had between 10 and 15 matching tokens. Within these clusters, we analyzed how token similarity in prompts correlates with the mean pairwise similarity of their image embeddings, thus assessing the influence of linguistic structure on visual similarity. In the second part of the experiment, we examined the same clusters from a semantic standpoint. Specifically, we extracted CLIP text embeddings (using the pre-trained \texttt{clip-vit-base-p32} model~\cite{eslami2021does}) to investigate whether semantic relationships among prompts correspond to higher similarity in the generated images. Together, these analyses reveal how both linguistic and semantic factors can shape the visual consistency of text-to-image outputs. 

\begin{figure}[htbp]
    \centering
    \includegraphics[width=\linewidth]{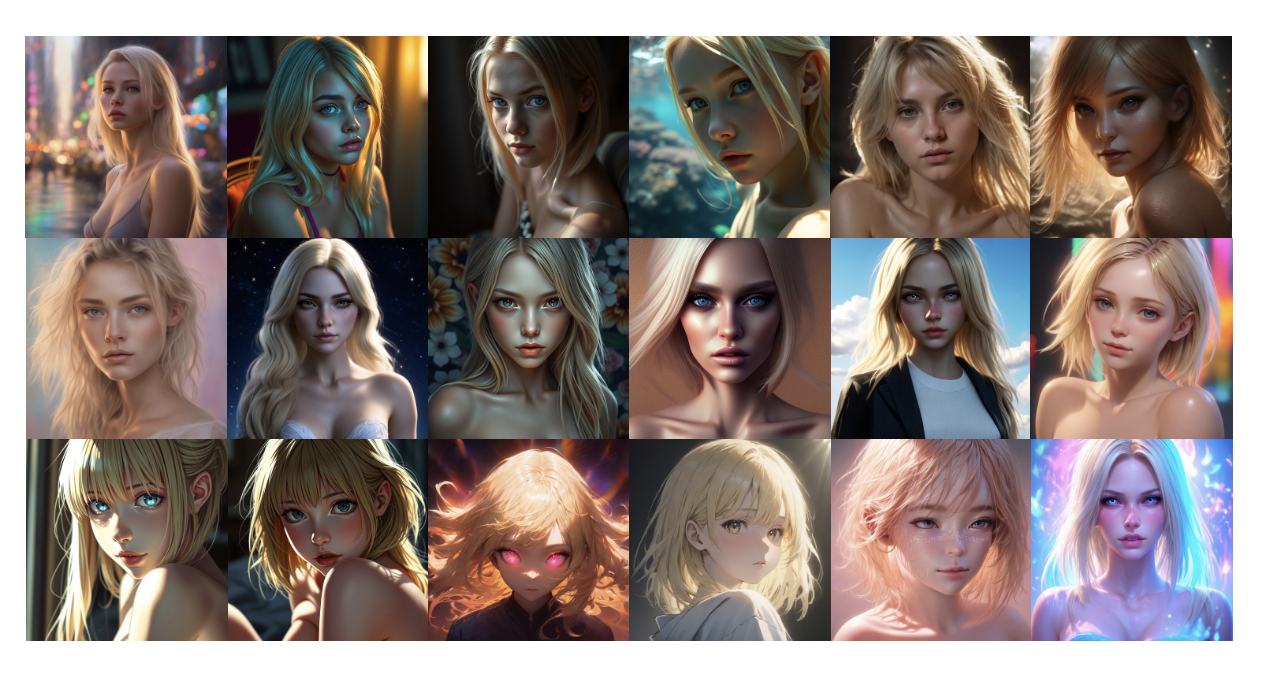} % Replace with your image file path
    \caption{Images from the Civiverse dataset, generated using variations of the most common prompt pattern: \textbf{realistic, atmospheric scene, masterpiece, best quality, detailed face, detailed skin texture, ultradetailed, body cinematic light, 1girl, blonde hair, 8k.}}
    \label{fig:civiverse-images} % Updated label for referencing
\end{figure}

\begin{figure}[htbp]
  \centering
  % First subfigure for token_images.png
  \begin{subfigure}[b]{0.45\textwidth}
    \centering
    \includegraphics[width=\textwidth]{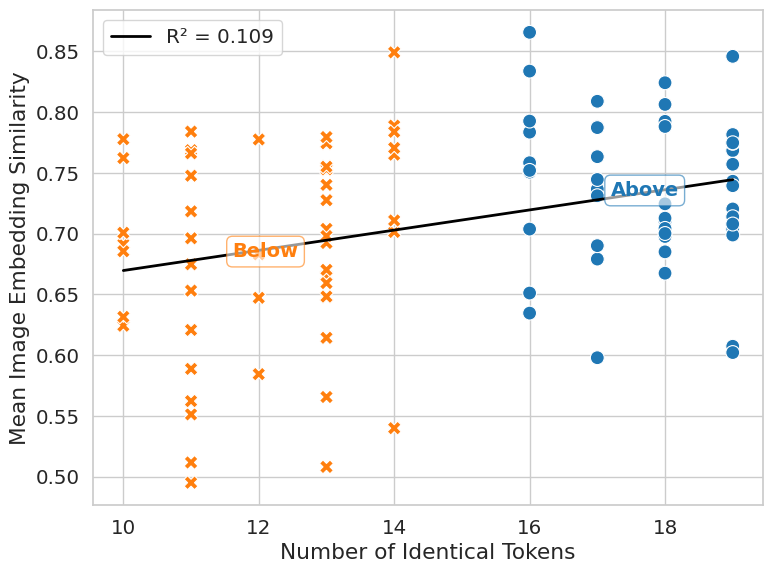}
    \caption{\textbf{Token-Based Image Similarity.}}
    \label{fig:token_images}
  \end{subfigure}
  \hfill
  % Second subfigure for text_images.png
  \begin{subfigure}[b]{0.45\textwidth}
    \centering
    \includegraphics[width=\textwidth]{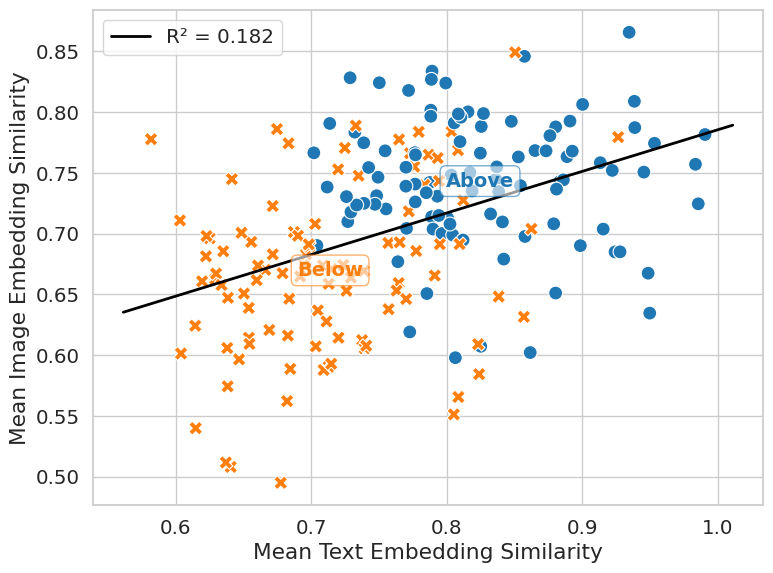}
    \caption{\textbf{Semantic-Based Image Similarity.}}
    \label{fig:text_images}
  \end{subfigure}
  \caption{Comparison of token-based and text-based image similarity across clusters. The graphs depict the impact of token and semantic similarities on image embedding similarity. The blue points represent cluster with high token similarity (Above 0.8), while the orange points represent cluster below this threshold (Between 0.5 and 0.7 token similarity). The R-squared values indicate the strength of the relationship in each plot.}
  \label{fig:side_by_side}
\end{figure}

Prompts were grouped into clusters based on token similarity, and for each cluster, we calculated the mean pairwise cosine similarity of the corresponding image embeddings. Clusters with high token similarity (\(\geq 0.8\)) consistently produced more visually homogeneous outputs, while those with moderate token similarity (0.5--0.7) exhibited greater visual diversity. These findings, illustrated in Fig.~\ref{fig:token_images}, demonstrate that a higher proportion of identical tokens in prompts correlates with reduced visual variation in the generated images. A Pearson correlation coefficient of 0.33 and an \(R^2\) value of 0.182 indicate a moderate positive relationship between token and image similarity. To further examine the link between linguistic patterns and image outputs, we analyzed the mean pairwise cosine similarity of CLIP text embeddings within the same clusters. The results mirrored the previous experiment: clusters with higher token overlap showed greater semantic alignment. Additionally, when comparing text similarity to image similarity, we observed a Pearson correlation coefficient of 0.43 and an \(R^2\) value of 0.109, as shown in Fig.~\ref{fig:text_images}, highlighting the reinforcing role of semantic similarity in shaping the connection between prompt language and image homogeneity.

Figure~\ref{fig:civiverse-images} presents examples of images from the Civiverse dataset, generated using variations of the most common prompt pattern: \textit{"realistic, atmospheric scene, masterpiece, best quality, detailed face, detailed skin texture, ultradetailed, body cinematic light, 1girl, blonde hair, 8k"}. These images illustrate the consistency in visual features, such as lighting, facial structure, subject characteristics (e.g., female, blonde hair), and overall style, which are shaped by prompts with high token and semantic similarity. However, the outputs also reveal inherent biases within the model, as features like blue eyes frequently appear even though they are not specified in the prompt, while the subjects are often depicted as young and with light skin tones. This demonstrates how formulaic language with specific descriptors not only amplifies these biases but also promotes a lack of diversity by reinforcing stereotypical and homogenized representations in TTI outputs.

\section{Limitations and Future Work}

Several factors limit the scope of these findings. First, although Civiverse is one of the largest available datasets for prompt analysis, its focus on the CivitAI platform restricts the generalizability of the results. Prompting behaviors may differ significantly across other TTI communities, such as Midjourney, DALL·E, or proprietary corporate environments, where variations in platform norms, model characteristics, and user demographics could affect the degree and nature of lexical convergence. Second, while we establish a link between repeated language and image homogeneity, this study does not evaluate the subjective impact of formulaic tokens on the perceived quality, creativity, or user satisfaction of the generated outputs. Future work could incorporate human evaluations, through crowdsourced ratings, expert critiques, or comparative analyses, to assess whether repeated descriptors genuinely enhance outcomes or simply reinforce entrenched stylistic norms. Third, the distinction between productive creative iteration and rote repetition remains inherently ambiguous. Repeated phrasing often serves as a necessary tool for refining specific aesthetic goals, but identifying when this repetition transitions into stifling creativity requires further investigation.

\section{Discussion}

In this study, we examined a seven-month record of user interactions from the Civiverse dataset to investigate how prompt language influences the creative outputs of TTI systems. Despite the platform's rapid user growth, our findings reveal a paradoxical trend: the linguistic and thematic diversity of submitted prompts is increasingly constrained. This phenomenon results from a combination of factors, including the widespread use of repeated stylistic patterns, the establishment of community "best practices," and the persistence of stable thematic categories. Below, we discuss the implications of these findings, highlighting both the benefits and limitations of evolving prompt behaviors.

Our analysis indicates that the language used in TTI prompts becomes progressively homogenized over time, as communities gravitate toward a restricted set of high-frequency descriptors and formulaic prompt patterns influenced by community tags. This trend offers several advantages: it enables users to achieve consistent, high-quality outputs and assists both novices and experienced individuals in navigating model idiosyncrasies more effectively. However, it also has significant drawbacks, including reduced lexical diversity and inhibited creative exploration. Notably, the majority of prompts on the platform are generated by users with minimal linguistic experimentation, referred to as “consistent repeaters", wo frequently rely on standardized tokens such as “masterpiece” or “best quality,” as well as specific rating system tags like "score\_9."

On a semantic level, the topics users prompt about remain relatively stable over time, reflecting a preference for achieving specific, consistent aesthetics rather than exploring diverse semantic aspects. Users predominantly focus on portraying particular subjects, especially female ones, as well as specific aesthetics, lighting techniques, artistic mediums, and photographic angles. Despite their substantial contribution to the dataset, consistent repeaters engage with a limited range of topics, concentrating heavily on experimenting with NSFW imagery. This narrow thematic focus reinforces the stability of subject matter within the community, further contributing to the homogenization of both language and content in TTI prompts.

A key finding of our analysis is the strong correlation between linguistic overlap in prompts and the visual homogeneity of generated outputs. The observed decline in Vendi Scores over time corresponds with a reduction in lexical diversity, as prompts with higher token or embedding similarity tend to produce images that converge on a common aesthetic. This pattern suggests that while reducing linguistic experimentation in prompts can yield specific and consistent outputs, it also fosters a TTI culture focused on narrow aesthetic goals, reinforcing the pre-existing biases embedded in the generative models' training data. Moreover, users contribute to this dynamic by favoring similar linguistic structures, which unintentionally introduce an additional layer of bias, constrain the range of visual variation across the platform, and diminish opportunities for creative exploration and innovation. Although factors such as model checkpoints, seed values, and hyperparameters influence output variability, our findings highlight linguistic repetition in prompts as a significant driver of stylistic convergence. Ultimately, the interaction between inherent model biases and user input choices establishes a feedback loop that perpetuates visual uniformity, limiting the potential for diverse and imaginative generative outcomes.

These findings underscore how both technical and social factors—including algorithmic biases, community norms, and users’ reliance on repetitive prompt behaviors—can shape TTI outputs. They suggest that the creative potential of TTI systems is not solely determined by the model’s latent space but is also significantly influenced by the cyclical and community-driven evolution of prompt language. While prompt guidelines and community tags are helpful for achieving high-resolution and aesthetically pleasing images, fostering greater creativity and mitigating unintended homogenization require future research and platform design to implement mechanisms that promote innovative prompts, encourage user experimentation, and diversify training corpora. By addressing the interplay between user-driven inputs and model-level capabilities, TTI systems can better support both broad creative exploration and the reliability of high-quality, formulaic prompts.

\section{Conclusion}

This study highlights the significant influence of user-driven behaviors on the outputs of TTI systems, emphasizing how linguistic homogenization in prompts contributes to visual uniformity in generated content. Our findings reveal that while formulaic language and community-driven prompt patterns enable consistent, high-quality results, they also stifle creative exploration and reinforce pre-existing biases in generative models. By analyzing prompt language, semantic trends, and visual diversity over time, we demonstrate the critical role of user input in shaping the cultural and aesthetic outputs of TTI systems. To unlock the full creative potential of these models, future research and platform design must prioritize mechanisms that encourage linguistic and thematic experimentation, promote inclusivity, and diversify training corpora. Only by addressing the complex interplay between user input and model capabilities can TTI systems truly foster innovation and diversity in AI-generated imagery.

\section*{Acknowledgements}
This research was supported by the Swiss National Science Foundation (SNSF) under the Ambizione Grant Scheme, Grant No. 216104.

\bibliographystyle{ACM-Reference-Format}
\bibliography{sample-base}


\newpage
\appendix

\section{Dataset Overview}

The relationship between prompt counts in datasets with and without duplicates is shown in following figure. The strong linear correlation (\(r = 0.9953\)) demonstrates that duplicate prompts significantly contribute to the dataset's overall volume while maintaining consistent trends. This highlights the importance of duplicates in reflecting user behavior, particularly in iterative refinement and experimentation with prompt submissions.

\begin{figure}[h]
    \centering
    \includegraphics[width=0.70\linewidth]{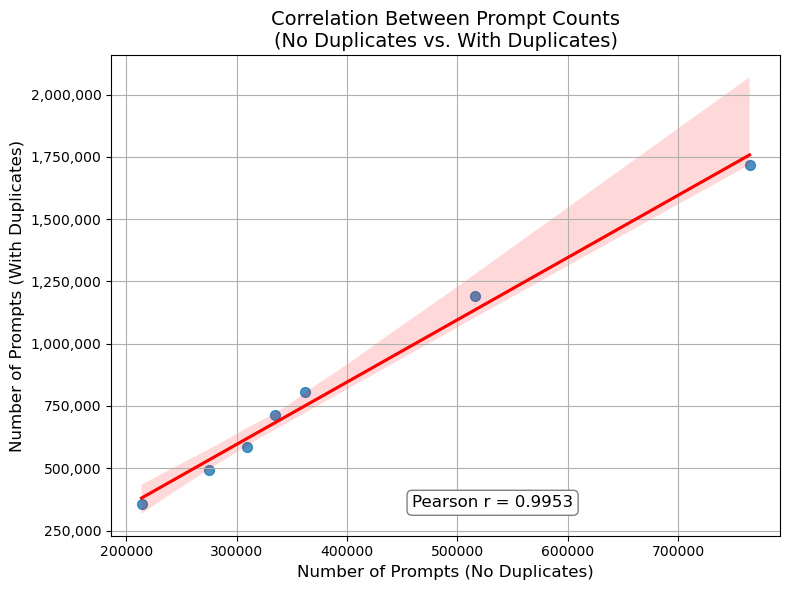}
    \caption{Correlation between Prompt Counts (No Duplicates vs. With Duplicates)}
    \label{fig:correlation_prompt_count}
\end{figure}

The following figure illustrates the proportion of duplicate prompts over time, which remains stable between 40\% and 50\% throughout the analyzed period. This stability indicates that users consistently engage in resubmitting prompts as part of their interaction with TTI systems, reflecting a regular pattern of prompt refinement and exploration. Together, these insights emphasize the dual role of duplicates in understanding user engagement and shaping the dataset’s structure.

\begin{figure}[h]
    \centering
    \includegraphics[width=0.75\linewidth]{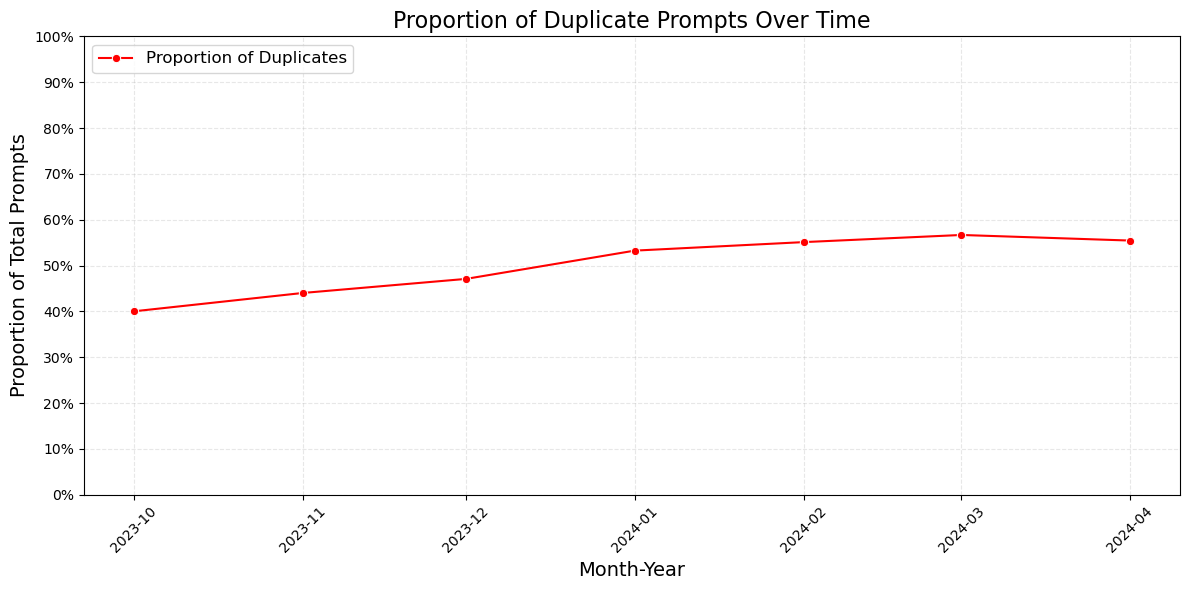}
    \caption{Proportion of Duplicate Prompts Over Time. The proportion of duplicate prompts remains consistently between 40\% and 50\% throughout the analyzed period, reflecting stable user behavior involving iterative prompt submissions and experimentation.}
    \label{fig:proportion_of_duplicates}
\end{figure}

\newpage
Table~\ref{tab:non_english_prompts} presents a breakdown of non-English prompts found in the initial Civiverse dataset. It lists the 29 most common languages along with their respective prompt counts, which were excluded from the analysis in this paper as they comprised only 3\% of the overall dataset.

\begin{table}[h]
    \centering
    \caption{Breakdown of Non-English Prompts in the Civiverse Dataset}
    \label{tab:non_english_prompts}
    \begin{tabular}{l r | l r}
        \toprule
        \textbf{Language (Code)} & \textbf{Count} & \textbf{Language (Code)} & \textbf{Count} \\
        \midrule
        Italian (it)  & 10,046  & Finnish (fi)   & 1,279 \\
        Romanian (ro) & 9,473   & Somali (so)    & 1,206 \\
        French (fr)   & 5,505   & Turkish (tr)   & 1,089 \\
        Afrikaans (af)& 5,204   & Portuguese (pt)& 1,051 \\
        Norwegian (no)& 4,424   & Polish (pl)    & 891 \\
        Catalan (ca)  & 3,970   & Swahili (sw)   & 623 \\
        Welsh (cy)    & 3,031   & German (de)    & 586 \\
        Swedish (sv)  & 3,008   & Lithuanian (lt)& 578 \\
        Estonian (et) & 2,995   & Croatian (hr)  & 484 \\
        Danish (da)   & 2,860   & Slovak (sk)    & 483 \\
        Spanish (es)  & 2,690   & Albanian (sq)  & 318 \\
        Tagalog (tl)  & 2,627   & Czech (cs)     & 269 \\
        Indonesian (id)& 2,437  & Hungarian (hu) & 188 \\
        Dutch (nl)    & 2,356   & Latvian (lv)   & 128 \\
        Slovenian (sl)& 1,509   &                &      \\
        \bottomrule
    \end{tabular}
\end{table}

\section{User Behavior Overview}

Table~\ref{tab:user_repetition_summary} provides an overview of user interaction patterns with TTI systems through repeated prompt submissions, categorizing users into three distinct groups. \textbf{Consistent Repeaters} represent the largest cohort, contributing a total of 1.95 million repeats, with an average of 76.6 repeats per user and a maximum of 1,200 repeats by a single individual. This group demonstrates a strong iterative approach, often experimenting extensively with specific ideas or styles, making them key contributors to the dataset's repeated content. \textbf{Occasional Repeaters}, while fewer in number, account for 250,000 repeats in total, averaging 35.2 repeats per user, with a maximum of 500. These users exhibit more sporadic repetition, indicating a balance between exploration and refinement in their engagement with TTI systems. \textbf{Non-Repeaters} make up the second-largest group, consisting of 21,835 users who never resubmit identical prompts. This behavior suggests either a one-time interaction with the system or a preference for crafting unique prompts without repetition.

This breakdown highlights the diversity in user interaction patterns, where consistent repeaters dominate repeated prompts, shaping much of the dataset's trends and structure. In contrast, occasional repeaters and non-repeaters reflect more exploratory or casual usage. The significant contribution of consistent repeaters emphasizes their role in refining and stabilizing certain thematic or stylistic outputs, which may influence broader community trends. 

\onecolumn
\begin{table}[h]
\centering
\caption{User Repetition Behavior Summary}
\label{tab:user_repetition_summary}
\begin{tabular}{lrrrr}
\toprule
Group & User Count & Total Repeats & Avg Repeats/User & Max Repeats \\
\midrule
Consistent Repeaters & 25,449 & 1,950,000 & 76.6 & 1,200 \\
Occasional Repeaters & 7,093 & 250,000 & 35.2 & 500 \\
Non-Repeaters & 21,835 & 0 & 0.0 & 0 \\
\bottomrule
\end{tabular}
\end{table}

\begin{table} [h]
   \centering
    \caption{Monthly Growth of Prompts and Users. The table summarizes the monthly count of unique and total prompts submitted, alongside the breakdown of unique user categories actively engaging with the Civiverse platform.}
    \begin{tabular}{lrrrrrr}
        \toprule
        \textbf{Month} & \textbf{Unique Prompts} & \textbf{Total Prompts} & \textbf{Consistent Repeater} & \textbf{Non-Repeater} & \textbf{Occasional Repeater} & \textbf{Unique Users} \\
        \midrule
        October 2023   & 213,779 & 356,368 & 2,933 & 2,782 & 2,605 & 8,320 \\
        November 2023  & 274,999 & 491,156 & 3,445 & 2,759 & 3,137 & 9,341 \\
        December 2023  & 309,649 & 585,146 & 4,146 & 3,062 & 3,344 & 10,552 \\
        January 2024   & 334,407 & 715,328 & 4,915 & 3,486 & 3,659 & 12,060 \\
        February 2024  & 362,012 & 806,377 & 5,590 & 3,548 & 3,633 & 12,771 \\
        March 2024     & 516,410 & 1,191,718 & 7,060 & 4,041 & 4,099 & 15,200 \\
        April 2024     & 764,838 & 1,716,634 & 11,618 & 5,742 & 3,769 & 21,129 \\
        \bottomrule
    \end{tabular}
\end{table}

\section{Lexical Diversity of Prompts}

\subsection{Full Dataset vs. Sampled Dataset Results}
Understanding the differences between the full dataset and a balanced monthly sample allows us to isolate trends in lexical diversity without the influence of fluctuating dataset sizes. The following tables compare total and unique word counts between these two scenarios to illustrate whether diversity metrics are robust across varying data volumes.

\begin{table}[ht]
\caption{Comparison of Full and Sampled Dataset Results}
\centering
\begin{minipage}[t]{0.48\textwidth}
\centering
\caption*{Full Dataset Results}
\begin{tabular}{lrr}
\toprule
\textbf{Month}      & \textbf{Total Words} & \textbf{Unique Words} \\ \midrule
October 2023        & 7,853,856            & 271,470               \\
November 2023       & 10,369,448           & 299,850               \\
December 2023       & 11,707,249           & 318,429               \\
January 2024        & 12,409,098           & 363,054               \\
February 2024       & 14,033,532           & 373,789               \\
March 2024          & 21,190,910           & 421,592               \\
April 2024          & 35,969,178           & 492,400               \\ \bottomrule
\end{tabular}
\end{minipage}
\hfill
\begin{minipage}[t]{0.48\textwidth}
\centering
\caption*{Sampled Dataset Results}
\begin{tabular}{lrr}
\toprule
\textbf{Month}      & \textbf{Total Words} & \textbf{Unique Words} \\ \midrule
October 2023        & 7,343,248            & 210,132               \\
November 2023       & 7,531,337            & 225,503               \\
December 2023       & 7,565,204            & 223,874               \\
January 2024        & 7,415,139            & 238,715               \\
February 2024       & 7,753,822            & 232,047               \\
March 2024          & 8,210,859            & 226,742               \\
April 2024          & 9,411,784            & 201,413               \\ \bottomrule
\end{tabular}
\end{minipage}
\end{table}

\subsection{Top 10 Words and 4-Grams Per Month}

To capture the most commonly used descriptors and linguistic structures, we analyzed the frequency of individual words and 4-grams within the dataset. The tables below summarize the top-ranking words and phrases for each month, highlighting the dominant trends and their persistence over time.

% October 2023
\begin{table}[H]
    \centering
    \begin{minipage}[t]{0.48\textwidth}
        \centering
        \caption{Top 10 Words in October 2023}
        \begin{tabular}{@{}ll@{}}
            \toprule
            Rank & Word (\%) \\
            \midrule
            1 & hair (73.66) \\
            2 & detailed (72.05) \\
            3 & quality (51.08) \\
            4 & eye (47.47) \\
            5 & best (44.60) \\
            6 & masterpiece (40.34) \\
            7 & high (30.70) \\
            8 & face (29.79) \\
            9 & breast (27.99) \\
            10 & 1girl (27.46) \\
            \bottomrule
        \end{tabular}
    \end{minipage}%
    \hfill
    \begin{minipage}[t]{0.48\textwidth}
        \centering
        \caption{Top 10 4-Grams in October 2023}
        \begin{tabular}{@{}ll@{}}
            \toprule
            Rank & 4-Gram (\%) \\
            \midrule
            1 & tom cruise tom cruise (22.29) \\
            2 & cruise tom cruise tom (22.29) \\
            3 & detailed cg unity 8k (21.61) \\
            4 & high quality film grain (21.48) \\
            5 & cg unity 8k wallpaper (20.32) \\
            6 & soft lighting high quality (20.24) \\
            7 & lighting high quality film (20.12) \\
            8 & extremely detailed cg unity (17.02) \\
            9 & masterpiece best quality high (13.79) \\
            10 & best quality high quality (11.57) \\
            \bottomrule
        \end{tabular}
    \end{minipage}
\end{table}

% November 2023
\begin{table}[H]
    \centering
    \begin{minipage}[t]{0.48\textwidth}
        \centering
        \caption{Top 10 Words in November 2023}
        \begin{tabular}{@{}ll@{}}
            \toprule
            Rank & Word (\%) \\
            \midrule
            1 & hair (76.40) \\
            2 & detailed (75.59) \\
            3 & quality (49.28) \\
            4 & eye (46.76) \\
            5 & best (40.17) \\
            6 & masterpiece (37.30) \\
            7 & high (33.66) \\
            8 & breast (31.12) \\
            9 & face (27.75) \\
            10 & long (27.14) \\
            \bottomrule
        \end{tabular}
    \end{minipage}%
    \hfill
    \begin{minipage}[t]{0.48\textwidth}
        \centering
        \caption{Top 10 4-Grams in November 2023}
        \begin{tabular}{@{}ll@{}}
            \toprule
            Rank & 4-Gram (\%) \\
            \midrule
            1 & high quality film grain (25.96) \\
            2 & detailed cg unity 8k (25.57) \\
            3 & cg unity 8k wallpaper (22.20) \\
            4 & extremely detailed cg unity (21.15) \\
            5 & 8k uhd dslr high (20.91) \\
            6 & uhd dslr high quality (19.91) \\
            7 & dslr high quality film (19.91) \\
            8 & detailed skin11 8k uhd (17.75) \\
            9 & skin11 8k uhd dslr (17.70) \\
            10 & shiny glossy translucent clothing11 (16.69) \\
            \bottomrule
        \end{tabular}
    \end{minipage}
\end{table}

% December 2023
\begin{table}[H]
    \centering
    \begin{minipage}[t]{0.48\textwidth}
        \centering
        \caption{Top 10 Words in December 2023}
        \begin{tabular}{@{}ll@{}}
            \toprule
            Rank & Word (\%) \\
            \midrule
            1 & detailed (77.48) \\
            2 & hair (72.96) \\
            3 & quality (47.35) \\
            4 & eye (47.21) \\
            5 & best (40.18) \\
            6 & masterpiece (38.11) \\
            7 & high (33.50) \\
            8 & body (26.70) \\
            9 & skin (26.61) \\
            10 & long (26.52) \\
            \bottomrule
        \end{tabular}
    \end{minipage}%
    \hfill
    \begin{minipage}[t]{0.48\textwidth}
        \centering
        \caption{Top 10 4-Grams in December 2023}
        \begin{tabular}{@{}ll@{}}
            \toprule
            Rank & 4-Gram (\%) \\
            \midrule
            1 & 8k uhd dslr high (29.31) \\
            2 & detailed skin11 8k uhd (29.11) \\
            3 & skin11 8k uhd dslr (28.11) \\
            4 & detailed cg unity 8k (26.83) \\
            5 & masterpiece best quality highres (19.69) \\
            6 & extremely detailed cg unity (18.60) \\
            7 & cg unity 8k wallpaper (18.55) \\
            8 & masterpiece best quality 1girl (18.28) \\
            9 & realistic32k masterpiece12high detailed skin11 (17.21) \\
            10 & ultra realistic32k masterpiece12high detailed (13.20) \\
            \bottomrule
        \end{tabular}
    \end{minipage}
\end{table}

% January 2024
\begin{table}[H]
    \centering
    \begin{minipage}[t]{0.48\textwidth}
        \centering
        \caption{Top 10 Words in January 2024}
        \begin{tabular}{@{}ll@{}}
            \toprule
            Rank & Word (\%) \\
            \midrule
            1 & hair (71.45) \\
            2 & detailed (68.88) \\
            3 & eye (43.92) \\
            4 & quality (42.35) \\
            5 & best (37.85) \\
            6 & masterpiece (36.01) \\
            7 & high (31.90) \\
            8 & breast (28.27) \\
            9 & skin (25.77) \\
            10 & body (24.97) \\
            \bottomrule
        \end{tabular}
    \end{minipage}%
    \hfill
    \begin{minipage}[t]{0.48\textwidth}
        \centering
        \caption{Top 10 4-Grams in January 2024}
        \begin{tabular}{@{}ll@{}}
            \toprule
            Rank & 4-Gram (\%) \\
            \midrule
            1 & masterpiece best quality 1girl (32.10) \\
            2 & detailed cg unity 8k (31.99) \\
            3 & cg unity 8k wallpaper (30.62) \\
            4 & extremely detailed cg unity (30.37) \\
            5 & highly detailed high budget (28.35) \\
            6 & masterpiece best quality highres (24.33) \\
            7 & ultra realistic32k masterpiece12high detailed (24.27) \\
            8 & realistic32k masterpiece12high detailed skin11 (23.27) \\
            9 & masterpiece12high detailed skin11 high (23.26) \\
            10 & detailed skin11 high quality11 19.26) \\
            \bottomrule
        \end{tabular}
    \end{minipage}
\end{table}

% February 2024
\begin{table}[H]
    \centering
    \begin{minipage}[t]{0.48\textwidth}
        \centering
        \caption{Top 10 Words in February 2024}
        \begin{tabular}{@{}ll@{}}
            \toprule
            Rank & Word (\%) \\
            \midrule
            1 & hair (75.23) \\
            2 & detailed (65.58) \\
            3 & eye (46.75) \\
            4 & quality (41.99) \\
            5 & best (34.14) \\
            6 & masterpiece (33.57) \\
            7 & breast (32.77) \\
            8 & high (31.65) \\
            9 & body (26.97) \\
            10 & black (26.38) \\
            \bottomrule
        \end{tabular}
    \end{minipage}%
    \hfill
    \begin{minipage}[t]{0.48\textwidth}
        \centering
        \caption{Top 10 4-Grams in February 2024}
        \begin{tabular}{@{}ll@{}}
            \toprule
            Rank & 4-Gram (\%) \\
            \midrule
            1 & score9 score8up score7up score6up (39.58) \\
            2 & score8up score7up score6up score5up (39.19) \\
            3 & score7up score6up score5up score4up (38.90) \\
            4 & best quality high quality (31.76) \\
            5 & masterpiece best quality 1girl (28.51) \\
            6 & detailed cg unity 8k (28.44) \\
            7 & masterpiece12high detailed skin11 high (28.39) \\
            8 & ultra realistic32k masterpiece12high detailed (19.38) \\
            9 & realistic32k masterpiece12high detailed skin11 (18.38) \\
            10 & detailed skin11 high quality11 (18.32) \\
            \bottomrule
        \end{tabular}
    \end{minipage}
\end{table}

% March 2024
\begin{table}[H]
    \centering
    \begin{minipage}[t]{0.48\textwidth}
        \centering
        \caption{Top 10 Words in March 2024}
        \begin{tabular}{@{}ll@{}}
            \toprule
            Rank & Word (\%) \\
            \midrule
            1 & hair (80.18) \\
            2 & detailed (57.64) \\
            3 & eye (48.81) \\
            4 & breast (39.15) \\
            5 & quality (34.56) \\
            6 & black (31.70) \\
            7 & long (29.07) \\
            8 & high (28.64) \\
            9 & best (28.14) \\
            10 & masterpiece (27.97) \\
            \bottomrule
        \end{tabular}
    \end{minipage}%
    \hfill
    \begin{minipage}[t]{0.48\textwidth}
        \centering
        \caption{Top 10 4-Grams in March 2024}
        \begin{tabular}{@{}ll@{}}
            \toprule
            Rank & 4-Gram (\%) \\
            \midrule
            1 & score9 score8up score7up score6up (39.04) \\
            2 & score8up score7up score6up score5up (39.11) \\
            3 & score7up score6up score5up score4up (39.19) \\
            4 & score9 score8up score7up sourceanime (39.78) \\
            5 & score9 score8up score7up 1girl (37.68) \\
            6 & best quality masterpiece highres (28.59) \\
            7 & style sdxllorapony diffusion v6 (27.53) \\
            8 & quality masterpiece highres solo (27.46) \\
            9 & masterpiece12high detailed skin11 high (22.12) \\
            10 & detailed skin11 high quality11 (20.12) \\
            \bottomrule
        \end{tabular}
    \end{minipage}
\end{table}

% April 2024
\begin{table}[H]
    \centering
    \begin{minipage}[t]{0.48\textwidth}
        \centering
        \caption{Top 10 Words in April 2024}
        \begin{tabular}{@{}ll@{}}
            \toprule
            Rank & Word (\%) \\
            \midrule
            1 & hair (92.97) \\
            2 & eye (60.40) \\
            3 & breast (59.40) \\
            4 & detailed (54.97) \\
            5 & score8up (46.47) \\
            6 & score9 (46.11) \\
            7 & black (40.01) \\
            8 & face (39.87) \\
            9 & score7up (38.06) \\
            10 & body (36.96) \\
            \bottomrule
        \end{tabular}
    \end{minipage}%
    \hfill
    \begin{minipage}[t]{0.48\textwidth}
        \centering
        \caption{Top 10 4-Grams in April 2024}
        \begin{tabular}{@{}ll@{}}
            \toprule
            Rank & 4-Gram (\%) \\
            \midrule
            1 & score9 score8up score7up score6up (41.42) \\
            2 & score8up score7up score6up score5up (41.70) \\
            3 & score7up score6up score5up score4up (41.47) \\
            4 & score9 score8up score7up 1girl (39.44) \\
            5 & score9 score8up score7up sourceanime (38.32) \\
            6 & style sdxllorapony diffusion v6 (36.85) \\
            7 & best quality highly detailed (32.12) \\
            8 & break score9 score8up score7up (31.07) \\
            9 & score6up score5up score4up sourceanime (27.00) \\
            10 & twilight style sdxllorapony diffusion (26.94) \\
            \bottomrule
        \end{tabular}
    \end{minipage}
\end{table}

\subsection{Word Frequency Analysis}

A closer look at word frequency counts illustrated in Tab.~\ref{tab:top_words}, reinforces the observed differences in linguistic behaviors among user categories. While common terms like "hair," "detailed," and "quality" appear across all groups, their use is especially pronounced among consistent repeaters, with "hair" included in 91\% of their submissions, compared to 79\% for occasional repeaters and 50\% for non-repeaters. Similarly, disparities arise for "breast" and "face," which occur more often in prompts by consistent repeaters. In contrast, non-repeaters display a broader range of word usage, consistent with their higher TTR and ENW values.

\begin{table}[ht]
\centering
\begin{multicols}{3}

\subsection*{Consistent Repeaters}
\begin{tabular}{@{}rlrr@{}}
\toprule
\# & Word        & Count   & Prompt (\%) \\ \midrule
1    & hair        & 83{,}178  & 91.26\%     \\
2    & detailed    & 68{,}646  & 75.43\%     \\
3    & eye         & 55{,}309  & 60.75\%     \\
4    & breast      & 47{,}591  & 52.29\%     \\
5    & quality     & 46{,}277  & 50.82\%     \\
6    & face        & 39{,}590  & 43.45\%     \\
7    & high        & 39{,}424  & 43.27\%     \\
8    & best        & 38{,}976  & 42.76\%     \\
9    & black       & 38{,}503  & 42.34\%     \\
10   & masterpiece & 38{,}213  & 41.92\%     \\ \bottomrule
\end{tabular}

\subsection*{Occasional Repeaters}
\begin{tabular}{@{}rlrr@{}}
\toprule
\# & Word        & Count   & Prompt (\%) \\ \midrule
1    & hair        & 72{,}165  & 79.27\%     \\
2    & detailed    & 58{,}651  & 64.33\%     \\
3    & eye         & 47{,}232  & 51.81\%     \\
4    & quality     & 34{,}829  & 38.31\%     \\
5    & best        & 31{,}037  & 34.12\%     \\
6    & breast      & 30{,}481  & 33.46\%     \\
7    & masterpiece & 28{,}935  & 31.82\%     \\
8    & black       & 25{,}666  & 28.21\%     \\
9    & girl        & 25{,}289  & 27.78\%     \\
10   & long        & 23{,}865  & 26.19\%     \\ \bottomrule
\end{tabular}

\subsection*{Non-Repeaters}
\begin{tabular}{@{}rlrr@{}}
\toprule
\# & Word        & Count   & Prompt (\%) \\ \midrule
1    & hair        & 45{,}193  & 49.66\%     \\
2    & detailed    & 35{,}963  & 39.53\%     \\
3    & eye         & 31{,}776  & 34.89\%     \\
4    & quality     & 25{,}770  & 28.29\%     \\
5    & breast      & 19{,}043  & 20.93\%     \\
6    & black       & 16{,}447  & 18.08\%     \\
7    & best        & 16{,}354  & 17.95\%     \\
8    & masterpiece & 14{,}523  & 15.96\%     \\
9    & girl        & 12{,}997  & 14.28\%     \\
10   & long        & 11{,}816  & 13.00\%     \\ \bottomrule
\end{tabular}

\end{multicols}
\caption{Top Common Words in Prompts Across User Categories and \% of Their Appearance in Prompts.}
\label{tab:top_words}
\end{table}

Taken together, these findings suggest that frequent re-submission of familiar descriptors can contribute to a narrower overall vocabulary and the reinforcement of certain linguistic or thematic templates. By contrast, one-off or experimental prompts may introduce greater variety, although they may have less cumulative influence on the platform’s collective lexicon. Moreover, some consistent repeaters appear to iterate heavily on a small set of base prompts—re-submitting nearly identical text with minor token adjustments—until they obtain their ideal output. While this strategy can be effective in refining prompt outcomes, it can also amplify reliance on a narrower set of descriptors within the community.

\subsection{Lexical Diversity Among User Categories}

The following Fig. presents the evolution of Self-Repetition Score (SRP) and Effective Number of Words (ENW) for each user category over time, highlighting distinct patterns in lexical behavior. Although SRP generally rises (and ENW declines) across all categories over the observed months, \emph{Non-Repeaters} consistently show higher ENW than the other groups, suggesting that, despite submitting fewer total prompts, they draw upon a broader or more varied vocabulary in their short-term usage.

\begin{figure}[h]
    \centering
    \includegraphics[width=0.80\linewidth]{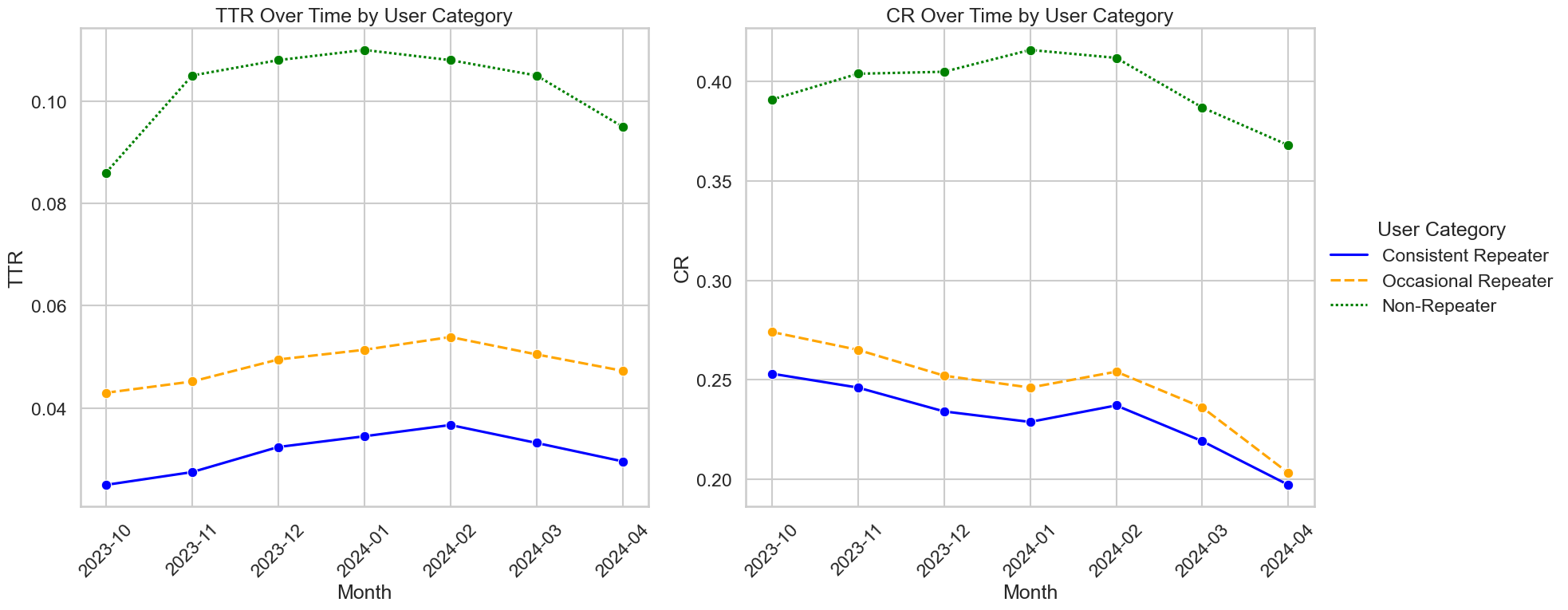}
    \label{fig:correlation_prompt_count}
\end{figure}

\begin{figure}[h]
    \centering
    \includegraphics[width=0.80\linewidth]{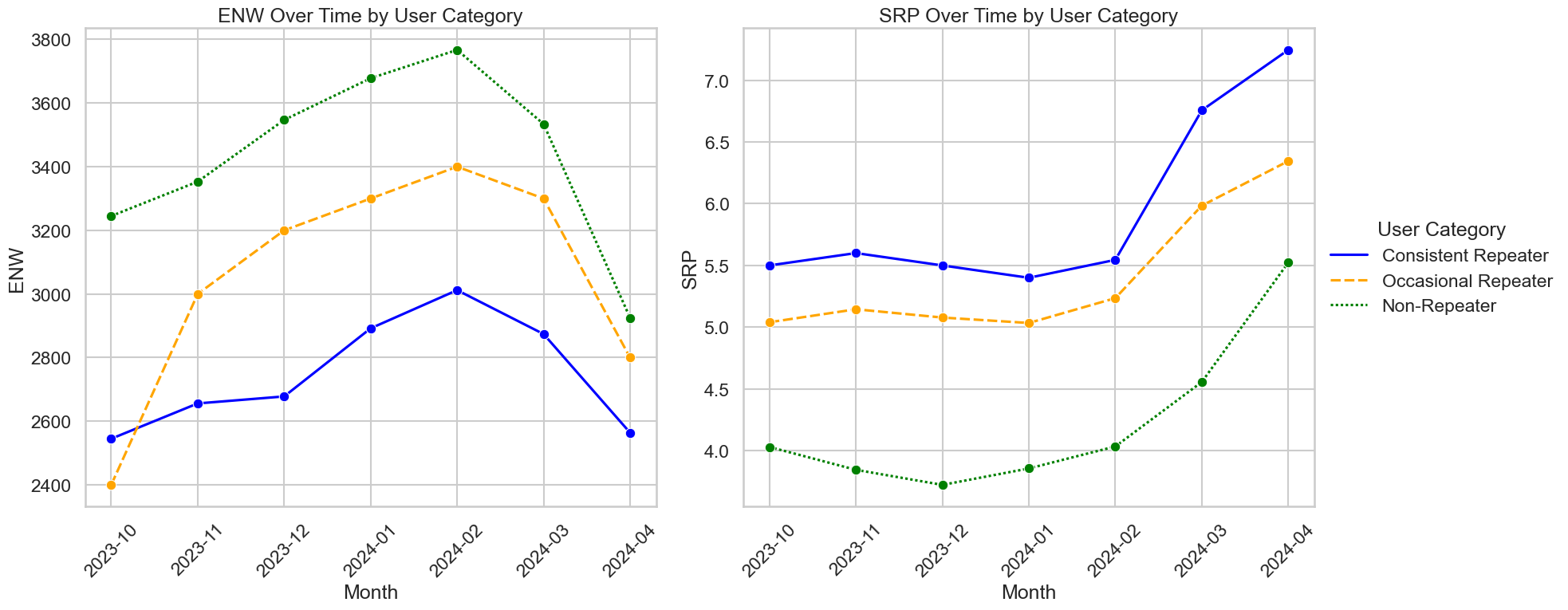}
    \caption{TTR, CR, SRP and ENW Scores Over Time per User Category.}
    \label{fig:proportion_of_duplicates}
\end{figure}

\newpage
\subsection{Analysis of Prompt Similarities and Repetitive Templates}

This table presents the remaining prompt patterns within specific clusters, detailing the total prompts, unique users, and proportional contributions of consistent repeaters, occasional repeaters, and non-repeaters for each cluster.

\begin{table}[ht]
\centering
\small
\begin{tabular}{p{7cm}rcccccc}
\toprule
\textbf{Prompt Pattern} 
& \textbf{\# Prompts} 
& \textbf{Unique Users} 
& \textbf{\% Consist.} 
& \textbf{\% Occ.} 
& \textbf{\% Non-Rep.} 
& \textbf{Emergence} \\ 
\midrule
\textbf{score\_9 score\_8\_up score\_7\_up, best quality11, masterpiece12, realistic, ejaculation, sperm, 1girl, detailed background, looking at viewer, seductive expression} 
    & 30,575   & 3,723 & 95.02\% & 3.47\%  & 1.51\% & February 2024 \\ \hline
\textbf{european woman, closeup, no pants, necklace, perfect face, purple iris, pale skin, no pores, depth field} 
    & 28,198   & 308   & 100.00\% & 0.00\% & 0.00\% & January 2024 \\ \hline
\textbf{masterpiece12, best quality12, 8k, hdr, ultra detailed, photorealistic, professional light, cinematic lighting, fashion photography, loragothic\_outfit06, 1girl, looking viewer} 
    & 24,170   & 4,656 & 4.19\% & 95.67\% & 0.14\% & April 2024 \\ \hline
\textbf{realistic13, detailed quality, sexy, sitting, shy, masterpiece12, photorealistic, nipples, naked, intricate details, pussy wet} 
    & 23,102   & 6,898 & 97.76\% & 1.24\%  & 1.00\% & March 2024 \\ \hline
\textbf{woman, in christmaspark, 11wearing furcoat114k, christmas tree, raw photo, best quality, depth field, ultra high resolution, masterpiece, ultradetailed12} 
    & 17,916   & 2,234 & 2.40\%  & 97.55\% & 0.05\% & December 2024 \\ \hline
\textbf{cinematic film, still life, loraquiron\_tarrawhite\_v1\_lora087, depth field, vignette, highly detailed, epic, gorgeous, film grain} 
    & 13,651   & 2,397 & 59.18\% & 37.80\% & 3.03\% & March 2024 \\ \hline
\textbf{full body, 2genital, gloss12, plump fat, overweight, chubby165, completely nude, nipplewhite, horns, body15, masterpiece, male anthro} 
    & 12,506   & 2,890 & 100.00\% & 0.00\% & 0.00\% & February 2024 \\ 
\bottomrule
\end{tabular}
\caption{Remaining Near-Duplicate Prompt Clusters.}
\label{tab:remaining_prompt_clusters}
\end{table}

\subsection{Prompt Similarity Over Time}

This table provides an overview of prompts with a similarity score above 0.8, detailing the number of prompts, unique users, and the proportional contributions of consistent repeaters, occasional repeaters, and non-repeaters over time. From October 2023 to April 2024, there is a steady rise in the number of prompts and unique users, with consistent repeaters becoming increasingly dominant, climbing from 68.40\% to 83.89\%. Meanwhile, the contributions of occasional and non-repeaters decrease over the same period, highlighting a trend toward more formulaic and refined prompt patterns.

\begin{table}[ht]
\label{tab:month_above08}
\centering
\caption{Monthly Prompts Above~0.8 Similarity}
\begin{tabular}{lrrrrr}
\toprule
Month      & \# Prompts & Unique Users & Consist. (\%) & Occ.(\%) & Non-Rep. (\%) \\
\midrule
2023-10-01 & 124,416    & 5,046        & 68.40         & 28.54    & 3.06          \\
2023-11-01 & 164,697    & 5,724        & 70.69         & 27.32    & 1.99          \\
2023-12-01 & 188,124    & 6,402        & 73.14         & 25.20    & 1.66          \\
2024-01-01 & 200,773    & 7,079        & 72.27         & 26.12    & 1.61          \\
2024-02-01 & 227,976    & 7,643        & 77.03         & 21.32    & 1.65          \\
2024-03-01 & 350,780    & 9,768        & 79.79         & 18.82    & 1.39          \\
2024-04-01 & 564,970    & 15,304       & 83.89         & 14.51    & 1.60          \\
\bottomrule
\end{tabular}
\end{table}

This table focuses on prompts with a similarity score below 0.8, presenting similar metrics to the first table. The number of prompts and unique users increases significantly over time, but the proportional contributions of consistent repeaters also grow, from 54.93\% to 70.25\%. Occasional repeaters and non-repeaters contribute more in this group than in the above-0.8 category, with occasional repeaters accounting for up to 36.62\% initially, though their share declines to 23.39\% by April 2024. Non-repeaters consistently contribute around 6–8\%, underscoring their role in maintaining some diversity within this cohort.

\begin{table}[ht]
\label{tab:month_below08}
\centering
\caption{Monthly Prompts Below~0.8 Similarity}
\begin{tabular}{lrrrrr}
\toprule
Month      & \# Prompts & Unique Users & Consist. (\%) & Occ.(\%) & Non-Rep. (\%) \\
\midrule
2023-10-01 & 89,388     & 7,301        & 54.93         & 36.62    & 8.44          \\
2023-11-01 & 110,333    & 8,248        & 55.33         & 38.12    & 6.55          \\
2023-12-01 & 121,538    & 9,191        & 57.96         & 36.25    & 5.79          \\
2024-01-01 & 133,644    & 10,643       & 59.84         & 34.56    & 5.60          \\
2024-02-01 & 134,048    & 11,145       & 63.10         & 30.82    & 6.07          \\
2024-03-01 & 165,645    & 13,075       & 66.64         & 27.76    & 5.59          \\
2024-04-01 & 199,886    & 17,345       & 70.25         & 23.39    & 6.36          \\
\bottomrule
\end{tabular}
\end{table}

\onecolumn
\section{Semantic Diversity of Prompts}
\label{appendix:semantic_diversity_over_time}

\subsection{Topic Modeling Results by Month}
This appendix presents the topics identified through our topic modeling process for each month, highlighting key areas of focus. The topics are grouped by month and displayed in the following sections for clarity and reference.

%============================================================
\subsection*{October 2023 (71 topics, 6904 specifiers)}

\begin{multicols}{2}
\raggedcolumns
\begin{enumerate}
    \item Anime/Cosplay \& Characters         % was: Anime/Cosplay & Character References
    \item Outdoor Scenes                      % was: City/Outdoor Scenes
    \item Stable Diffusion Lora References, Lingerie
    \item Lighting \& Reflections
    \item Food Elements
    \item Hair \& Styling
    \item Age \& Generational Descriptors
    \item Fantasy \& Magic
    \item Aquatic Scenes
    \item Male Personal Names                 % was: Male Personal Names (no change needed, but consistent)
    \item Explicit Sexual Content (NSFW)      % was: Explicit Sexual Content (NSFW) References
    \item Weather Elements
    \item Animal Hybrids
    \item Colorful \& Vibrant
    \item Body \& Muscle/Prosthesis
    \item Scale \& Size Descriptors
    \item Photorealism \& Ultra Detail
    \item Composition
    \item Orientation \& Angles
    \item Geometry \& Symmetry
    \item Mood \& Emotions
    \item Cyberpunk \& Tech
    \item Portrait Illustrations
    \item Ethnic \& Cultural Descriptors
    \item Touch/Hand Details
    \item High-Resolution \& Pixel
    \item Positional \& (Some) Nude Angles
    \item Encoded Lora/Stable Diffusion Details
    \item Materials (Wood/Metal/Stone)
    \item Quality/Resolution Descriptors
    \item X-ray
    \item Camera Gear
    \item Viewpoint \& Illustration
    \item Nuclear/Destruction
    \item Skin \& Pores
    \item Nipples \& Breasts (NSFW)           % was: NSFW: Large Breasts
    \item Eyes \& Eye Details
    \item Music \& Performance
    \item Positive/Complimentary Adjectives
    \item Playful/Pointy Clothing
    \item Gaming \& Combat (Lovecraftian/Minecraft/etc.)
    \item Mouth/Lip/Fang Details
    \item Photography \& Photo Editing
    \item Angles \& High/Low Cuts
    \item Titles \& Relationships (Lord, Maid, etc.)
    \item Crime
    \item Female Names                        % was: Female Personal Names
    \item Historical \& Retro Aesthetics
    \item Power \& Transformation
    \item Facial Expressions (Wink/Laugh)
    \item Sex-Related (Male/Female Body Parts)
    \item Gender References
    \item Fire
    \item Times of Day (Morning/Noon/Night)
    \item Weapons \& Tools
    \item Animal/Rabbit Ears
    \item Legwear \& Thigh-High Clothing
    \item Sports \& Athletics                % was: Sports & Athletic Activities
    \item Sci-Fi (Star Wars/Marvel)
    \item Looking at Viewer / POV Angles
    \item Textiles \& Lines
    \item Well-Proportioned Textures
    \item Public/Unseen \& Revealing
    \item Cold/Warm Contrasts
    \item Academic/University \& Research
    \item Tension
    \item Ties \& Whips
    \item Mystery \& Unexpected
    \item Background Details
    \item Painting Styles
    \item Cinematic Effects
\end{enumerate}
\end{multicols}

%============================================================
\subsection*{November 2023 (73 topics, 8633 specifiers)}

\begin{multicols}{2}
\raggedcolumns
\begin{enumerate}
    \item Anime/Cosplay \& Characters        % was: Anime/Cosplay & Character References
    \item Biomedicine \& Biomedical
    \item LED
    \item Polaroid Effects
    \item Hair \& Hairstyles
    \item Beard \& Mustache
    \item Male Personal Names                % was: Male Names (Liam, Tom, etc.)
    \item Explicit Sexual Content (NSFW)     % was: Explicit Sexual Content (NSFW) References
    \item Ethnic \& Cultural Descriptors
    \item Mythical Themes
    \item Quality Masterpiece
    \item Numbers
    \item Linear \& Origami Geometry
    \item Emotional States
    \item Lingerie
    \item Minimalistic Proportions
    \item Organic Items
    \item Animals (Leopard, Rabbit, Raccoon)
    \item Adapted Lora Models
    \item Undersized \& Pose (Peak Turn)
    \item Lovecraft
    \item Symphony \& Music
    \item Photoreal \& Ultra Detailed
    \item Sweet \& Magnificent
    \item Marine \& Nautical Scenes
    \item Depth of Field
    \item Ultra Detailed Fractal
    \item Nipples \& Breasts (NSFW)          % was: Medium Breasts & Nipples (NSFW)
    \item Obese \& Shabby/Plump
    \item Eyes
    \item Sunglasses
    \item Liquid, Rain \& Moist
    \item Lips, Mouth \& Skin
    \item Tongue \& Piercings
    \item Female Names                       % was: Female Names (Mary, Dorothea, etc.)
    \item Grayscale \& Palette
    \item UE5 \& R5 Configurations
    \item Medieval \& Renaissance Themes
    \item Looking at Viewer \& Perspective
    \item Overdetailed \& Face Details, Leave
    \item Letter
    \item Symbol \& Font
    \item Composition
    \item Lighting \& Sunset
    \item Refined \& Raw, Nouveau
    \item Supervillain References
    \item Malevolent \& Supergirl
    \item Military Combat
    \item Mercury
    \item Jewelry
    \item LoraAdd Detail \& Sharpness
    \item Supernova \& Galactic
    \item Tactical Weapons
    \item Physically-Based Body
    \item LoraJacket \& LoraPixel
    \item PhotoUltra \& Nikon
    \item Prettify \& Style
    \item Muscular \& Physique
    \item LLBreton
    \item POM \& Perky
    \item Legwear
    \item Tummy \& Pelvis
    \item Photoshop \& Tumblr
    \item Low \& Highrise Angles
    \item Pastell \& Pastel Insertion
    \item Plane \& Pilot Flight
    \item Suspenseful \& Fear
    \item Pregnant \& Family
    \item Age References
    \item Color Palette
    \item Background Details
    \item Painting Styles
    \item Cinematic Effects
\end{enumerate}
\end{multicols}

%============================================================
\subsection*{December 2023 (73 topics, 9479 specifiers)}

\begin{multicols}{2}
\raggedcolumns
\begin{enumerate}
    \item LED
    \item Ninja References
    \item Anime/Cosplay \& Characters       % was: Anime/Cosplay & Character References
    \item Ethnic \& Cultural Descriptors
    \item Room Interiors
    \item Explicit Sexual Content (NSFW)    % was: Explicit Sexual Content (NSFW) References
    \item Lingerie
    \item Hair \& Hairstyles
    \item Phoenix Trees
    \item Pony Pets
    \item Cafe \& Food
    \item P30 Numbering
    \item Mechanics \& Research
    \item Makeup
    \item Length \& Oversize Proportions
    \item Hair Colors
    \item Lovely \& Remarkable
    \item Body Types
    \item Nipples \& Breasts (NSFW)         % was: Nipple & Covered Nipples (NSFW)
    \item Lightrealistic \& Photoreal
    \item Materials (Leather, Titan)
    \item Sniper Trigger, Photo1girl
    \item Myth \& Magical
    \item Female Names
    \item Modeunique \& Quality
    \item Nikke Photography \& Tumblr
    \item Composition
    \item Officers \& Police
    \item Male Focus \& Optic
    \item Low \& Overexposure
    \item UE5 \& R5
    \item Painful Expressions
    \item Harmony \& Peaceful
    \item Neckerchief \& Rabbit Ears
    \item Lightroom \& Recording
    \item Tracer \& Tracing
    \item LoraAdd Details \& Bulma
    \item Lift \& Resistance
    \item Legwear \& Nike
    \item Marketplace \& Merchant
    \item Ragged \& Loose
    \item Ugly Facial Traits
    \item Looking at Viewer
    \item Moon \& Stars
    \item Stardust
    \item Puff \& Smoke
    \item Haute Couture
    \item Melting \& Snowball
    \item Facial Details
    \item Frightening Expressions
    \item Musi \& Opera
    \item LU1 \& 4DCG
    \item Passage \& Transportation
    \item Leggings
    \item Prettified Details
    \item Training \& Cheerleading
    \item Lettering \& Logo
    \item Timeless \& Century Past
    \item Mortal \& Reaper
    \item Lora Sliders (Age/Weight/Detail)
    \item Portraying \& Presentation
    \item Orgasm References (NSFW)
    \item Mirrorless Details
    \item Tornado \& Meteor
    \item Malevolent \& Mischievous
    \item Medieval \& Renaissance Themes
    \item Male Personal Names              % was: Male Names (Liam, Tom, etc.)
    \item Mythical Themes
    \item Life \& Lively
    \item Color Palette
    \item Background Details
    \item Painting Styles
    \item Cinematic Effects
\end{enumerate}
\end{multicols}

%============================================================
%============================================================
\subsection*{January 2024 (78 topics, 10036 specifiers)}

\begin{multicols}{2}
\raggedcolumns
\begin{enumerate}
    \item Life \& Lively
    \item Exquisite Emotions
    \item Lora \& Model References
    \item Legacy
    \item Librarian
    \item Postcard
    \item Lightroom \& Photographical
    \item Ethnic \& Cultural Descriptors, Liam
    \item Male Personal Names            % was: Male Names
    \item Explicit Sexual Content (NSFW)
    \item Erotic Positions
    \item Lingerie
    \item Ohwx FHD \& R5 CG, Mohawk
    \item Beards \& Mustache
    \item Painting Styles                % was: Painting Style
    \item Illustration \& Artanime
    \item Length \& Sizes
    \item Minimalism \& Proportions
    \item LED
    \item Law \& Order
    \item Rebellion
    \item LoraQuality \& Masterpiece
    \item Receding \& Outer
    \item Color Palette
    \item Lifesizebody \& Physicallybased
    \item Queer Community
    \item Nipples \& Breasts (NSFW)      % was: Nipple & Nipples Exposed (NSFW)
    \item Mage
    \item Demonic \& Priestess
    \item Looking at Viewer
    \item LightRealistic \& Photohyperrealism
    \item Makeup
    \item Jewelleries
    \item Horns
    \item Rabbit Ears
    \item Pixelated \& Hightech
    \item Female Names                   % was: Female Names (no change, but consistent)
    \item lora quality models
    \item Artist Names
    \item Planet \& Space
    \item Origami
    \item Passage \& Transportation
    \item Music \& Instruments
    \item Genitalia (NSFW)
    \item Perspective \& Observing
    \item Letters \& Fonds
    \item Times of Day (Morning/Noon/Night)  % was: Time of the Day
    \item Textures \& Fabrics
    \item Outdoor Scenes                 % was: Outdoor Places
    \item Exposure
    \item Food \& Drinks
    \item Obese \& Plump
    \item Plane \& Jet Pilot
    \item Food Materials
    \item Skin \& Texture
    \item Lora Sliders (Age/Weight/Detail)
    \item Sports \& Athletics            % was: Sports & Athletics (unchanged if already consistent)
    \item Legs \& Leggins
    \item Nuance \& Mimic
    \item Outfits \& Clothing            % unify if matching “Clothing \& Outfits”
    \item Animals
    \item Guns \& Pistols
    \item Mirror \& Reflection
    \item Hair Colors
    \item Lighting Details
    \item Naturalism
    \item Outdoor Scenes                 % was: Outdoor Spaces
    \item Eyes \& Pupils
    \item Face Details
    \item Topwear \& Nike
    \item Moobs \& Boho Mane
    \item Pit
    \item Aesthetic Adjectives           % was: Aeshetic Adjectives
    \item Ugly Expressions
    \item Anime/Cosplay \& Characters    % was: Anime/Cosplay & Character References
    \item Background Details
    \item Painting Styles
    \item Cinematic Effects
\end{enumerate}
\end{multicols}

%============================================================
\subsection*{February 2024 (81 topics, 11056 specifiers)}

\begin{multicols}{2}
\raggedcolumns
\begin{enumerate}
    \item Anime/Cosplay \& Characters      % was: Anime/Cosplay & Character References
    \item Indoor Spaces
    \item Parks
    \item Male Personal Names             % was: Male Names
    \item Hair \& Hairstyles              % was: Hairstyles (unified with “Hair & Hairstyles”)
    \item Ribbon \& Twintails
    \item Liquid
    \item Rainy \& Foggy
    \item Explicit Sexual Content (NSFW)
    \item Lingerie
    \item Outdoor Scenes                  % was: Outdoor Spaces
    \item Nature \& Natural Environment
    \item Clothing \& Outfits
    \item Noisy Sounds
    \item Linear \& Isometric
    \item Textures
    \item Animals
    \item Materials (Nylon)
    \item Ethnic \& Cultural Descriptors
    \item Radiant \& Luminous
    \item Length \& Proportions
    \item Nipples \& Breasts (NSFW)
    \item Color Palette
    \item Sea Animals
    \item Flowers
    \item Female Names
    \item Peaceful Expressions
    \item Emotions \& Moods
    \item Electronics
    \item Robots
    \item Photoreal \& Ultra Detailed     % was: LightingMasterpiece & Ultra Quality (if you’re unifying these)
    \item Outdoor Scenes                  % was: Outdoor Locations
    \item Exhibit \& Representation
    \item Nervous \& Rebellious
    \item Photogenic Shots
    \item P30 \& 38
    \item Leotard \& Primordial Polaroid
    \item Female References
    \item Light \& Lighting
    \item Queer Community
    \item Battle
    \item Drinks \& Food
    \item Eyes \& Pupils
    \item Makeup
    \item Skin Details
    \item Pores \& Skin
    \item Looking at Viewer
    \item Cinematic Effects               % was: Cinematic Details
    \item Genitalia (NSFW)
    \item Myth \& Mystic
    \item Times of Day (Morning/Noon/Night)
    \item Maker \& Design
    \item Tongue \& Piercing
    \item Birthday
    \item Background Details
    \item Numbers
    \item Legs \& Leggins
    \item Muscles \& Physique
    \item Past \& Oldfashioned
    \item Guns \& Combat
    \item Company \& Product
    \item Logo
    \item Marker \& Print
    \item Palm \& Hand
    \item Repair \& Patch
    \item Projection \& Perspective
    \item Mural Artwork
    \item Artist Names
    \item Quarter \& Multitone
    \item Color Palette
    \item Prison \& Redemption
    \item Aesthetic Adjectives            % was: Aesthetics Adjectives
    \item Sports \& Athletics             % was: Sports & Atheltics
    \item Outdoor Scenes                  % was: Outdoor Places
    \item Overdetailed \& High Resolution
    \item Nvidia \& GPUS
    \item Perspective \& Observing
    \item Frightening Expressions
    \item Emotional States
    \item Body Details
    \item Painting Styles
\end{enumerate}
\end{multicols}

%============================================================
\subsection*{March 2024 (76 topics, 14024 specifiers)}

\begin{multicols}{2}
\raggedcolumns
\begin{enumerate}
    \item Anime/Cosplay \& Characters     % was: Anime/Cosplay & Character References
    \item Lighting Details
    \item Male Personal Names            % was: Male Names
    \item Female Names
    \item Genitalia (NSFW)               % was: Genitalia
    \item Explicit Sexual Content (NSFW)
    \item Outdoor Scenes                 % was: Outdoor Locations
    \item Fruits
    \item Food \& Drinks
    \item Psychedelic
    \item Lens \& Photographic Nikon
    \item Numbers
    \item Textures
    \item Liquids
    \item Clothing \& Outfits
    \item Color Palette
    \item Rain \& Umbrella
    \item Hair \& Hairstyles             % was: Hairstyles & Haircolors
    \item Ethnic \& Cultural Descriptors
    \item Lengths \& Sizes
    \item Religion
    \item Phase \& Overview
    \item Guns \& Pistols
    \item Teeth \& Open Mouth
    \item Eyes \& Pupils                 % was: Eyes & Pupilrs
    \item Frightening Expressions
    \item Emotional States               % was: Emotional State
    \item Relieved Facial Expressions
    \item Panicking \& Disgusting Regret
    \item Photoreal \& Realistic 8K
    \item Player \& Poker Deck
    \item Refreshing \& Resilient (Repair/Remove)
    \item PhotoMasterpiece \& QualityHyper
    \item PhotoFull \& Artisanal Rendering
    \item Legend \& Retrofuturism
    \item Obscured \& Reflection
    \item Read \& Generation Producing
    \item Spaces \& Planets
    \item Marketplaces
    \item Legs \& Leggins
    \item Skin Texture
    \item Robot
    \item Explicit Sexual Content (NSFW) % was: Rating_ExplicitNSFW, or unify if desired
    \item Posing \& Postures
    \item Looking at Viewer
    \item Ears
    \item Muscles \& Physique
    \item Makeup
    \item Jewelleries
    \item Body Sizes
    \item Lingerie
    \item Background Details
    \item Hightech
    \item Swimsuits
    \item Letter \& Font
    \item Plump \& Overweight
    \item Rage \& Psychologically Upset
    \item Hands \& Palms
    \item Moods \& Emotional State        % if unifying “Moods & Emotional State” with “Emotional States”
    \item Aesthetic Adjectives
    \item Times of Day (Morning/Noon/Night)
    \item Policemen
    \item Painting Styles
    \item Artstation
    \item R6 Xentaiposat \& R5 DSLr
    \item Music \& Sounds
    \item Genitalia (NSFW)               % unify if separate references to "Genitalia"
    \item Shadows \& Lighting
    \item Pupils \& Eyes                 % unify if consistent
    \item Smiley Expressions
    \item Resolution \& Quality
    \item Age References
    \item Transportation Means
    \item Sports \& Athletics            % was: Sports & Athetlics
    \item Poisonous \& Deadly
    \item Cinematic Effects
\end{enumerate}
\end{multicols}

%============================================================
\subsection*{April 2024 (83 topics, 17839 total specifiers)}

\begin{multicols}{2}
\raggedcolumns
\begin{enumerate}
    \item Anime/Cosplay \& Characters    % was: Anime/Cosplay & Character References
    \item Lighting Details
    \item Male Personal Names            % was: Male Names
    \item Female Names
    \item Genitalia (NSFW)
    \item Explicit Sexual Content (NSFW)
    \item Hands \& Palms
    \item Color Palette
    \item Ethnic \& Cultural Descriptors
    \item Hair \& Hairstyles             % was: Hairstyles & Haircolors
    \item Outdoor Scenes                 % was: Nature & Outdoors
    \item R6 \& XT3
    \item Food \& Drinks
    \item Nipples \& Breasts (NSFW)      % was: Breast & Nippls (NSFW)
    \item Posing \& Posture
    \item Illumination \& Radiance
    \item Offer \& Trade
    \item Rain \& Umbrellas              % was: Raind & Umbrellas
    \item Lingerie
    \item Clothing \& Outfits            % was: Outfits and Clothing
    \item Revealing Clothes
    \item Masterpiece \& High Resolution
    \item Photo \& Photograph
    \item Ghosts \& Spirits
    \item Mystic \& Myths
    \item Cinematic Effects              % was: Cinematics Effects
    \item Looking at Viewer
    \item Guns \& Battles
    \item Animals
    \item Robots
    \item Space \& Planets
    \item Playfulness \& Whimsy
    \item Plump \& Overweight
    \item Punishment
    \item Price \& Contract Pay
    \item Region NC \& Phoenix
    \item verexposure
    \item Phinnherz
    \item Reclining \& Liner Stitched
    \item Photoreal \& QualityRealistic
    \item Noon \& Nocturnal Daylight
    \item Pupils \& Eyes
    \item Outstanding \& Finest
    \item Jewelleries
    \item LowAngle \& OverDetailed
    \item Print \& Marker Representing
    \item Artist Names
    \item Face Details
    \item Skin Details
    \item Oscar
    \item Pulitzer Professor
    \item Muscles \& Physique
    \item Sports \& Athletics
    \item Makeup
    \item High Resolution
    \item Kingdoms
    \item OldSchool Vibes
    \item Perverted Consent \& Prostitution (NSFW)
    \item Aesthetic Adjectives
    \item Tongue \& Piercing
    \item Times of Day (Morning/Noon/Night)
    \item Furniture
    \item OnlyFans (NSFW)
    \item Music \& Sounds
    \item Texture \& Patterns
    \item Materials
    \item Background Details
    \item Rabbit Ears
    \item Robots
    \item High Tech
    \item Fantasy \& Mystique
    \item Post Apocalyptic
    \item Focu \& Light
    \item Pleasurable \& Charm
    \item Furs
    \item Poisonous \& Deadly
    \item Smiley Expressions
    \item Moods \& Emotional State
    \item Explicit Sexual Content (NSFW)  % was: Rating_ExplicitNSFW (if unifying)
    \item Nervous \& Rebellious
    \item Photogenic Shots
    \item Fonts \& Letters
    \item Numbers
\end{enumerate}
\end{multicols}

\subsection{Semantic Diversity Across User Categories}
\label{sec:semantic_diversity_user_categories}

\begin{figure}[h] % 'h' to place it here, you can also use 't' for top, 'b' for bottom
        \centering
        \includegraphics[width=1.1\linewidth]{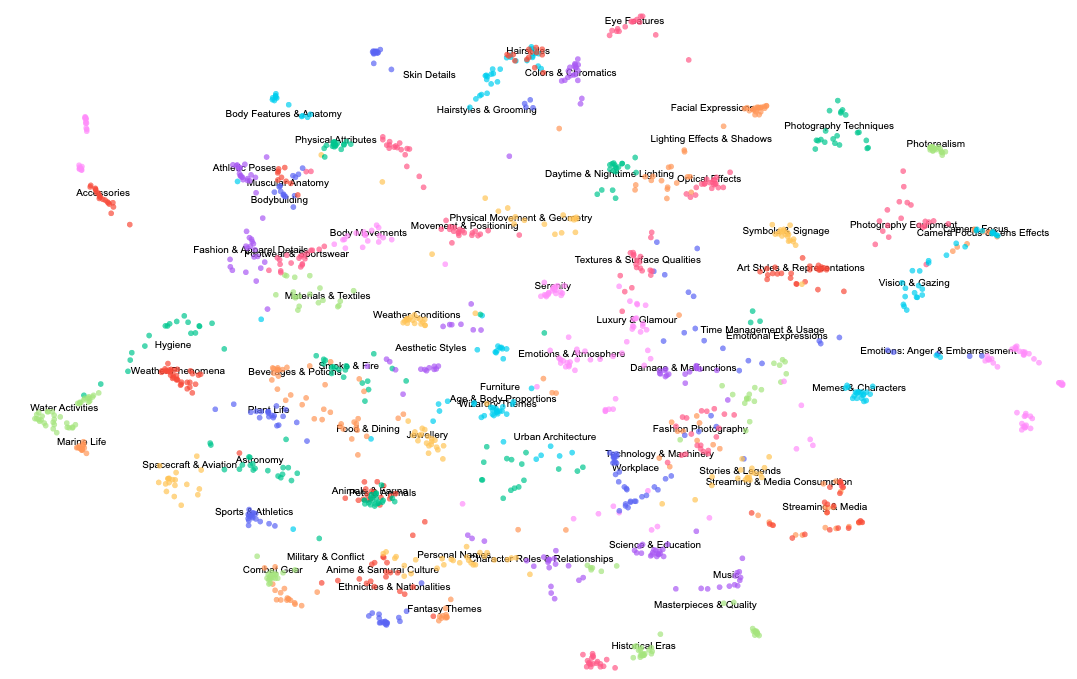} % Second image
        \caption{UMAP visualization of MiniLM-L6-v2 embeddings of prompt specifiers and HDBSCAN-identified topics for the \textbf{Non-Repeaters}  user category}
        \label{fig:non_repeaters}
\end{figure}

\twocolumn
\subsection{Clustering Patterns Across User Categories} This clustering analysis was conducted as a subexperiment to explore the semantic patterns in prompt usage across user categories by examining the embedding space to assess how semantically similar prompts were grouped, with proximity in the embedding space indicating similarity and greater distances signifying diversity.

\begin{figure}[h]
    \centering
    \begin{minipage}[t]{0.32\textwidth}
        \centering
        \includegraphics[width=\linewidth]{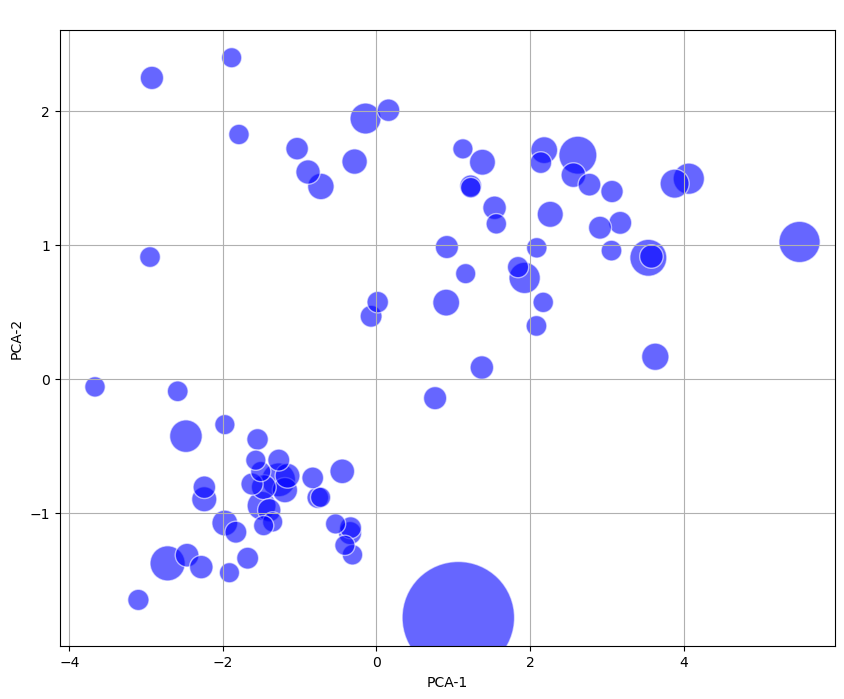}
        \caption{Clusters of Similar Text Embeddings (Threshold \(\geq 0.8\)) for \textbf{Consistent Repeaters}.}
        \label{fig:consistent_repeaters}
    \end{minipage}
    \hfill
    \begin{minipage}[t]{0.32\textwidth}
        \centering
        \includegraphics[width=\linewidth]{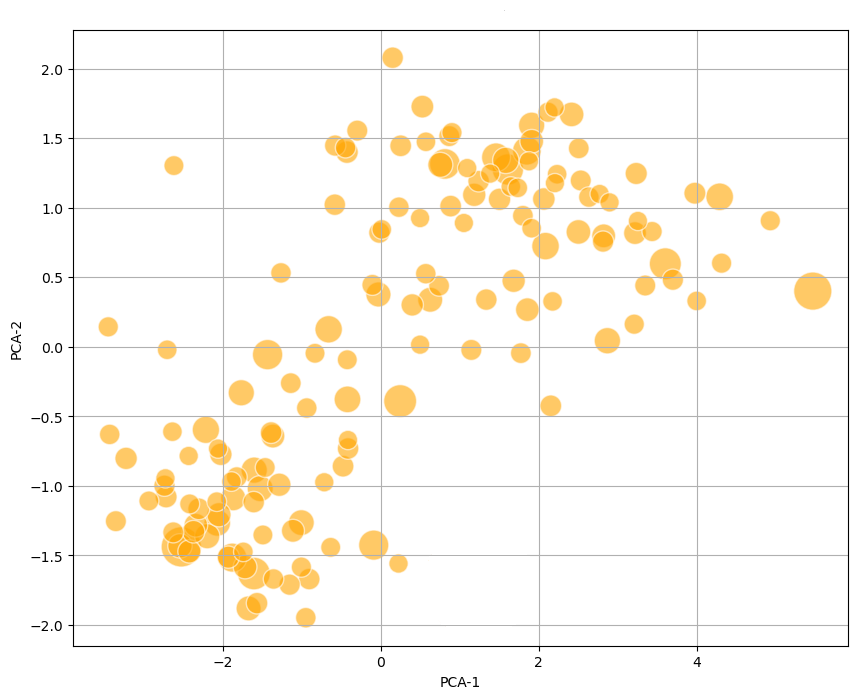}
        \caption{Clusters of Similar Text Embeddings (Threshold \(\geq 0.8\)) for \textbf{Occasional Repeaters}.}
        \label{fig:occasional_repeaters}
    \end{minipage}
    \hfill
    \begin{minipage}[t]{0.32\textwidth}
        \centering
        \includegraphics[width=\linewidth]{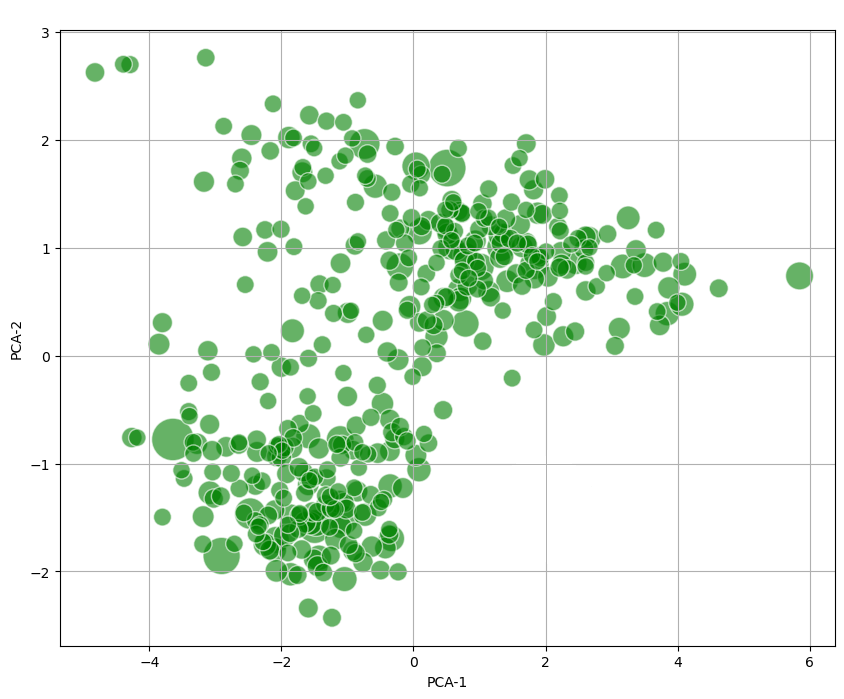}
        \caption{Clusters of Similar Text Embeddings (Threshold \(\geq 0.8\)) for \textbf{Non-Repeaters}.}
        \label{fig:non_repeaters}
    \end{minipage}
\end{figure}

To calculate the CLIP text embeddings, the pre-trained CLIP model, specifically the \texttt{clip-vit-base-p32} variant \cite{eslami2021does}, was utilized due to its proven effectiveness in extracting high-quality multimodal embeddings. Initially, an equal number of unique prompts per user category was parsed and converted into numerical tensors using these CLIP text embeddings. Clustering was then performed based on a cosine similarity threshold of 0.8, grouping only prompts with substantial semantic similarity to identify the extent to which semantically similar clusters could be formed. Additionally, a minimum cluster size of 20 was imposed to ensure meaningful groupings, while representative subsets were sampled from larger user categories to preserve computational efficiency without compromising the integrity of the analysis. For visualization, Principal Component Analysis (PCA) \cite{kurita2021principal} was applied to reduce the dimensionality of the embeddings to two components, preserving the data’s overall variance while enabling a clear depiction of the cluster structures. Each group was then visualized as a bubble on a two-dimensional plot, with the bubble size indicating the number of prompts within the cluster.

Distinct clustering patterns were revealed across user categories. For consistent repeaters, 36,450 prompts were clustered into 79 groups, including a prominent cluster of 15,192 prompts primarily associated with NSFW content. Occasional repeaters had 31,223 prompts clustered into 166 groups, while non-repeaters, with 22,802 clustered prompts and 237 groups, exhibited the highest number of distinct clusters and the greatest semantic diversity.

\section{Visual Diversity in Prompts}

\subsection{Weighted vs. Unweighted Regression}
We conducted additional analyses comparing unweighted and weighted regressions to account for (1) token similarity in prompts and (2) the absolute number of identical tokens within prompts.

\paragraph{Token Similarity}
An unweighted regression using token similarity produced a slope of 0.3426, with an \(R^2\) of 0.1821, indicating that roughly 18\% of the variance in image similarity is explained by token-level resemblance. This relationship was statistically significant (p-value \(\approx 2.97 \times 10^{-10}\)). When weighting by cluster size, the \(R^2\) dropped slightly to 0.145, suggesting that larger clusters—those with more repeated vocabulary—exert a stronger pull on the overall trend. Both models confirm a strong positive relationship between token similarity and image similarity.

\paragraph{Number of Identical Tokens}
An unweighted regression using the absolute count of identical tokens yielded a slope of 0.0083 and an \(R^2\) of 0.109, demonstrating that approximately 11\% of the variance in image similarity can be attributed to exact token matches. The weighted regression was slightly lower (\(R^2=0.962\)), indicating minimal impact from cluster size. The positive slope (p-value \(\approx 7.91 \times 10^{-4}\)) confirms that a higher number of identical tokens corresponds to modest increases in image similarity.

\onecolumn
\begin{figure}[htbp]
    \centering
    \begin{tabular}{cccc}
        \includegraphics[width=0.19\textwidth]{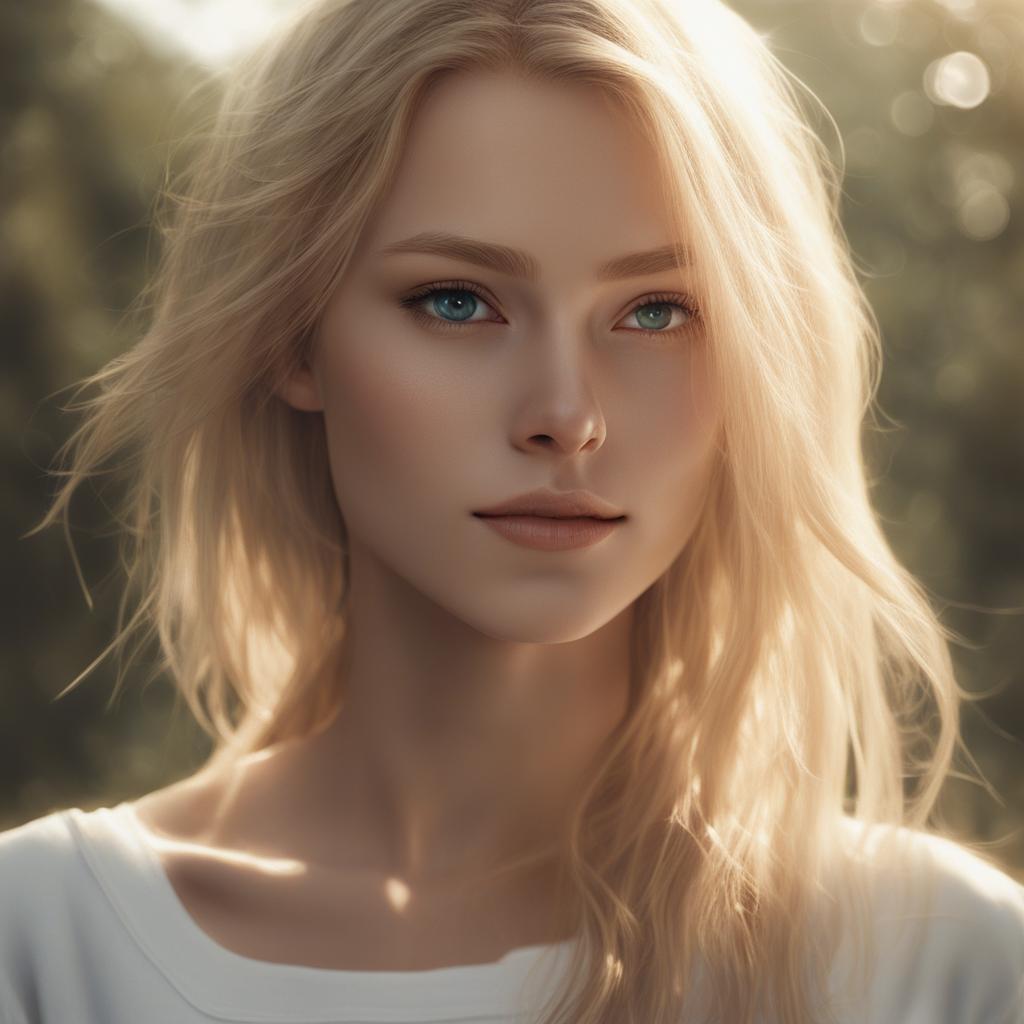} &
        \includegraphics[width=0.19\textwidth]{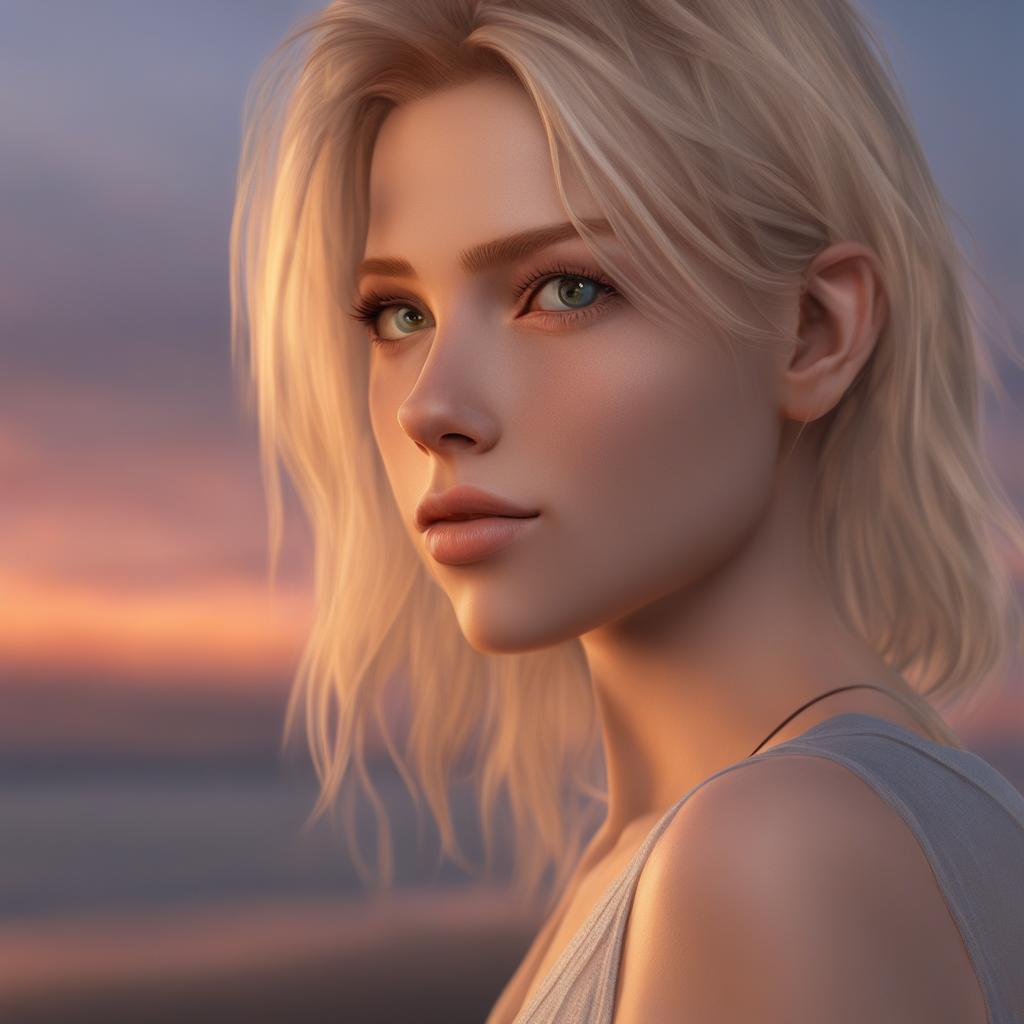} &
        \includegraphics[width=0.19\textwidth]{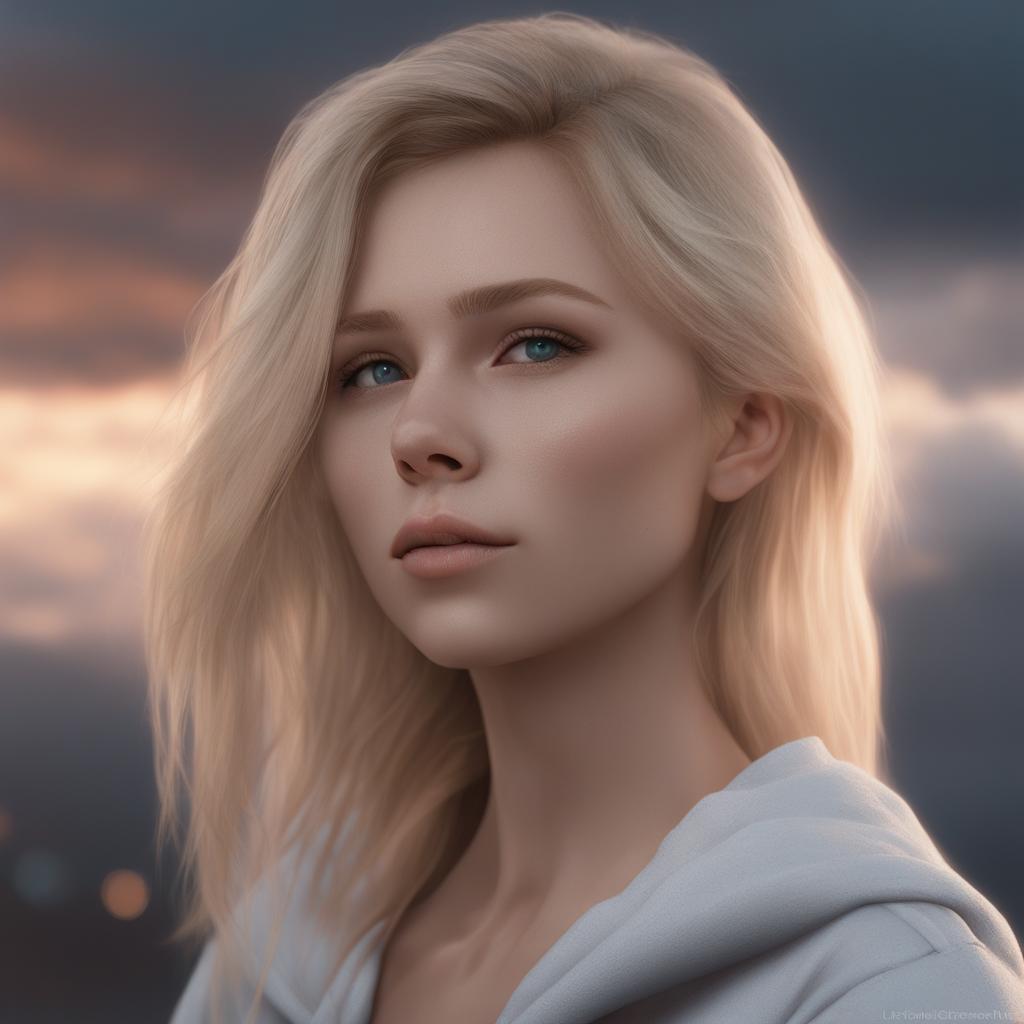} &
        \includegraphics[width=0.19\textwidth]{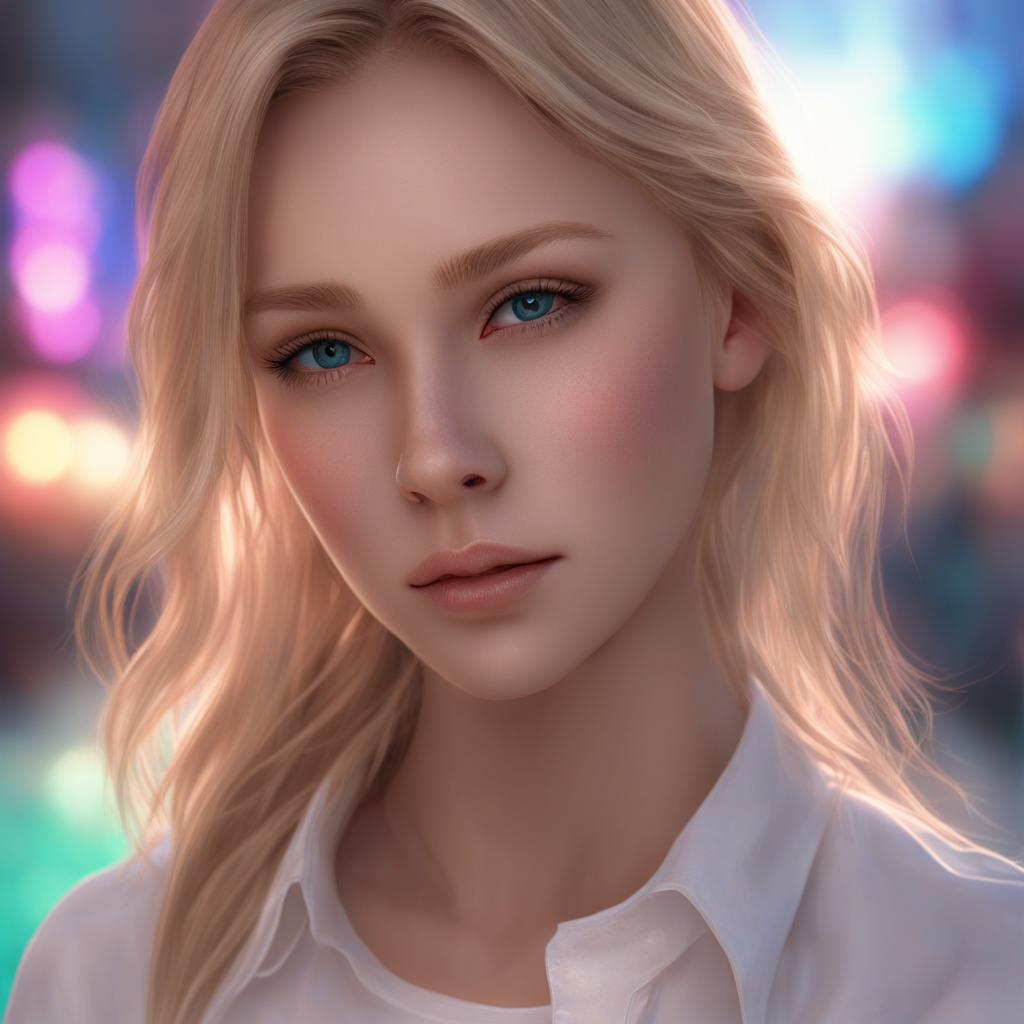} \\
        \textbf{Nature} (SD XL) & \textbf{Golden Sunset} (SD XL) & \textbf{Sky Cloud} (SD XL) & \textbf{Colorful} (SD XL) \\
        
        \includegraphics[width=0.19\textwidth]{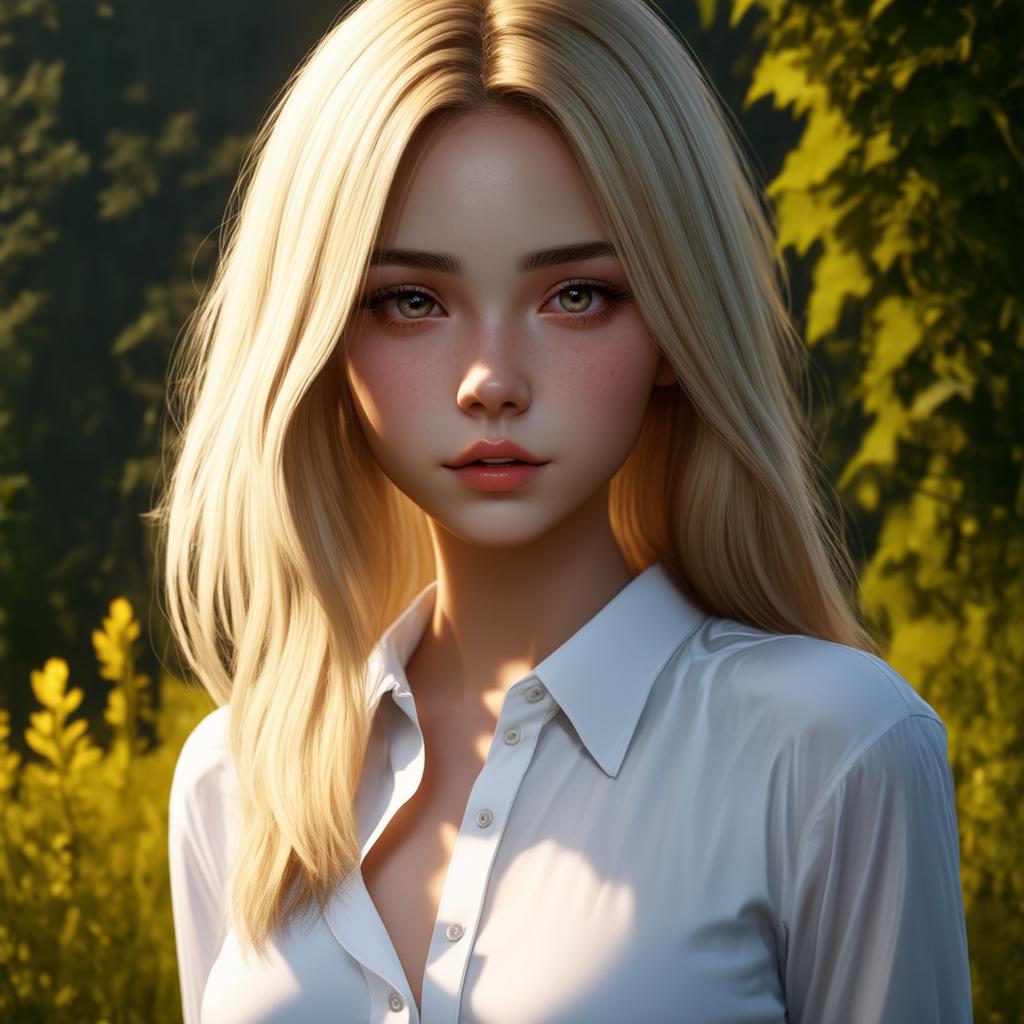} &
        \includegraphics[width=0.19\textwidth]{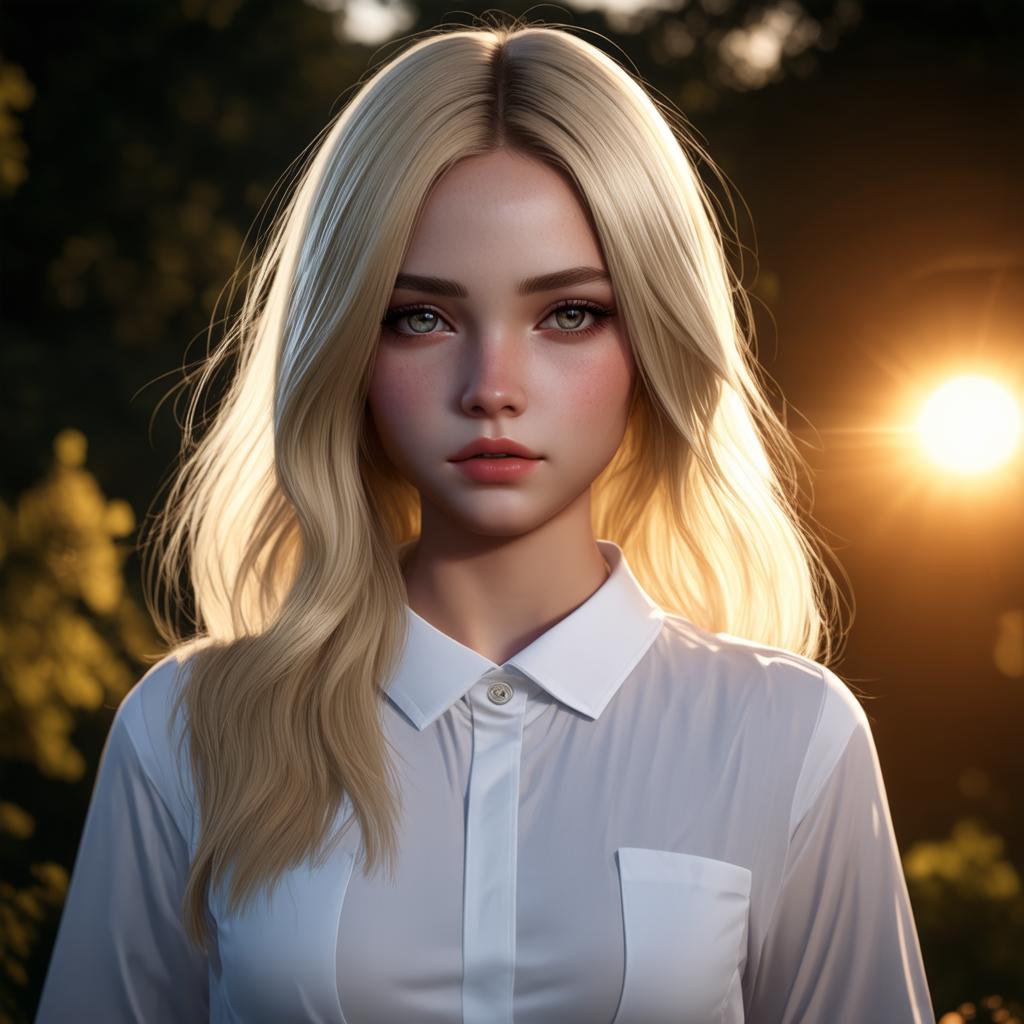} &
        \includegraphics[width=0.19\textwidth]{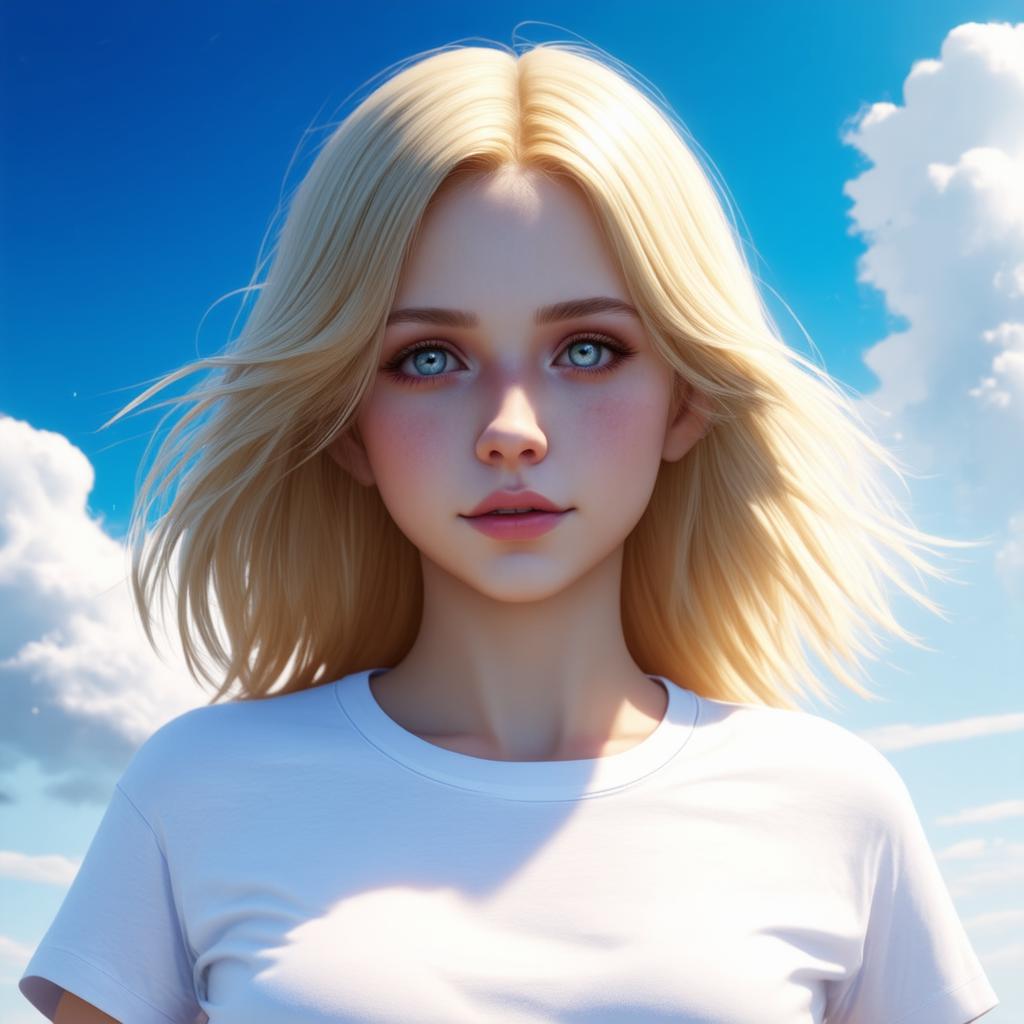} &
        \includegraphics[width=0.19\textwidth]{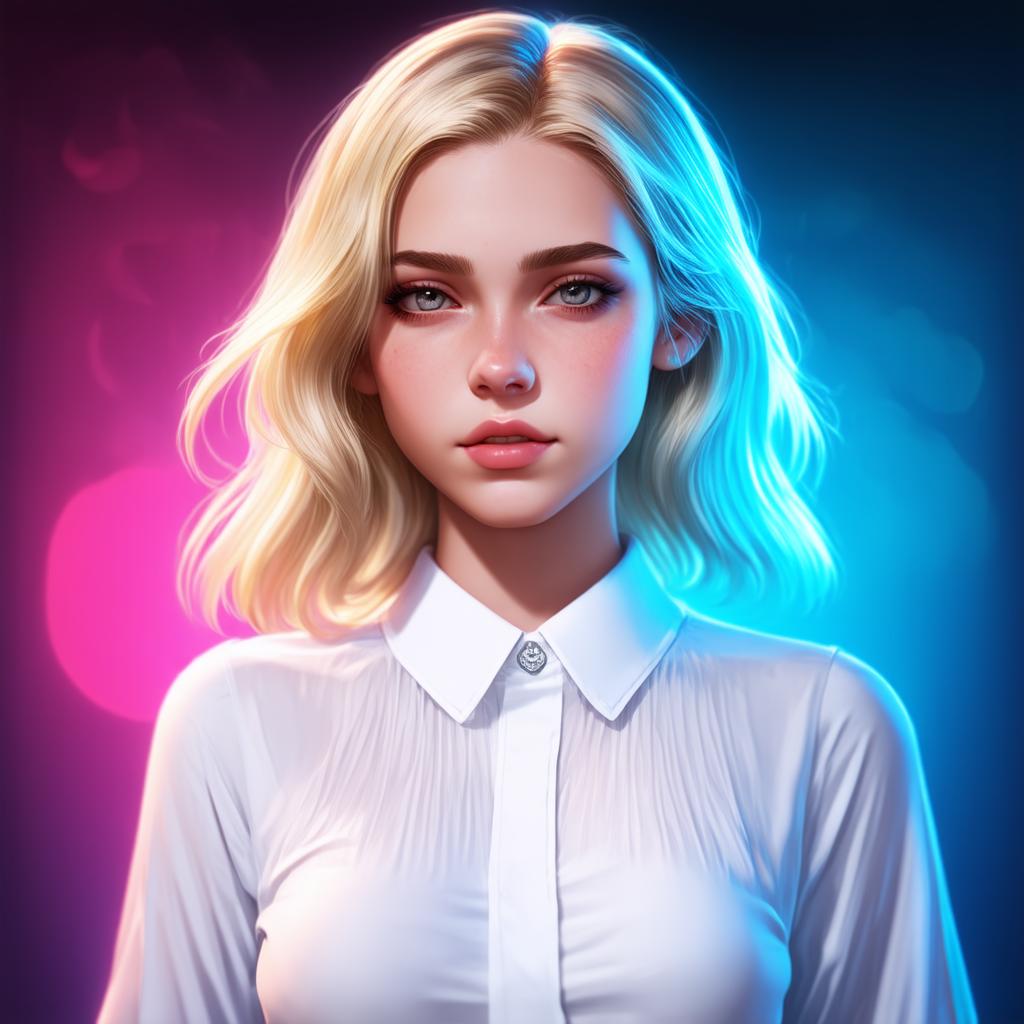} \\
        \textbf{Nature} (SD 3) & \textbf{Golden Sunset} (SD 3) & \textbf{Sky Cloud} (SD 3) & \textbf{Colorful} (SD 3) \\
        
        \includegraphics[width=0.19\textwidth]{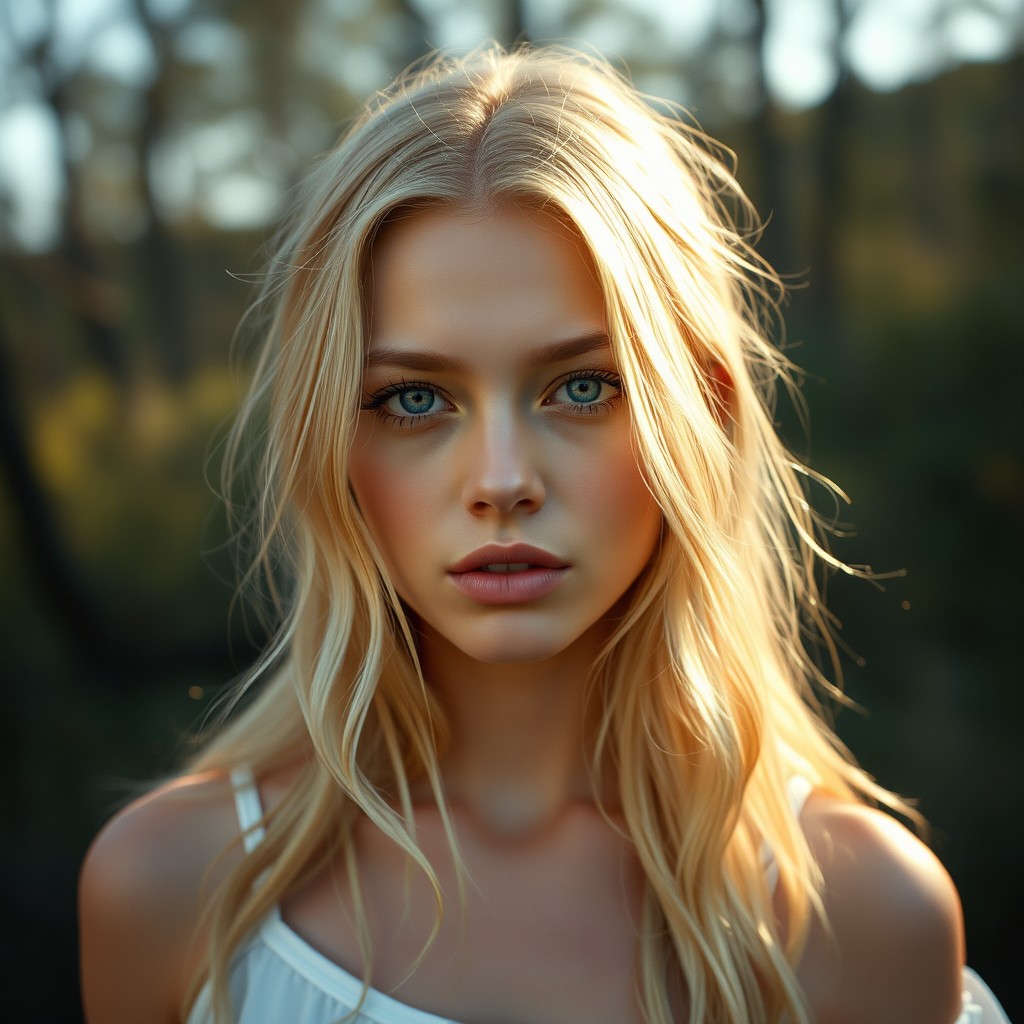} &
        \includegraphics[width=0.19\textwidth]{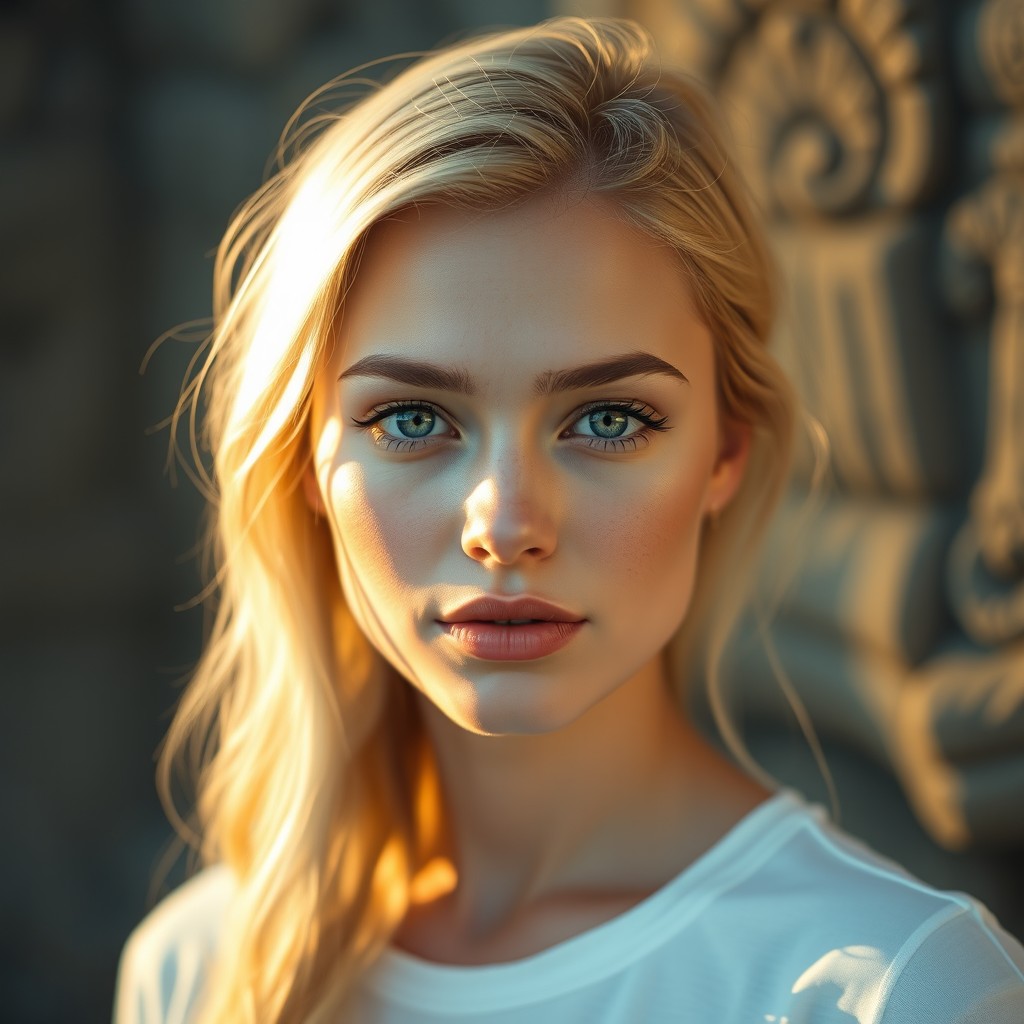} &
        \includegraphics[width=0.19\textwidth]{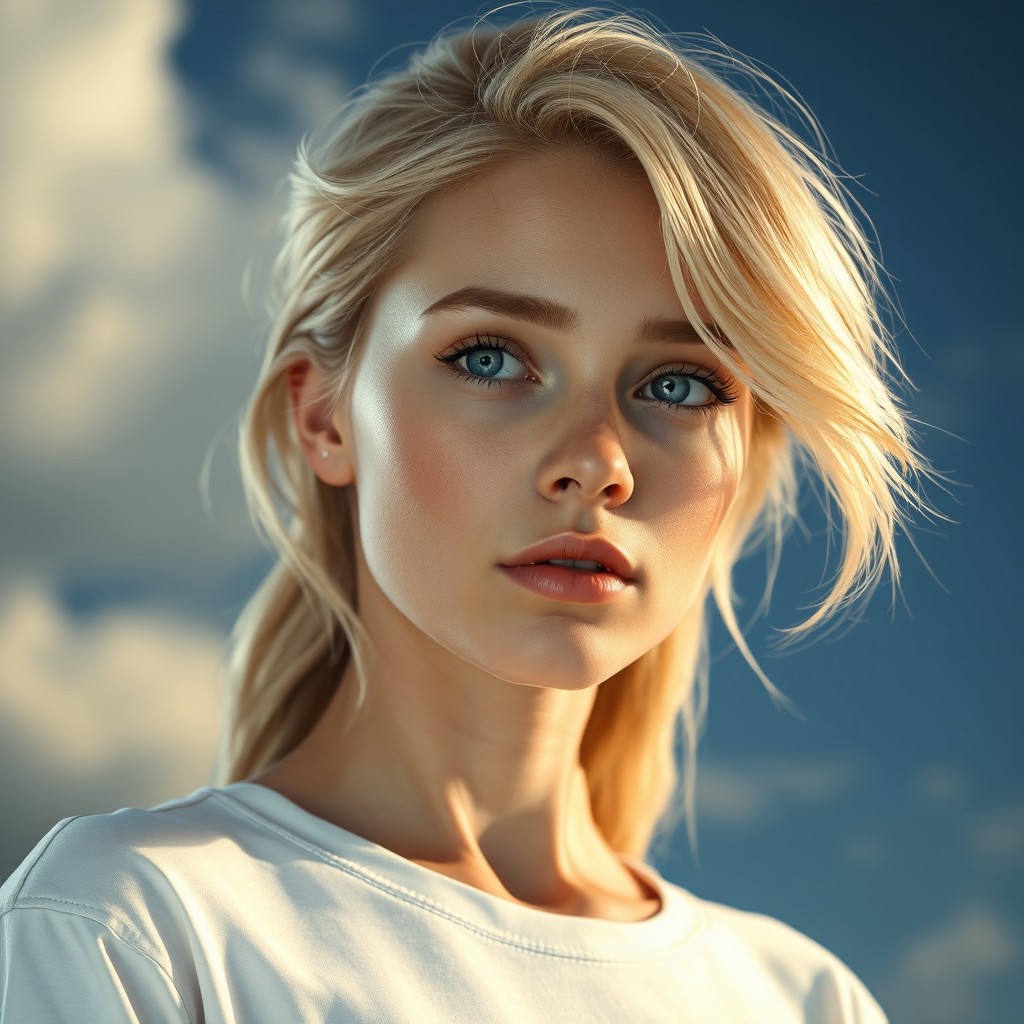} &
        \includegraphics[width=0.19\textwidth]{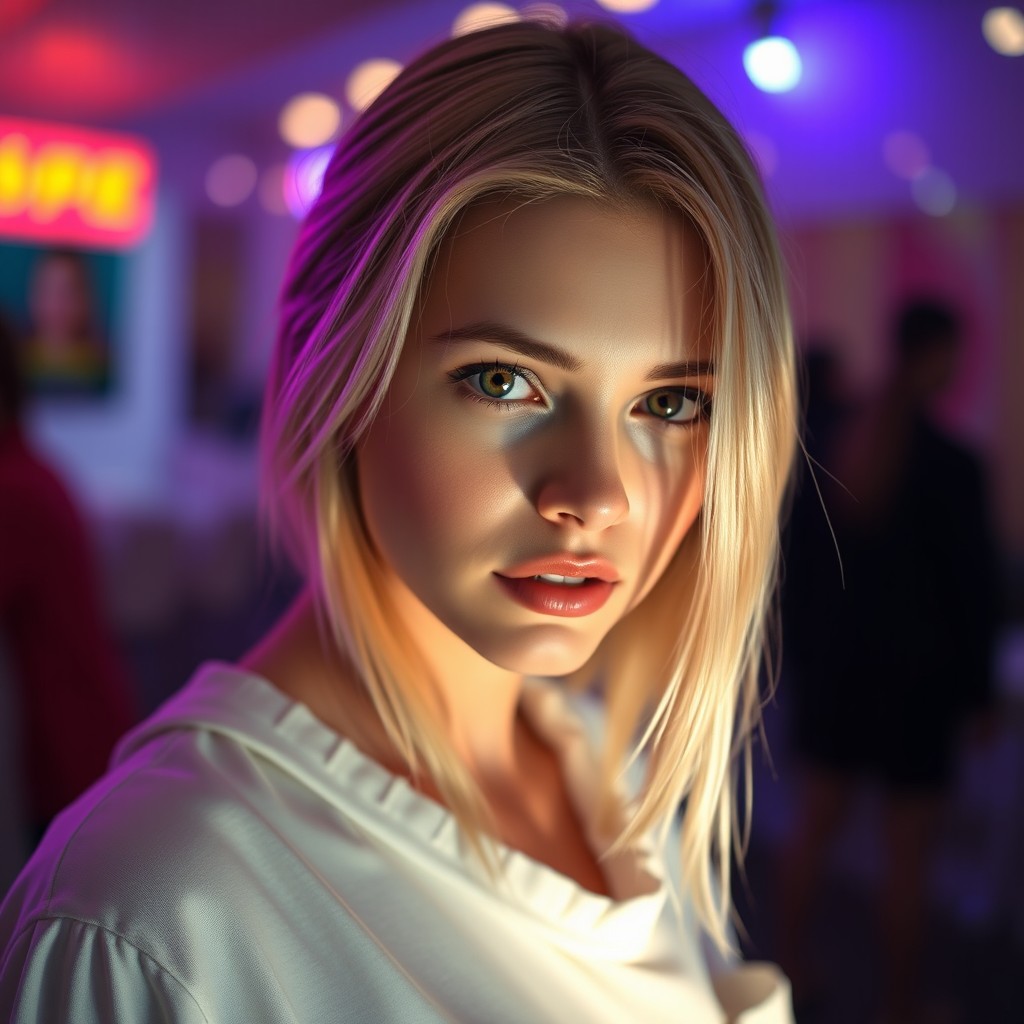} \\
        \textbf{Nature} (FLUX.1) & \textbf{Golden Sunset} (FLUX.1) & \textbf{Sky Cloud} (FLUX.1) & \textbf{Colorful} (FLUX.1) \\
    \end{tabular}
    
   \caption{Comparison of outputs from three models (Stable Diffusion XL, Stable Diffusion 3, and FLUX.1) using the most common high-similarity prompt pattern. The only variation is the background descriptor (\textit{nature}, \textit{golden sunset}, \textit{sky cloud}, \textit{colorful}) while other specifiers (e.g., \textit{realistic, atmospheric scene, ultradetailed, best quality}) remain constant. Despite this small change, high token overlap yields homogeneous and visually similar results across models.}
    \label{fig:comparison_SD_XL_SD3_Flux}
\end{figure}

\begin{figure}[htbp]
    \centering
    % First row: SD XL
    \begin{subfigure}{0.23\textwidth}
        \centering
        \includegraphics[width=\textwidth]{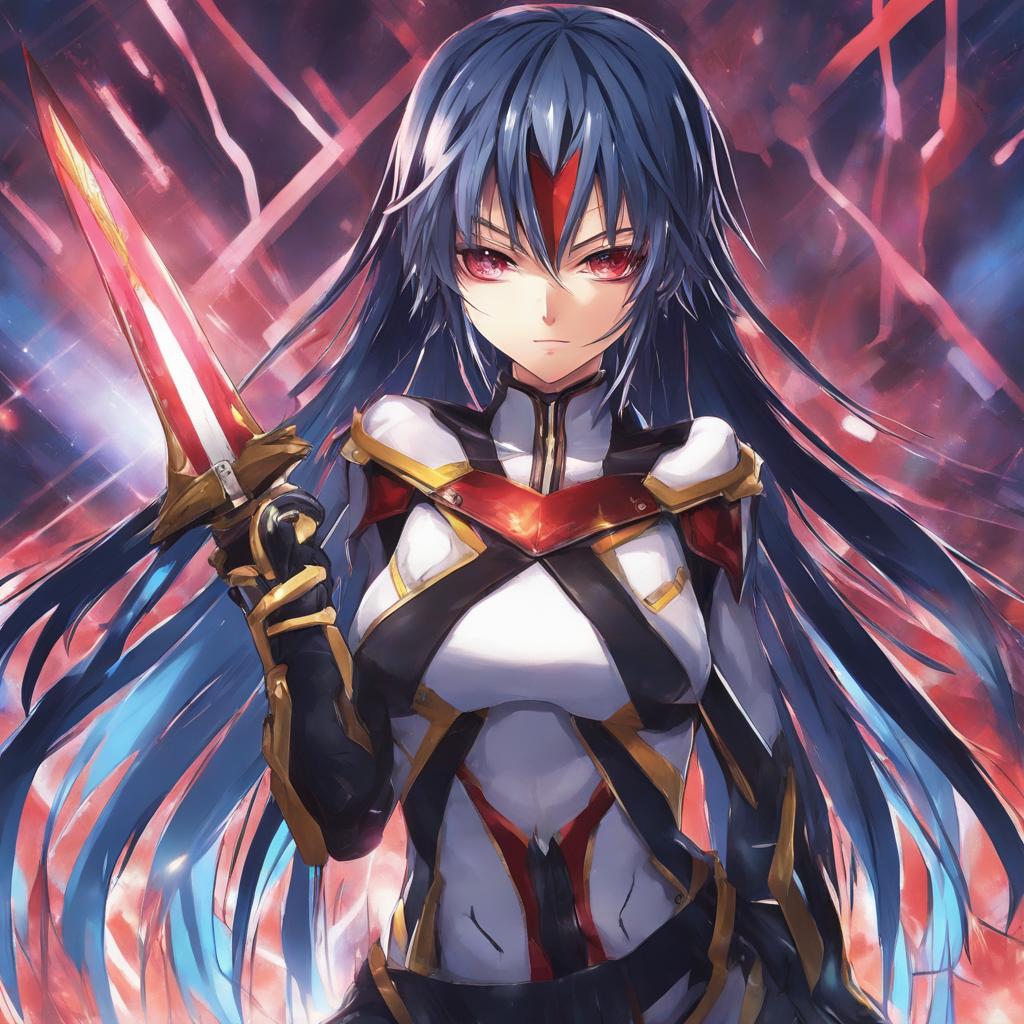}
        \caption{Medaka Kurokami (SD XL)}
        \label{fig:medaka_XL}
    \end{subfigure}
    \begin{subfigure}{0.23\textwidth}
        \centering
        \includegraphics[width=\textwidth]{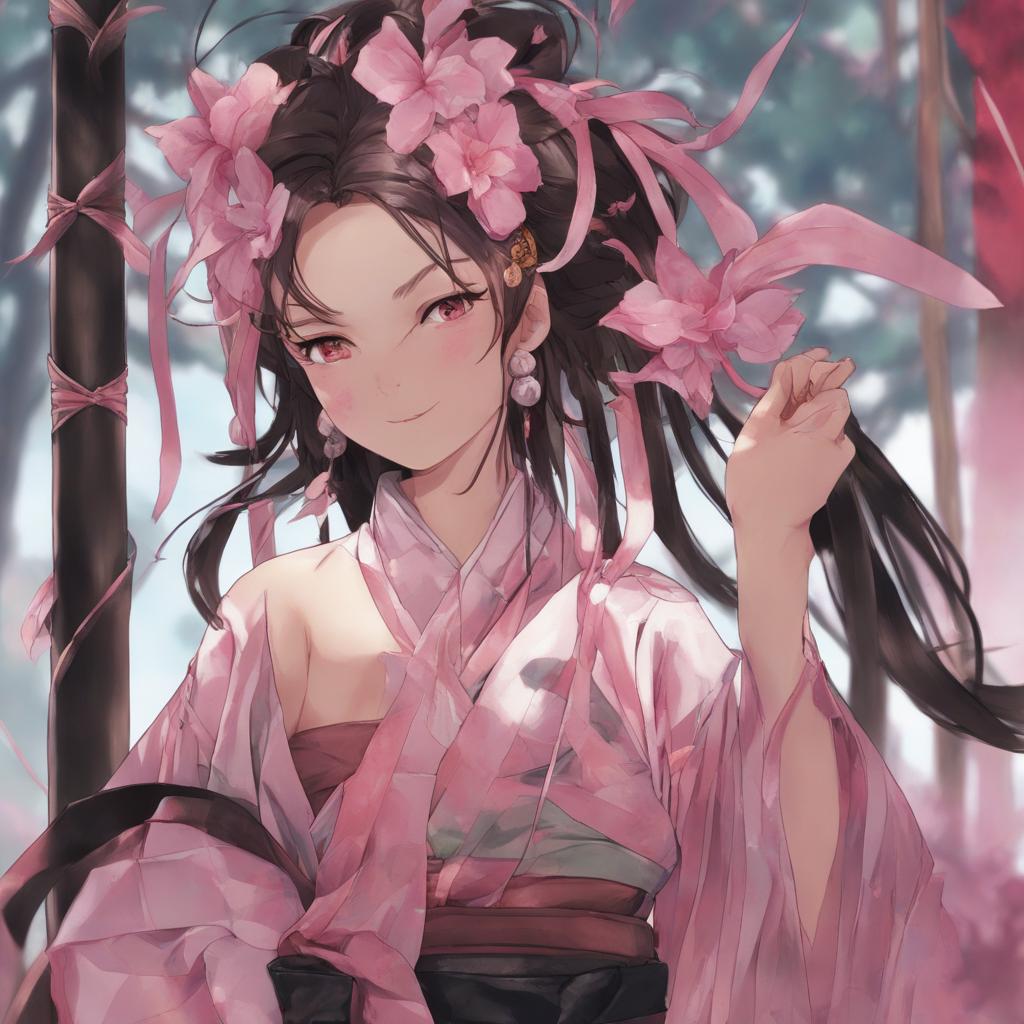}
        \caption{Nezuko Kamado (SD XL)}
        \label{fig:nezuko_XL}
    \end{subfigure}
    \begin{subfigure}{0.23\textwidth}
        \centering
        \includegraphics[width=\textwidth]{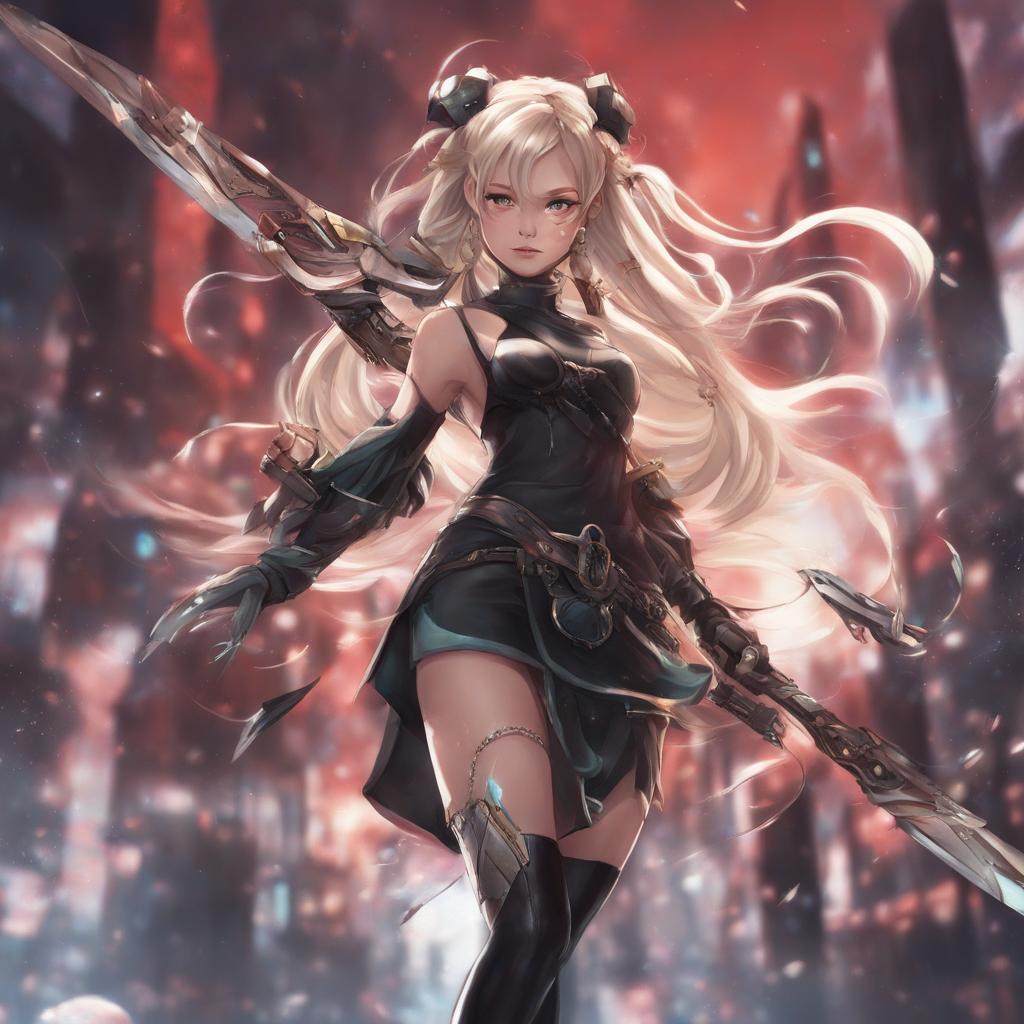}
        \caption{Blade MK3 (SD XL)}
        \label{fig:blade_XL}
    \end{subfigure}
    \begin{subfigure}{0.23\textwidth}
        \centering
        \includegraphics[width=\textwidth]{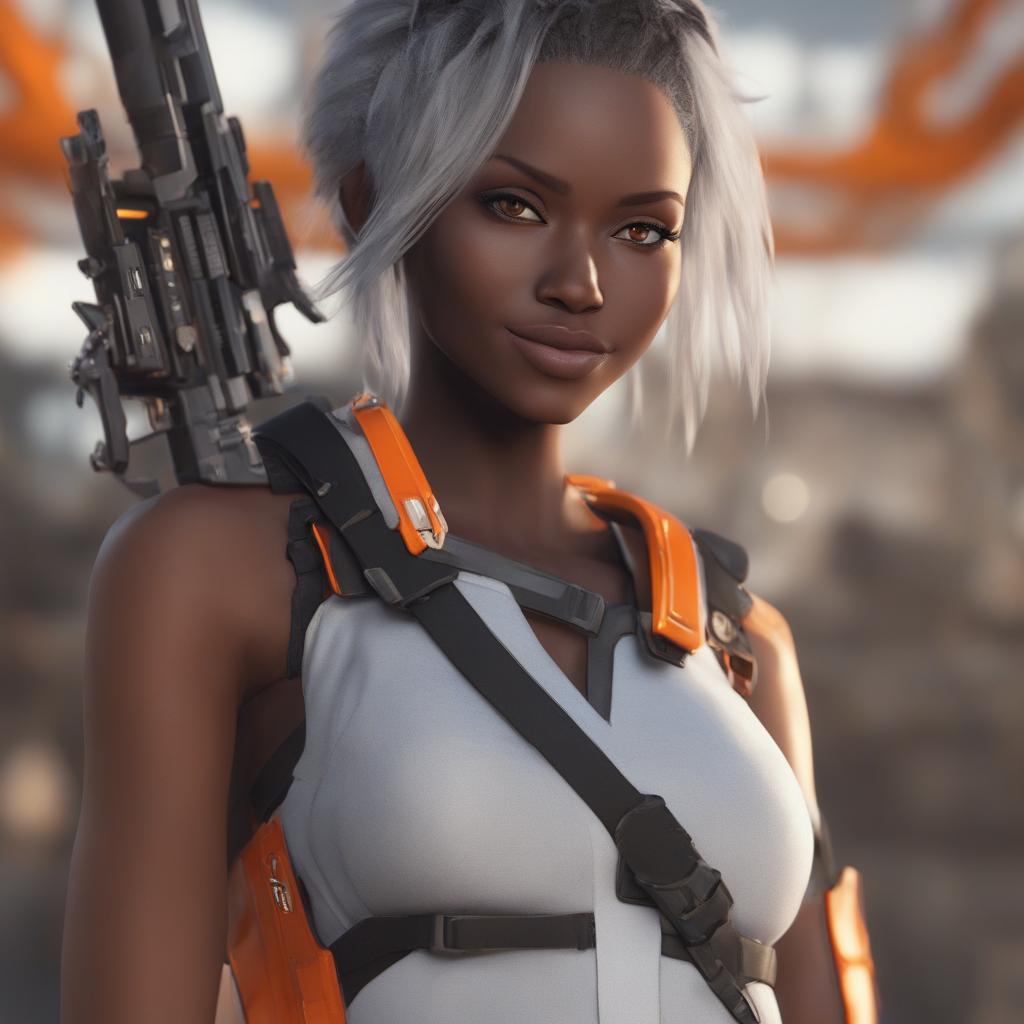}
        \caption{Dark-Skinned Female (SD XL)}
        \label{fig:dark_skin_XL}
    \end{subfigure}
    
    % Second row: SD 3
    \vspace{1em} % Space between rows
    \begin{subfigure}{0.23\textwidth}
        \centering
        \includegraphics[width=\textwidth]{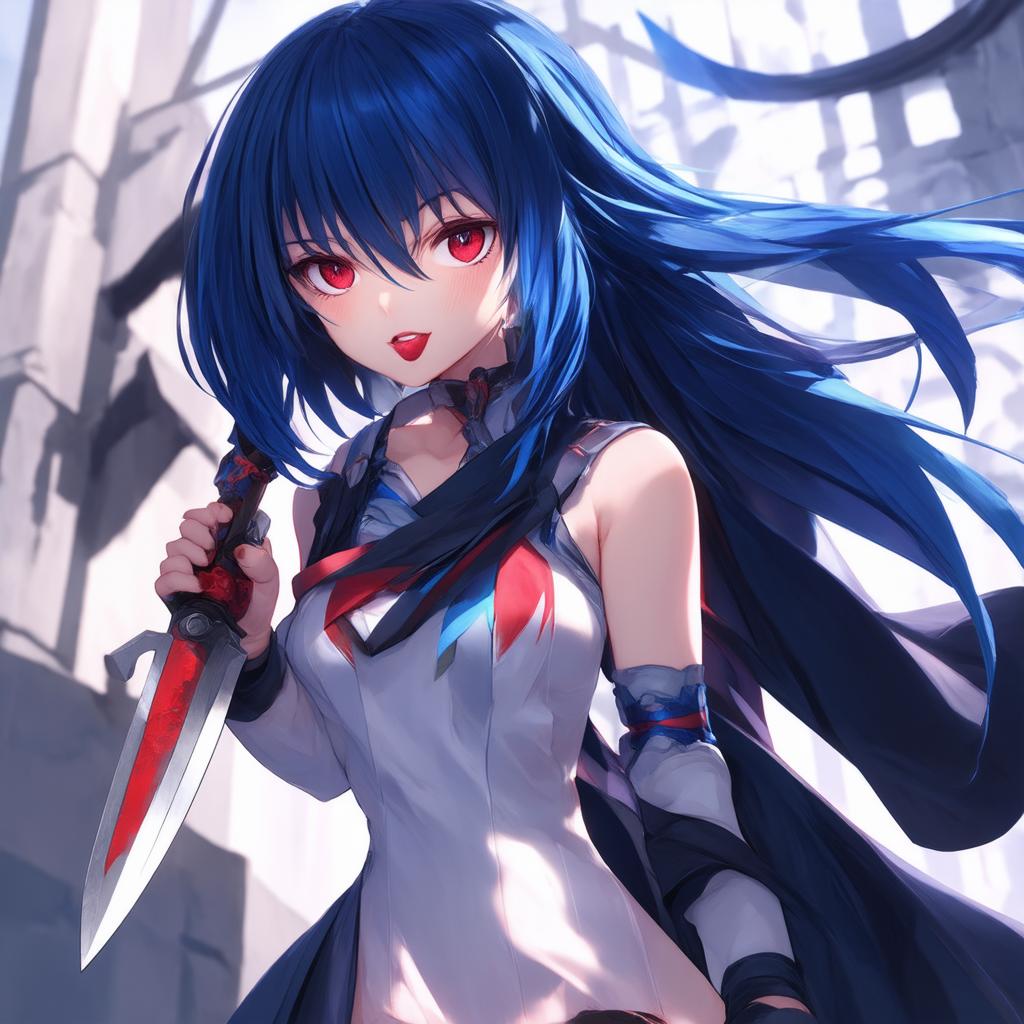}
        \caption{Medaka Kurokami (SD 3)}
        \label{fig:medaka_S3}
    \end{subfigure}
    \begin{subfigure}{0.23\textwidth}
        \centering
        \includegraphics[width=\textwidth]{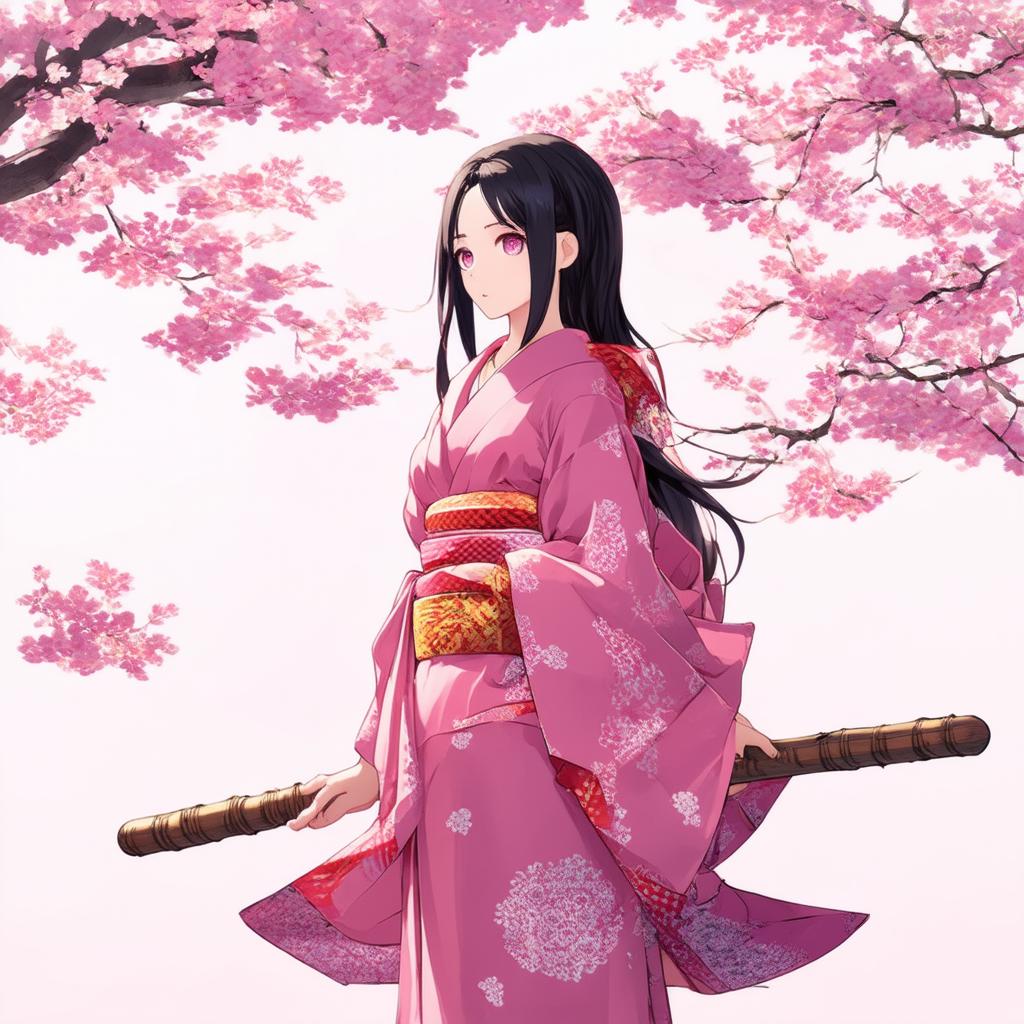}
        \caption{Nezuko Kamado (SD 3)}
        \label{fig:nezuko_S3}
    \end{subfigure}
    \begin{subfigure}{0.23\textwidth}
        \centering
        \includegraphics[width=\textwidth]{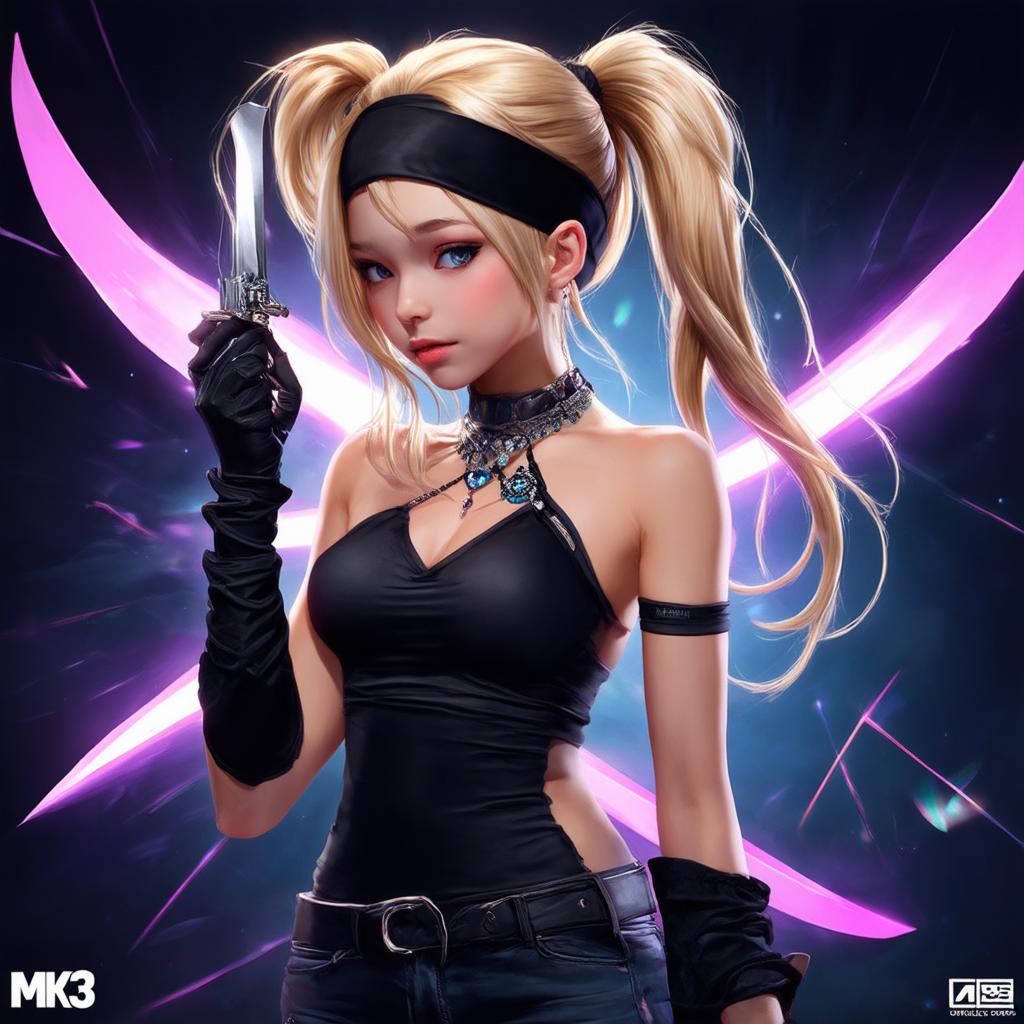}
        \caption{Blade MK3 (SD 3)}
        \label{fig:blade_S3}
    \end{subfigure}
    \begin{subfigure}{0.23\textwidth}
        \centering
        \includegraphics[width=\textwidth]{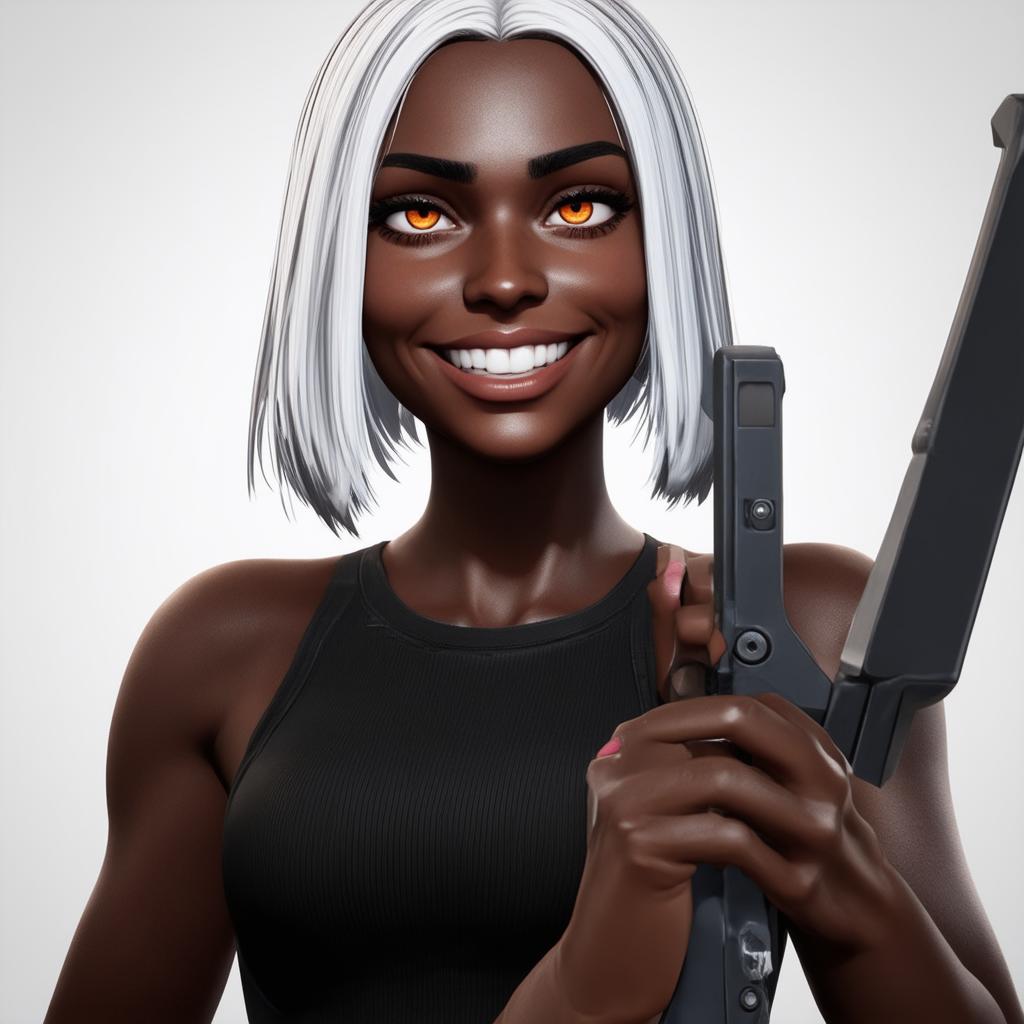}
        \caption{Dark-Skinned Female (SD 3)}
        \label{fig:dark_skin_S3}
    \end{subfigure}
    
    % Third row: FLUX.1
    \vspace{1em} % Space between rows
    \begin{subfigure}{0.23\textwidth}
        \centering
        \includegraphics[width=\textwidth]{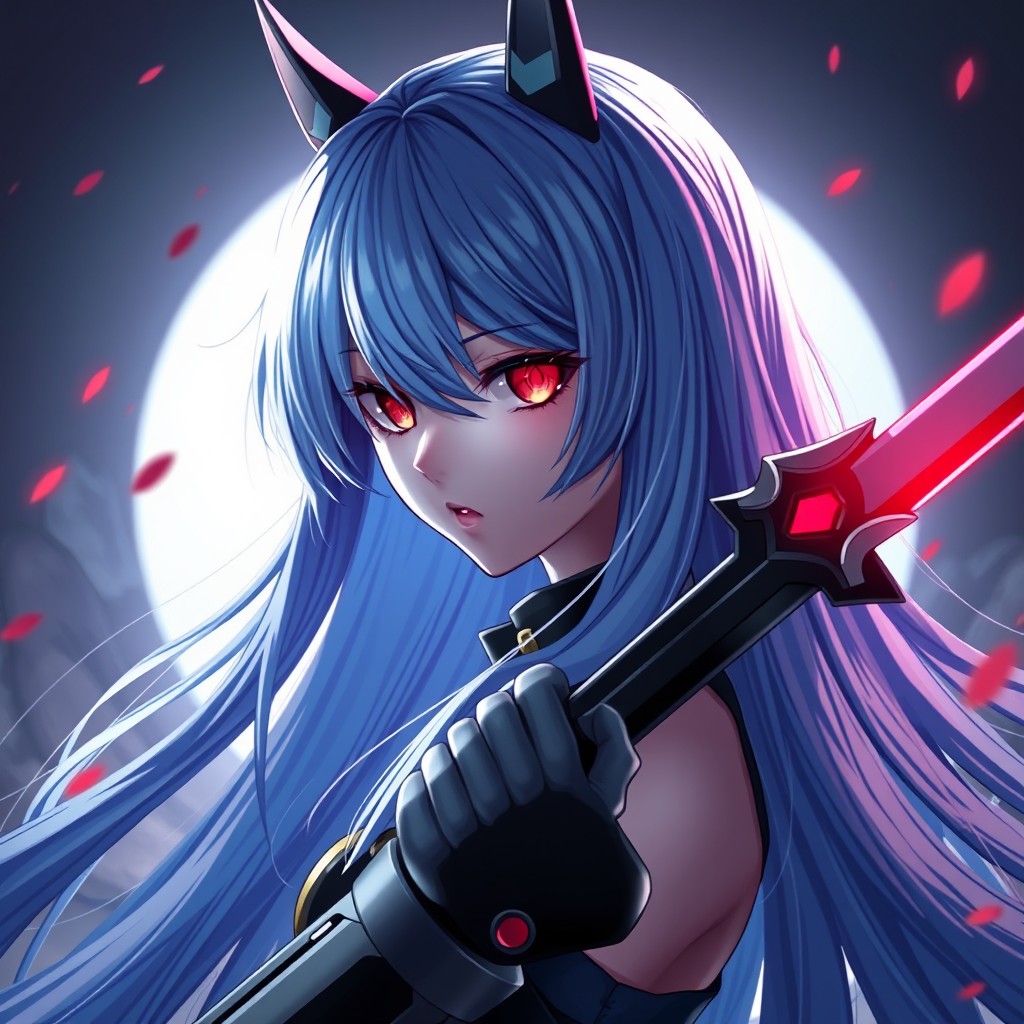}
        \caption{Medaka Kurokami (FLUX.1)}
        \label{fig:medaka_FLUX}
    \end{subfigure}
    \begin{subfigure}{0.23\textwidth}
        \centering
        \includegraphics[width=\textwidth]{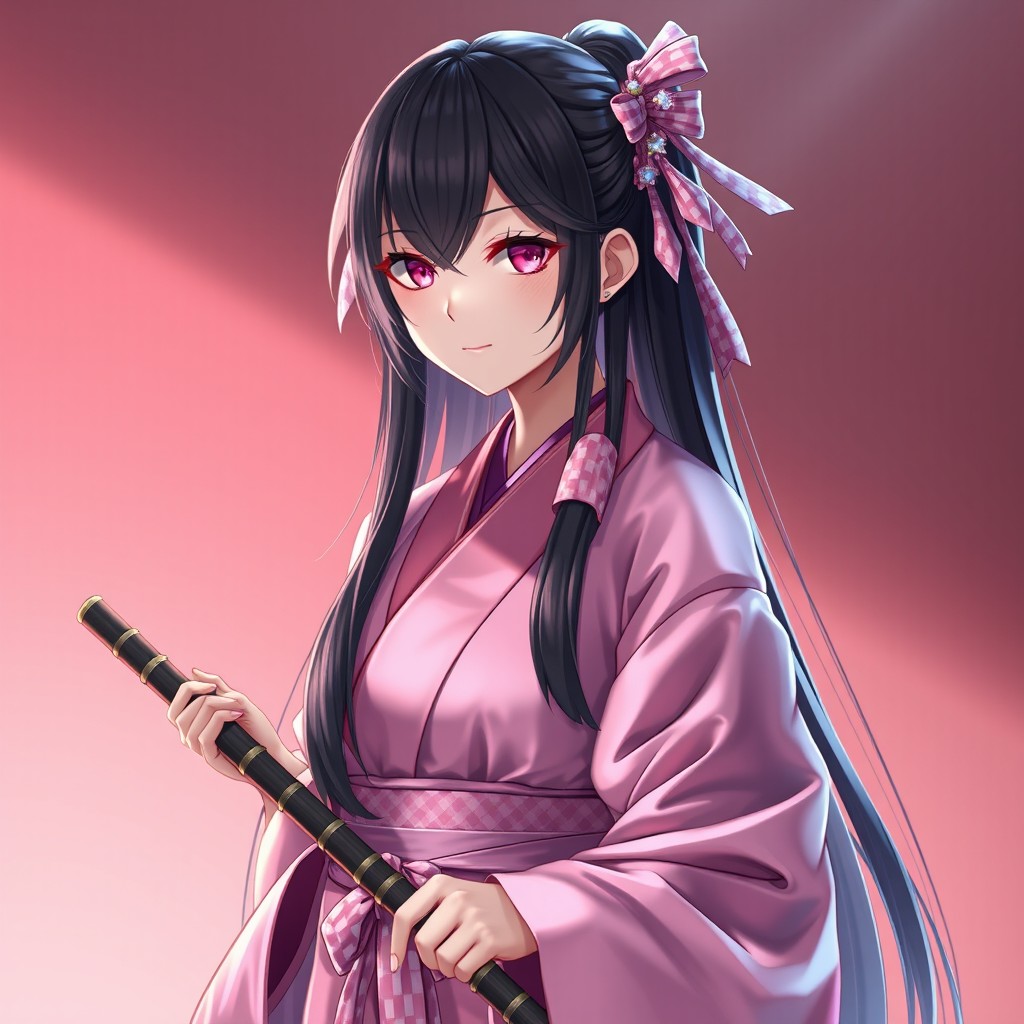}
        \caption{Nezuko Kamado (FLUX.1)}
        \label{fig:nezuko_FLUX}
    \end{subfigure}
    \begin{subfigure}{0.23\textwidth}
        \centering
        \includegraphics[width=\textwidth]{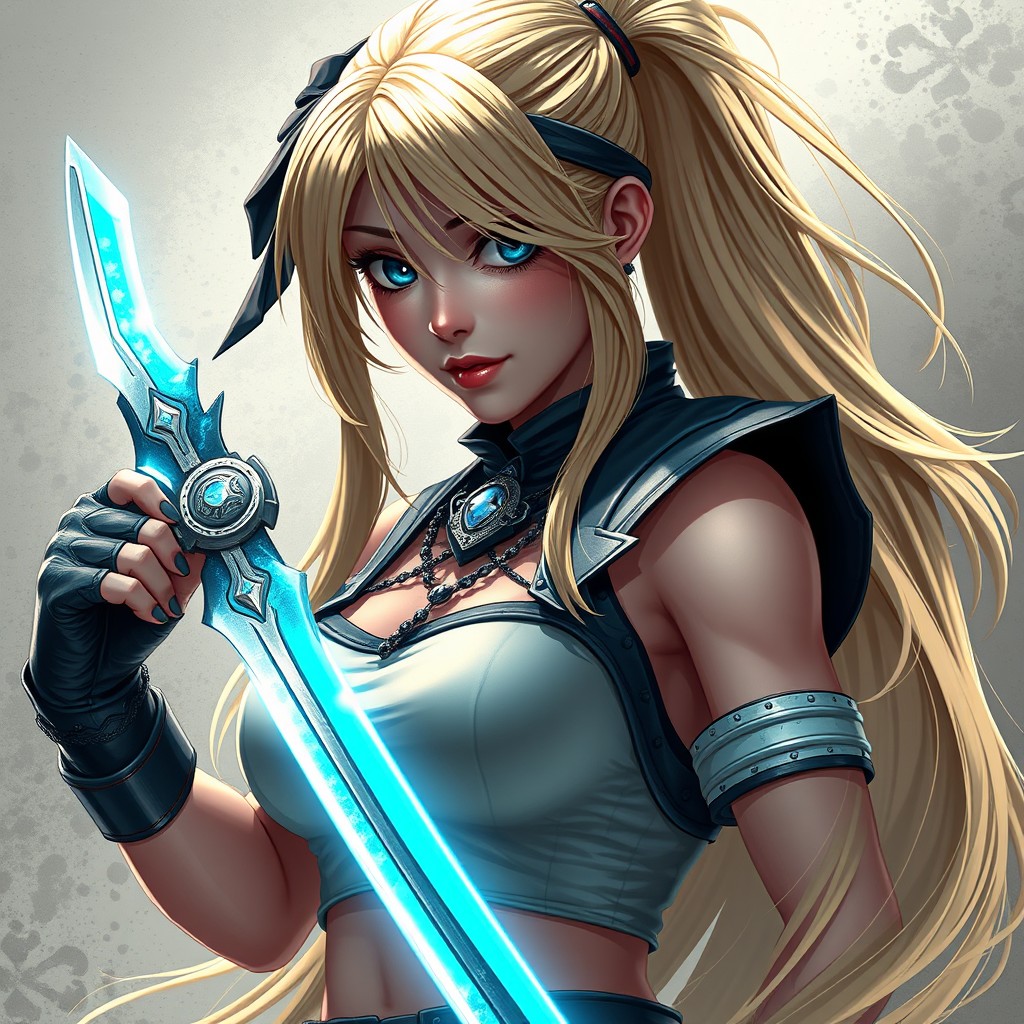}
        \caption{Blade MK3 (FLUX.1)}
        \label{fig:blade_FLUX}
    \end{subfigure}
    \begin{subfigure}{0.23\textwidth}
        \centering
        \includegraphics[width=\textwidth]{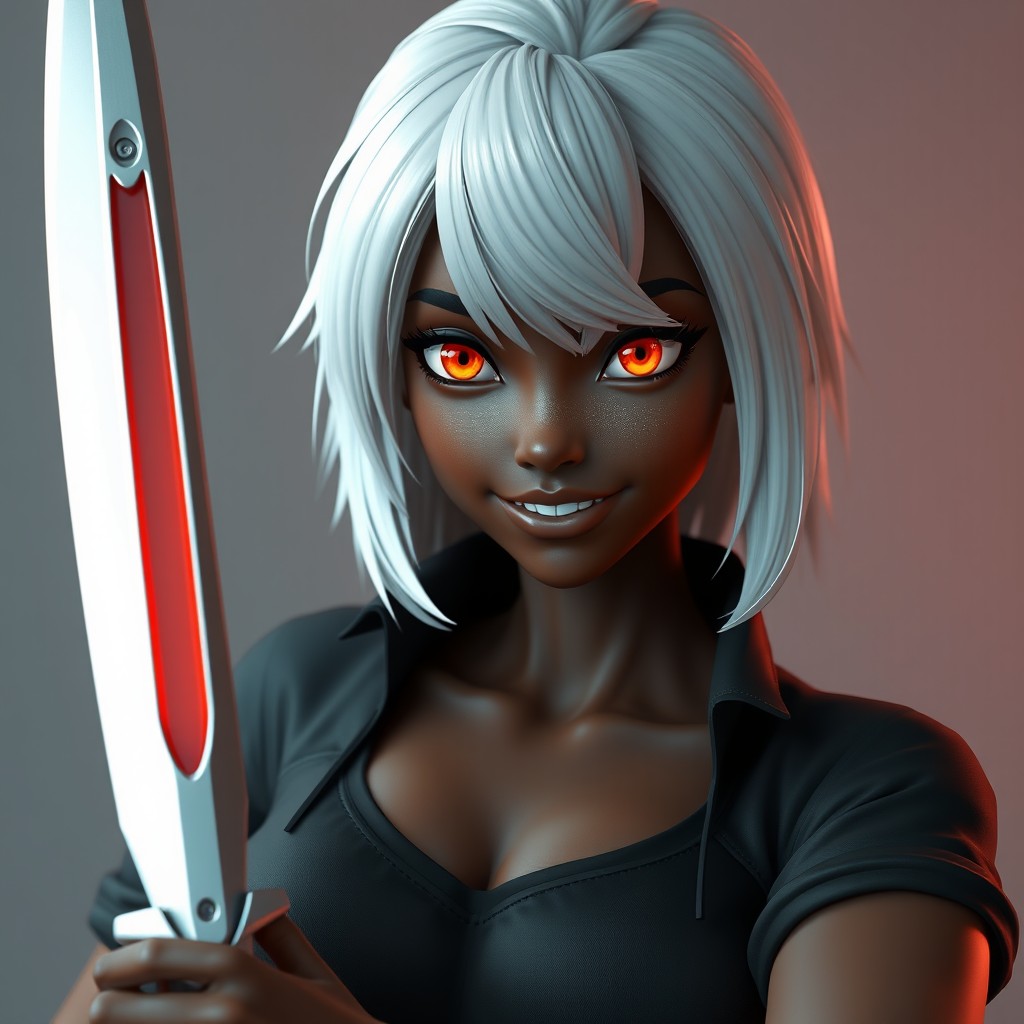}
        \caption{Dark-Skinned Female (FLUX.1)}
        \label{fig:dark_skin_FLUX}
    \end{subfigure}
    
    \caption{Comparison of generated images across Stable Diffusion XL (Row 1), Stable Diffusion 3 (Row 2), and FLUX.1 (Row 3) for the prompt structure: 
    \textit{"\textbf{extremely detailed, cg unity, 4k, wallpapermasterpiece, ultra quality, qualityultradetailedbest, illustrationbest}"}. 
    Each model's output substitutes specific elements into the shared structure, showcasing variations such as \textbf{Medaka Kurokami} (blue hair, red eyes, holding blade), \textbf{Nezuko Kamado} (pink kimono, bamboo), \textbf{Blade MK3} (blonde hair, gloves, ponytail), and a \textbf{dark-skinned female} (white hair, uniform, weapon). This illustrates how models adapt shared prompts to generate character-specific outputs.}
    \label{fig:comparison_SD_XL_SD3_Flux_dif}
\end{figure}

\end{document}